\documentclass[aps,prd,epsf,onecolumn,floatfix,showpacs,groupedaddress,nofootinbib]{revtex4} 
\usepackage{graphicx}
\usepackage{bm}
\newcommand{\lsim}{\mathrel{\vcenter{\hbox{$<$}\nointerlineskip\hbox{$\sim$}}}}
\newcommand{\gsim}{\mathrel{\vcenter{\hbox{$>$}\nointerlineskip\hbox{$\sim$}}}}

\newcommand{\oln}{\overline}

\newcommand{\eye}{{\cal I}}
\newcommand{\half}{\frac{1}{2}}
\newcommand{\egzk}{E_{\rm GZK}}
\newcommand{\rarr}{\rightarrow}
\newcommand{\sig}{\sigma^{\rm CC}_{\nu N}}

\newcommand{\sigstd}{\sigma_{31}}
\newcommand{\wint}{w_{\rm int}}
\newcommand{\wdk}{w_{\rm dk}}
\newcommand{\wdkmax}{w_{\rm dk}^{\rm max}}
\newcommand{\wth}{w_{\rm th}}
\newcommand{\Enu}{E_\nu}
\newcommand{\Esh}{E^{\rm sh}_{\rm th}}
\newcommand{\Eth}{E^\tau_{\rm th}}
\newcommand{\Etau}{E_\tau}
\newcommand{\btau}{\beta_\tau}
\newcommand{\rfov}{R_{\rm FOV}}

\newcommand{\lmin}{l_{\rm min}}
\newcommand{\dvert}{d_{\rm vert}}
\newcommand{\dhor}{d_{\rm hor}}
\newcommand{\acc}{{\cal A}cc}
\newcommand{\toothin}{``too-thin'' }
\newcommand{\dmin}{d_{\rm min}}
\newcommand{\dmax}{d_{\rm max}}

\newcommand{\zdk}{z_{\rm dk}}
\newcommand{\zint}{z_{\rm int}}
\newcommand{\zthin}{z_{\rm thin}}
\newcommand{\zell}{z_{\rm elev}}
\newcommand{\zw}{z_{\rm w}}
\newcommand{\zu}{z_U}
\newcommand{\zl}{z_L}
\newcommand{\zb}{z_{B}}
\newcommand{\ze}{z_{E}}
\newcommand{\zc}{z_{\rm cloud}}
\newcommand{\zcc}{z^{\rm crit}_{\rm cloud}}

\newcommand{\zg}{z_{\rm ground}}
\newcommand{\zuhas}{z_{U}({\rm HAS})}
\newcommand{\zlhas}{z_{L}({\rm HAS})}
\newcommand{\zbhas}{z_{B}({\rm HAS})}
\newcommand{\zehas}{z_{E}({\rm HAS})}
\newcommand{\zhath}{{\hat z}({\rm HAS})}
\newcommand{\coshatuas}{\cos{\hat \theta}_n}
\newcommand{\coshathas}{\cos{\hat \theta}_z}
\newcommand{\zuuas}{z_{U} ({\rm UAS}) }
\newcommand{\zluas}{z_{L} ({\rm UAS}) }
\newcommand{\zbuas}{z_{B} ({\rm UAS}) }
\newcommand{\zeuas}{z_{E} ({\rm UAS}) }
\newcommand{\zhatu}{{\hat z} ({\rm UAS})}
\newcommand{\zhatugd}{{\hat z}({\rm UAS\otimes G})}
\newcommand{\zhathsp}{{\hat z}({\rm HAS\otimes S})}
\newcommand{\rint}{r_{\rm int}}
\newcommand{\Rearth}{R_\oplus}
\newcommand{\rb}{r_{\rm B}}

\newcommand{\rsr}{r_{\rm sr}}
\newcommand{\rmantle}{r_{\rm m}}
\newcommand{\rc}{r_{\rm c}}
\newcommand{\rhoearth}{\rho_{\rm earth}}
\newcommand{\rhoatm}{\rho_{\rm atm}}
\newcommand{\rhosr}{\rho_{\rm sr}}
\newcommand{\rhostd}{\rho_{2.65}}
\newcommand{\rhow}{\rho_{\rm w}}
\newcommand{\rhom}{\rho_{\rm m}}
\newcommand{\rhoc}{\rho_{\rm c}}
\newcommand{\betasr}{\beta_{19}^{\rm sr}}
\newcommand{\betaw}{\beta_{19}^{\rm w}}
\newcommand{\lwsr}{L_{\rm w-sr}}
\newcommand{\lsrm}{L_{\rm sr-m}}
\newcommand{\lmc}{L_{\rm m-c}}
\newcommand{\thz}{\theta_z}
\newcommand{\thn}{\theta_n}

\newcommand{\thhor}{\theta_{\rm hor}}

\newcommand{\dnu}{d_\nu}
\newcommand{\dtau}{d_\tau}
\newcommand{\dtot}{d_{\rm tot}}
\newcommand{\zp}{z^{\prime}}
\newcommand{\thp}{\theta^{\prime}}
\newcommand{\thhorp}{\theta_{\rm hor}^{\prime}}
\newcommand{\thnp}{\theta_n^{\prime}}
\newcommand{\zpdk}{z^{\prime}_{\rm dk}}
\newcommand{\zpu}{z^{\prime}_{U}}
\newcommand{\zpl}{z^{\prime}_{L}}

\newcommand{\zpuuas}{z^{\prime}_{U} ({\rm UAS}) }

\newcommand{\zpbuas}{z^{\prime}_{B} ({\rm UAS}) }
\newcommand{\zpeuas}{z^{\prime}_{E} ({\rm UAS}) }
\newcommand{\zphatu}{{\hat z}^{\prime} ({\rm UAS}) }
\newcommand{\hsp}{HAS$\otimes$S}
\newcommand{\hgd}{HAS$\otimes$G}
\newcommand{\ugd}{UAS$\otimes$G}
\newcommand{\usp}{UAS$\otimes$S}
\newcommand{\coshass}{\cos\theta^*_{\rm S}}
\newcommand{\cosuasg}{\cos\theta^*_{\rm G}}
\newcommand{\nubar}{\bar{\nu}}
\newcommand{\numu}{\nu_\mu}
\newcommand{\nutau}{\nu_\tau}
\newcommand{\nue}{\nu_e}
\newcommand{\nuebar}{\bar{\nu}_e}
\newcommand{\bwide}{\begin{widetext}}
\newcommand{\ewide}{\end{widetext}}
\newcommand{\beq}[1]{\begin{equation} \label{(#1)}}
\newcommand{\eeq}{\end{equation}}
\newcommand{\barr}{\begin{array}}
\newcommand{\earr}{\end{array}}
\newcommand{\ba}[1]{\begin{eqnarray} \label{(#1)}}
\newcommand{\ea}{\end{eqnarray}}

\newcommand{\rf}[1]{(\ref{(#1)})}
\begin{document}
\title
{Acceptances for Space-Based and Ground-Based Fluorescence Detectors,\\
and Inference of the Neutrino-Nucleon Cross-Section above $10^{19}$
eV}
\author{Sergio Palomares-Ruiz}
\email{sergio.palomares-ruiz@vanderbilt.edu}
\affiliation{Department of Physics and Astronomy, Vanderbilt
University, Nashville, Tennessee 37235-1807}
\author{Andrei Irimia}
\email{andrei.irimia@vanderbilt.edu}
\affiliation{Department of Physics and Astronomy, Vanderbilt
University, Nashville, Tennessee 37235-1807}
\author{Thomas J. Weiler}
\email{tom.weiler@vanderbilt.edu}
\affiliation{Department of Physics and Astronomy, Vanderbilt
University, Nashville, Tennessee 37235-1807}
\date{\today}
\begin{abstract}
Detection of ultra-high energy neutrinos will be useful for unraveling
the dynamics of the most violent sources in the cosmos and for
revealing the neutrino cross-section at extreme energy. If there
exists a Greisen-Zatsepin-Kuz'min (GZK) suppression of  
cosmic-ray events above $\egzk\sim 5 \times 10^{19}$~eV, as predicted
by theory, then the only messengers of energies beyond $\egzk$ are
neutrinos. Cosmic neutrino fluxes can initiate air-showers through
interaction in the atmosphere, or in the Earth.  Neutrino trajectories
will be downgoing to nearly horizontal in the former case, and
``Earth-skimming'' in the latter case. Thus it is important to know
the acceptances (event rate/flux) of proposed air-shower experiments 
for detecting both types of neutrino-initiated 
events. We calculate these acceptances for fluorescence detectors,
both space-based as with the EUSO and OWL proposals, and ground-based,
as with Auger, HiRes and Telescope Array. The neutrino cross-section
$\sig$ is unknown at energies above $5.2 \times 10^{13}$~eV. Although
the popular QCD extrapolation of lower-energy physics offers the
cross-section value of $0.54 \times 10^{-31}\,(E_\nu/10^{20}{\rm
  eV})^{0.36}\,{\rm cm}^2$, new physics could raise or lower this
value. Therefore, we present the acceptances of horizontal (HAS) and
upgoing (UAS) air showers as a function of $\sig$ over the range
$10^{-34}$ to $10^{-30}$ cm$^2$. The dependences of acceptances on
neutrino energy, shower-threshold energy, shower length, and shower
column density are also studied. We introduce a cloud layer, and study
its effect on rates as viewed from space and from the ground. For UAS,
we present acceptances for events over land (rock), and over the ocean
(water). Acceptances over water are larger by about an order of
magnitude, thus favoring space-based detectors. We revisit the idea of
Ref.~\cite{KW} to infer $\sig$ at $E_\nu\gsim 10^{20}$ from the ratio
of HAS-to-UAS events, and obtain favorable results. Included in our
UAS calculations are realistic energy-losses for taus, and
Earth-curvature effects. Most of our calculation is analytic, allowing
insight into the various subprocesses that collectively turn an
incident neutrino into an observable shower.
\end{abstract}
\maketitle

\section{Introduction and Purpose}

Detection of ultra-high energy ($E_\nu > 10^{18}$~eV~$\equiv$~EeV)
neutrinos is important for several reasons. First of all, neutrino
primaries are not deflected by magnetic fields and so should point
back to their cosmic sources. This contrasts with cosmic-rays, which
are charged and follow bent trajectories. Secondly, well above the
Greisen-Zatsepin-Kuzmin (GZK) energy of $\egzk\sim 5\times
10^{19}$~eV~\cite{G66,ZK66}, they may be the only propagating
primaries. As such, they may be the only messengers revealing the
ultimate energy-reach of extreme cosmic accelerators, generally
believed to be powered by black holes. Above $\egzk$, the GZK
suppression~\cite{G66,ZK66,Zayyad1,Letessier1} of cosmic-rays results
from the resonant process $N+\gamma_{\rm CMB}\rarr \Delta\rarr N
+\pi$; $\egzk$ is the lab-frame energy corresponding to the kinematic
threshold $\sqrt{s}=M_\Delta$ for excitation of the intermediate
$\Delta$ resonance. A handful of cosmic-ray events have been detected
with estimated energies exceeding $10^{20}$~eV. The record energy is
the famous Fly's Eye event at $3\times 10^{20}$~eV~\cite{FlysEye}. The
observable neutrino spectrum could extend to much higher
energies. Thirdly, in contrast to cosmic-rays and photons, neutrinos
are little affected by the ambient matter surrounding the central
engines of Nature's extreme accelerators. Accordingly, neutrinos may
carry information about the central engine itself, inaccessible with
other primaries. In principle, neutrinos may be emitted from close to
the black hole horizon, subject only to energy-loss due to
gravitational redshifting. An analogy can be made to solar studies
performed with photons versus neutrinos. The photons are emitted from
the outer centimeter of the Sun's chromosphere, while the neutrinos
are emitted from the central core where fusion powers the
Sun. Fourthly, neutrinos carry a quantum number that cosmic-rays and
photons do not have - flavor. Neutrinos come in electron, muon, and
tau flavors. One may think of this ``extra'' flavor degree of
information as the neutrino's superb analog to polarization for the
photon, or nucleon number $A$ for the cosmic-ray. Each of these
attributes, flavor, polarization, and nucleon number, carries
information about the nature and dynamics of the source, and about the
environment and pathlength of the inter-galactic journey. The flavor
ratios of cosmic neutrinos are observable~\cite{flavorID}. Several
papers have recently analyzed the benefits that neutrino flavor
identification offers for unraveling the dynamics of cosmic
sources~\cite{pimuchain}. The fifth reason why ultra-high energy
neutrino primaries traveling over cosmic distances are interesting is
that such travel allows studies of the fundamental properties of
neutrinos themselves. For studying some properties of the neutrino,
such as neutrino stability/lifetime~\cite{nulifetime}, pseudo-Dirac
mass patterns~\cite{pDirac}, it is the cosmic distance that is
essential; for other properties, it is the extreme energy that is
essential. A clear example of the latter is any attempt to determine
the neutrino cross-section at energies beyond the reach of our
terrestrial accelerators.

In this paper we will examine the potential for cosmic-ray experiments
designed to track ultra-high energy air-showers by monitoring their
fluorescence yield~\cite{fluoresence}, to detect horizontal
air-showers (HAS) and upgoing air-showers (UAS) induced by a cosmic
neutrino flux. We will also study the ability of these experiments to
infer the neutrino-nucleon cross-section $\sigma_{\nu N}$ at energies
above $10^{19}$~eV, from the ratio of their UAS and HAS events. Such
energies are orders of magnitude beyond the energies accessible to
man-made terrestrial accelerators. From the point of view of QCD, such
a cross-section measurement would be an interesting microscope into the
world of small-x parton evolution. The neutrino cross-section above
$10^{19}$~eV could agree with any of the various QCD-motivated
extrapolations that have been published~\cite{GQRS,sigma20}, or
not. The cross-section could also be quite different than the
extrapolations. For example, if a new threshold is crossed between
terrestrial neutrino energies $\sim 100$~GeV, and the extreme energies
reached by cosmic-rays, ~$\sim 10^{11}$~GeV, then the cross-section
could much exceed the QCD-extrapolations. On the other hand,
saturation effects can significantly reduce the total cross-section at
these very high energies~\cite{saturation}. The nine orders of
magnitude increase in lab energy reach corresponds to 4.5 orders of
magnitude increase in center-of-momentum energy reach. Even the
center-of-momentum energy at the $e-p$~HERA collider is more than
three orders of magnitude below the cosmic-ray reach. This remarkable
energy reach of cosmic-rays presents ample room for new physics beyond
our Standard Model. Proposals for new physics thresholds in this
energy region include low-scale unification with gravity, in which
neutrino-nucleon scattering produces mini-black holes~\cite{miniBH}
and/or brane-wraps~\cite{branewraps}, non-perturbative electroweak
instanton effects~\cite{EWinstanton}, compositeness
models~\cite{composite}, a low energy unification scale in string 
inspired models~\cite{string}, and Kaluza-Klein modes from compactified
extra dimensions~\cite{extradimensions}. All of these models produce a
strongly-interacting neutrino cross-section above the new threshold. 
Dispersion relations allow one to use low-energy elastic scattering to
place constraints on the high-energy
cross-section~\cite{disprelation}, but the constraints are quite weak.

For HAS and UAS, we provide analytical calculations of the event-rate
to flux ratio as a function of $\sigma_{\nu N}$. This ratio is known as
the ``instantaneous experimental acceptance'', with units of {\sl 
area$\times$solid angle}. The time-averaged acceptance includes an
experimental ``duty factor,'' the fraction of time that the experiment
is functioning. We will not include the duty factor in our calculations
of acceptances. We note that acceptances are also sometimes called 
``apertures.''

Experimental acceptance offers a very meaningful 
figure of merit for statistical reach. One has merely
to multiply an experiment's acceptance by Nature's flux to arrive at
an event rate for the experiment. Multiplying again by the
experiment's run time (including the duty factor), one obtains the 
total number of events.  Acceptance times run time is termed the
experimental ``exposure''.

The acceptances we calculate are scalable to large area experiments
such as HiRes, Auger, and in the near future Telescope Array, which are
anchored to the ground, and to super-large area experiments such as
EUSO and OWL, which are proposed to orbit the Earth from space. A
horizontal shower, deeply initiated, is the classic signature for a
neutrino primary. The weak nature of the neutrino cross-section means
that horizontal events begin where the atmospheric target is most
dense, low in the atmosphere.  In contrast, the ultra-high energy $pp$
cross-section exceeds 100~mb, so the air-nucleon cross-section exceeds
a barn! Even the vertical atmospheric column density provides hundreds
of interaction lengths for a nucleon, and so the cosmic-ray interacts 
high in the atmosphere. The weak nature of the neutrino cross-section
also means that the event rate for neutrino-induced HAS is
proportional to the neutrino-nucleon cross-section.

For an neutrino-induced UAS, the dependence on neutrino cross-section
is more complicated, and more interesting. The Earth itself is opaque
for neutrinos with energies exceeding about a PeV of energy.  However,
``Earth-skimming'' neutrinos, those with a short enough chord length
through the Earth, will penetrate and exit, or penetrate and interact.
In particular, there is much interest in the Earth-skimming process
$\nutau\rarr\tau$ in the shallow Earth, followed by $\tau$ decay in
the atmosphere to produce an observable shower. In Ref.~\cite{KW} it
was shown that the rate for the Earth-skimming process
$\nutau\rarr\tau$ is {\sl inversely} proportional to $\sigma_{\nu
  N}$. There it was emphasized that $\sigma_{\nu N}$ could be inferred
from a measurement of the ratio of HAS to UAS rates.~\footnote{The
  prospects of inferring the neutrino-nucleon cross-section in the
  energy range of 100 TeV - 100 PeV at neutrino telescopes such as
  IceCube, were studied in Ref.~\cite{Icecubecs}; prospects at higher
  energies were studied in Ref.~\cite{Augercs} for the 
  Auger observatory.} Of course, an implicit assumption is that there
is enough neutrino flux at extreme energies to generate HAS and UAS
event samples.

The inverse dependence of UAS rate on $\sigma_{\nu N}$ is broken by
the $\tau\rarr$~{\sl shower} process in the atmosphere. As the
cross-section decreases, the allowed chord length in the Earth
increases, and the tau emerges with a larger angle from the Earth's
tangent plane. This in turn provides a smaller path-length in air in
which the tau may decay and the resulting shower may evolve. This
effect somewhat mitigates the inverse dependence of the UAS on
$\sigma_{\nu N}$.

Ref.~\cite{KW} provided an approximate calculation of the whole UAS
process, and gave an approximate result for the dependence of the
HAS/UAS ratio on $\sigma_{\nu N}$. In this work, we improve upon
Ref.~\cite{KW} in several ways. We include the energy dependences of
the tau energy-losses in the Earth, and of the tau lifetime in the 
atmosphere. For the energy-losses, we distinguish between tau
propagation in earth rock and propagation in ocean water. These
calculations are carried out in Section~\ref{sec:rates}. On the issue
of shower development, we incorporate the dependence of atmospheric
density on altitude.  We also impose requirements on the resulting
shower such that a sufficiently long visible shower-length is
projected onto the Earth's tangent plane, thus meeting experimental
requirements for visibility. This is done in
Section~\ref{sec:visibleAS}. In the case of the upgoing showers, the 
pathlength of the pre-decayed tau may be so long that the Earth's
curvature enters into the altitude dependence. We include the
non-negligible correction from curvature in
Section~\ref{sec:visibleAS}. We include the partial loss of visibility
due to high cirrus or low cumulus cloud layers in
Section~\ref{sec:clouds}. It is estimated that clouds will obscure the
viewing area about 60-70\% of the time. For ground-based observation,
it is mainly  the low-lying cumulus clouds that limit visibility. For
space-based observation, it is mainly the high cirrus clouds that
limit visibility.~\footnote{In fact, low-lying cumulus clouds may aid
  in HAS identification for space-based observing. When the HAS hits
  the cloud layer, diffuse reflection of the forward Cerenkov cone can
  be seen as a one-time ``\v{C}erenkov flash''. The time of the flash
  and the measured height of the cloud then provide the absolute (t,z)
  coordinates of the shower.} And in
Section~\ref{sec:cloudsncurvature}, we combine the corrections from  
clouds with that from the Earth's curvature.

Our results are illustrated in a series of plots of acceptances, for
ground-based and space-based experiments, versus neutrino-nucleon
cross-section, in Section~\ref{sec:results}. Situations with and
without cloud layers are analyzed, as are events over solid earth and
over the ocean. Incident neutrino energies, energy thresholds for
experimental detection of the air-shower, and various shower-trigger
parameters are varied.  Earth-curvature effects are included 
in our UAS calculations. These reduce the event rate. Next comes the
discussion Section~\ref{sec:discussion}. It presents several small
issues, and includes a comparison of our work with prior work. A final
Section recaps our conclusions. Some of the more tedious but necessary
formulas are derived in an Appendix.

The reader who believes that a picture (or four) is a worth thousand
words may wish to jump to Section~\ref{sec:results}. Such a reader
especially may find it useful to reference
Tables~\ref{table:variables} and \ref{table:parameters}, where the
variables and parameters are defined.

Among our conclusions, we find that the HAS/UAS ratio is or order of
unity for cross-section values very near to the commonly extrapolated 
value of $0.5\times 10^{-31}\,{\rm cm}^2$ at $E_\nu\sim 10^{20}$~eV. 
This is fortunate, for it offers the best possibility that both HAS
and UAS rates can be measured, and a true cross-section inferred. We
display our HAS and UAS acceptance plots for a cross-section range
from superweak $10^{-34}\,{\rm cm}^2$ to a microbarn, $10^{-30}\,{\rm
  cm}^2$. This range includes the QCD-extrapolations of $\sig$, and
the region of the HAS/UAS cross-over. It also encompasses any effects
of new neutrino physics, either increasing  or decreasing $\sig$. The
highest energy for which the neutrino cross-section has been measured
is that at the HERA accelerator. The measurement is $\sig\sim 2 \times
10^{-34} \,\rm{cm}^2$ at $\sqrt{s}=314$ GeV~\cite{HERAcs}, the latter
corresponding to an energy on fixed target of $5.2 \times 10^{13}$~eV
(52~TeV). It is hard to imagine that $\sig$ at $10^{20}$~eV would not
have grown beyond the HERA value. Even so, the acceptances shown for
superweak cross-sections may have some relevance to a possible WIMP
flux~\cite{WIMPaccept}. Modeling of a WIMP event rate requires 
modifications in the shower development for HAS, and in the chain 
WIMP$\rarr$UAS, that we do not pursue here.

\begin{table}
\begin{tabular}{|l|l|} \hline
$L$ & chord length of $\nu$ trajectory through Earth 	\\ \hline
$\zint$ & vertical height (depth) of HAS (UAS) $\nu$ interaction \\
\hline 
$\zdk$ & altitude of upgoing $\nutau$ decay (no Earth-curvature) \\
\hline 
$\zpdk$ & altitude of upgoing $\nutau$ decay including
Earth-curvature\\ \hline 
$z_U$ & maximum visible shower altitude (HAS $\neq$ UAS)\\ \hline
$z_L$ & minimum visible shower altitude (HAS $\neq$ UAS)\\ \hline
$\zb$ & altitude where shower first attains threshold brightness 
  (HAS $\neq$ UAS)\\ \hline
$\ze$ & altitude where shower extinguishes (HAS $\neq$ UAS)\\ \hline
$\zcc$ & critical altitude for suppression from cloud layer\\ \hline
$\zpbuas$ & altitude where UAS attains threshold brightness,
  including Earth-curvature \\ \hline
$\zpeuas$ & altitude where UAS extinguishes, 
  including Earth-curvature \\ \hline
$\thz$ & zenith angle of HAS event   		\\ \hline
$\thn$ & nadir angle of UAS event (no Earth-curvature)\\ \hline
$\thhor=\frac{\pi}{2}-\theta_z$ & horizontal angle of UAS event\\
\hline 
$\thnp$ & nadir angle of UAS event including Earth-curvature \\ \hline
$\thhorp$ & horizontal angle of UAS event including Earth-curvature\\
\hline 
$\dtot$& total column along chord of Earth 		\\ \hline
$\dnu$ & column density of $\nu$ in the Earth		\\ \hline
$\dtau$& column density of $\tau$ in the Earth		\\ \hline
$\coshass$ & minimum shower angle, cloud-dependent,
	for space-observatory				\\ \hline
$\cosuasg$ & minimum shower angle, cloud-dependent,
	for ground-observatory				\\ \hline
$\zhath$ & maximum altitude from which initiated HAS
	can reach the ground				\\ \hline
$\zhatu$ & minimum altitude from which initiated UAS
	can reach $\zthin$				\\ \hline
$\zphatu$ & $\zhatu$ with Earth-curvature included      \\ \hline
$\zhatugd$ & $\zhatu$ modified for cloud layer above 	\\ \hline
$\zhathsp$& $\zhath$ modified for cloud layer below 	\\ \hline
\end{tabular}
\caption{\label{table:variables} List of variables and their meaning.
(Conversion between variables $z$ and $w$ is given by $w\cos\theta
=z$.)}
\end{table}

\begin{table}
\begin{tabular}{|l|l|l|} \hline
$h$ & scale height of the atmosphere & 8~km 	\\  \hline 
$\zg$ & ground altitude, kept as a symbol for later substitutions & 0
\\ \hline 
$\zthin$ & altitude beyond which air is too thin to fluoresce
significantly  
	& $3h$ \\ \hline
$\beta_{19}$& tau energy-attenuation constant at $E_\tau=10^{19}$~eV 
  &1.0 (0.55) $\times 10^{-6}\,{\rm cm}^2/{\rm g}$ for rock (water) \\
\hline 
$\alpha$ & exponent of the energy-dependence of $\beta_\tau$ & 0.2 \\
\hline 
$\dvert$& vertical atmospheric column density & 1,030~g/cm$^2$ \\
\hline
$\dhor$ & horizontal atmospheric column density & 36,100~g/cm$^2$ \\
\hline 
$\dmin$ & minimum acceptable shower column density & 
	300, {\bf 400}~g/cm$^2$ \\ \hline
$\dmax$ & maximum shower column density at extinction & 
	{\bf 1200}, 1500~g/cm$^2$\\ \hline
$\lmin$ & minimum acceptable shower length 
	projected on the Earth's surface & {\bf 10~km}, 5~km 	\\
\hline 
$\rfov$ & radius (or half-scale) of the experimental 
	field of view & 230~km 					\\
\hline 
$\zw$ & depth of ocean &  3.5~km \\ \hline
$\zc$ & altitude of cloud layer & {\bf 2}, 4, 8, 12~km \\ \hline
$E_\nu$ & incident neutrino energy & ${\bf 10^{20}}$, $10^{21}$~eV;
\\ \hline
$\Esh$ & detector threshold energy & 
	$10^{19}$, ${\bf 5\times 10^{19}}$ 
\\ \hline
$\Eth$ & tau threshold energy & $\frac{3}{2}\,(3)\times\Esh$ for
hadron (electron) mode \\  
\hline 
$\sig$ & neutrino (or WIMP) cross-section & $10^{-30}$,
	${\bf 10^{-31}}$, $10^{-32}$, $10^{-33}{\rm cm}^2$  \\ \hline
\end{tabular}
\caption{\label{table:parameters} List of parameters, their meaning, 
and their chosen value(s); the bold-faced value is the chosen
``canonical'' value.} 
\end{table}

\begin{table}
\begin{tabular}{|l|l|} \hline
HAS  & Horizontal Air-Shower 			\\ \hline
UAS  & Upgoing (``Earth-skimming'') Air-Shower 	\\ \hline
\hsp & HAS seen from space-based observatory 	\\ \hline
\usp & UAS seen from space-based observatory 	\\ \hline
\hgd & HAS seen from ground-based observatory 	\\ \hline
\ugd & UAS seen from ground-based observatory 	\\ \hline
\end{tabular}
\caption{\label{table:symbols} List of symbols and their meaning.}
\end{table}

\section{Air-Shower Rates}
\label{sec:rates}

The variables and parameters needed to describe UAS and HAS are
sufficiently numerous that we have collected many of them in 
Table~\ref{table:variables} and~\ref{table:parameters} for easy
reference. In Table~\ref{table:symbols} we explain the different
symbols used throughout this work. Many of the variables are best
explained by the three schematic diagrams in
Figs.~\ref{fig:schem1}--~\ref{fig:curvature}.

\subsection{Upgoing Air-Showers (UAS)}
\label{sec:UASrate}

Ultra-high energy (UHE) neutrinos are expected to arise from  the
decay of pions and subsequently muons produced in astrophysical
sources~\cite{pimuchain}. For this decay chain, the flavor mix at the
source is $\nue:\numu=1:2$. The maximal mixing between $\numu$ and
$\nutau$ inferred from terrestrial oscillation experiments then leads,
after propagation for many oscillations lengths and to a very good
approximation, to a flavor ratio at Earth of
$\nue:\numu:\nutau=1:1:1$, i.e., flavor democracy. Thus, a detector  
optimized for $\nue$ or $\numu$ or $\nutau$ can expect a healthy
signal from cosmic neutrinos.

It is useful to define the neutrino charged-current (CC) interaction
mean free path (MFP) as
\beq{MFP}
\lambda_{\nu} = \frac{1}{\sig\;{\oln \rho}} = \frac{63\,{\rm
    km}}{\rhostd\; \sigstd}
\eeq
where ${\oln \rho}$ is the mean number-density of the target-matter,
and $\rhostd$ is the mean density in units of the value for surface
rock $\rhosr= 2.65~{\rm g/cm}^3$.   Density is usually
expressed in units of g/cm$^3$, with the multiplicative factor of
$N_{\rm A}=6.022\times 10^{23}\,{\rm g}^{-1}$ implicitly
understood. The mean density of ocean water is $1.0~{\rm g/cm}^3$. 
The cross-section $\sigstd$ is the CC cross-section 
in units of $10^{-31}\,{\rm cm}^2$.
The commonly used high-energy
neutrino-nucleon CC cross-section extrapolated from QCD~\cite{GQRS} is
$0.54\times 10^{-31}\,(\Enu/10^{20}{\rm eV})^{0.363}{\rm cm}^2$.

We will ignore the NC contribution to the neutrino MFP for three
reasons. First, the NC cross-section is expected to be small compared
to the CC cross-section, as it is known to be at the lower energies of 
terrestrial accelerators. Secondly, the NC interaction does not
absorb the neutrino, but rather lowers the energy of the propagating
neutrino by a small amount; in the SM, the energy loss is only
$\langle y\rangle\sim$~20\%. And thirdly, the increase in complexity
of our calculation, when the NC MFP is included, seems unwarranted.
We also ignore multiple CC interactions due to the ``tau regeneration''
decay chain $\nutau\rarr\tau\rarr\nutau$. Here, it is the long decay
length of the tau that results from production at  $E_\nu >
10^{17}$~eV that makes tau regeneration negligible.

\begin{figure}
\includegraphics[width=.7\textwidth]{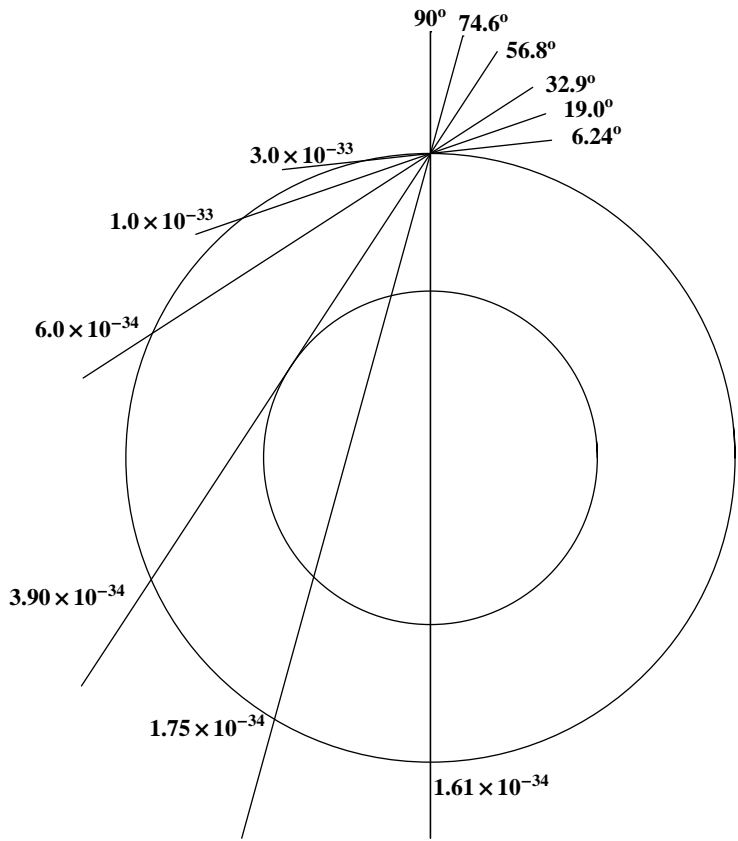}
\caption{ \label{fig:nuchord} 
  Shown are neutrino trajectories for which the interaction MFP
  matches the chord length through the Earth. The various trajectories
  are parameterized values of the neutrino cross-section. Also shown
  is the trajectory's angle with respect to horizontal.
}
\end{figure}

In Fig.~\ref{fig:nuchord} we show an interesting relation between the 
neutrino cross-section, the neutrino's MFP in the Earth, and roughly
speaking, the maximum horizontal angle for which the neutrino may
transit the Earth. 

In this figure, the Earth has been approximated according to the
two-shell model. There is a central core with mean density 12~g/cm$^3$
out to a radius of 3485.7~km, and a mantle with mean density
4.0~g/cm$^3$ out to the the Earth's radius of $\Rearth=6371$~km. The
point of this figure is that, although the Earth is marginally
transparent for neutrinos with the HERA cross-section of $2\times
10^{-34}\,{\rm cm}^2$, the Earth quickly becomes opaque at larger
cross-section. For the cross-section values extrapolated to $\sim
10^{20}$~eV, horizontal angles are very small, and the trajectories
are truly ``Earth skimming'' (cf.\ Eq.~\rf{thetahor}).

The tau energy-attenuation length is $\lambda_{\tau} = (\btau\,{\oln
  \rho})^{-1}$, where $\btau(E)$ is the coefficient giving a scale to
  tau energy loss:
\beq{dEdx}
d\Etau/dx= - \btau(E)\,\rho\,\Etau\,.
\eeq
The coefficient $\btau(E)$ is weakly energy-dependent. For the
energies of interest, $E_\nu > 10^{18}$ eV, tau energy losses are 
dominated by photo-nuclear processes, being the electromagnetic
mechanisms of ionization, bremsstrahlung and $e^+$-$e^-$ pair
production negligible~\cite{losses}. We find that the recent
calculations of $\beta_{\tau}(E)$~\cite{DRSS,BS,KLS} are well
fitted in the energy region of interest by a simple power
law~\cite{Bertou1}:~\footnote{In a very recent paper~\cite{DHR05}, a  
  logarithmic fit to $\beta_{\tau}(E)$ is presented. We find that our
  fit agrees quite well with that one in the region of our interest,
  $10^{18}\, \rm{eV} \le E_\nu \le 10^{21}\, \rm{eV}$. For our
  purposes, the power law fit is more useful in that it allows
  analytic integration of some energy-dependences.}
\beq{beta}
\btau(E)=
\beta_{19}\,\left(\frac{E_\tau}{10^{19}{\rm eV}}\right)^\alpha\,,
	\quad \alpha=0.2\,,
\eeq
with the constant pre-factor $\betasr=1.0\times 10^{-6}\,{\rm
cm}^2/{\rm g}$ for surface rock ($\langle A\rangle=22$, $\langle 
Z\rangle=11$), and $\betaw=0.55\times 10^{-6}\,{\rm cm}^2/{\rm g}$
for water ($\langle A\rangle=11.9$, $\langle Z\rangle=6.6$);
$\beta_{\tau}(E)$ scales as $\langle A\rangle$. The tau
energy-attenuation length at $\Etau=10^{19}$~eV is
$\lambda_\tau=3.8$~km in surface rock, and 18~km in water. The tau
decay MFP is $c\tau_{\tau} = 490\,(E_{\tau}/10^{19}$eV) km. For taus 
with energies at and above $10^{18}$ eV, the decay MFPs are much
longer than the energy-attenuation length. In this paper, we safely 
neglect the small probability of decay within the Earth for those taus
which would otherwise emerge from the Earth with energy above
$10^{18}$ eV. We have checked that the results we present in this work
are reduced by less than a few percent when the tau decay probability 
within the Earth is included. 

The muon energy-attenuation length is 7 times smaller than that of
the tau, and the electron energy-attenuation length is many times
smaller again (the $\mu$ decay length is $\sim 10^8$ times longer than
that of  the tau). Because the energy-attenuation length for a tau is
an order of magnitude longer than that of a muon, UAS events are
dominantly initiated by the CC interaction of tau neutrinos. 

The ratio of the tau energy-attenuation length to the neutrino MFP
$\lambda_\tau/\lambda_\nu = N_A \sig / \btau \sim (\sig/10^{-31}
\rm{cm}^2) \times 0.06\,(0.11)$ for rock (water), is independent of 
$\rho$ and only weakly dependent on tau energy. For $\sig \lsim 2
\times 10^{-30}{\rm cm}^2$, we expect most of the path length in Earth
(rock or water) to be neutrino; for $\sig \gsim 2\times 10^{-30}{\rm
  cm}^2$, we expect most of the path length to be tau. In detail, this
remark will also depend on the direction of the initial neutrino
through the total chord length, and on the threshold energy of the
detector (the minimum record-able tau energy).

\begin{figure}
\includegraphics[width=.56\textwidth]{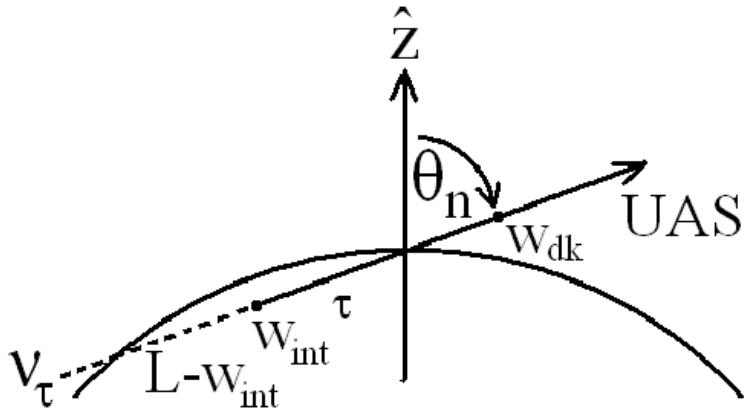}
\caption{\label{fig:schem1} 
  Coordinates describing UAS.
}
\end{figure}

Consider a tau produced in the Earth along a chord on the trajectory
of an incoming $\nu_{\tau}$. Label the chord length by $\mathit{L}$
and the distance between the interaction and the Earth's surface by
$\wint$, as shown in Fig.~\ref{fig:schem1}. In general, we will use
the variable $w$ to represent distance or location along the lepton
trajectory; when needed, $z$ will label the distance normal to the
Earth's surface, and $x$ and $y$ will label the Earth's tangent plane.
The chord length $L$ and the nadir-angle $\thn$ of the upgoing
neutrino trajectory characterize the same degree of freedom, coupled 
together by the geometric relation $L=2\,R_\oplus\,\cos\thn$, where
$R_\oplus=6371$~km is the radius of the Earth. 

We follow the calculation of the rate of UAS events as given in
Ref.~\cite{KW}, although with some important improvements. The rate is
\beq{UASrate}
R_{\nutau} ({\rm UAS}) = F_{\nu_{\tau}} \pi A \int_0^{2 R_{\oplus}} 
\frac{\mathit{L}\,d\mathit{L}} {2 R^2_{\oplus}}
\,P_{\nu_\tau\rightarrow\tau}(\mathit{L}) \otimes\, 
P_{\rm dk}(\mathit{L})\,, 
\eeq
where $P_{\nu_{\tau} \rightarrow \tau} (\mathit{L})$ is the
probability for a $\nu_{\tau}$ along a chord of length $\mathit{L}$ in 
Earth to produce a tau which exits the surface with a given threshold
energy $E_{\rm th}$, $P_{\rm dk}(\mathit{L})$ is the probability of
decay for the tau emerging into the air, and $F_{\nutau}$ is the
$\nutau$ differential flux in the usual dimensional units of $({\rm
  energy\cdot time\cdot area\cdot steradian})^{-1}$. The $\otimes$ 
symbol denotes coupling between the two probabilities, as shown in
detail below. In Eq.~\rf{UASrate} we have not considered the
possibility of the tau decaying inside the Earth, which for the
energies of interest in this study is a very good approximation. The
detector's field of view (FOV) is $A=\int dx\int dy$. In general, the
FOV includes some detector-efficiency weighting for shower
identification, as explained in Section~\ref{sec:visibleAS}. The 
operator $\pi A\int L\,dL/2\,R_\oplus^2$ is a convenient rewriting of
the integrals $\int d{\vec A}\cdot{\hat n}\int d\Omega$, obtained when
use is made of the relation $L=2\,R_\oplus\,\cos\thn$. The angular
dependence of the interaction and decay probabilities are therefore
implicit in the $L$-dependence.

The interaction probability is given by 
\beq{Pnutotau}
P_{\nu_{\tau} \rightarrow \tau} (\mathit{L}) = 
\int_0^{\min\{\mathit{L}\,,\wth\}}
\frac{d\wint}{\lambda_{\nu}(\wint)}\;
e^{-\sig \,\dnu(L,\wint)}\,,
\end{equation}
where $d\wint/\lambda_{\nu}$ is the probability for neutrino
conversion into tau lepton in the interval $[\wint, \wint + d\wint]$, 
and the exponential gives the survival probability of the neutrino
to reach the interaction point $\wint$. The column density traversed
by the neutrino is given by
\beq{dnu}
\dnu=\int^L_{\wint} dw\,\rhoearth (w)\,.
\eeq

If the density $\rhoearth$ were a constant, the exponential in
Eq.~\rf{Pnutotau} would be simply $e^{-(L-\wint)/\lambda_\nu}$, with 
$\lambda_\nu^{-1} =\sig \,\rhoearth$. Such will be the case
if the absorption in the Earth limits chord lengths to just the outer
layer of Earth-matter (water or surface rock). The angle of the
trajectory above the horizon is related to the chord length as
$\sin\thhor=L/2\,\Rearth$. Setting the chord length equal to the
neutrino MFP $\lambda_\nu$, we get for the typical angle
\beq{thetahor}
\thhor \simeq (2\,\Rearth\,\sig \,\rhoearth)^{-1}
= 0.28^\circ\times\sigma^{-1}_{31}\,\left(\frac{\rhosr}{\rho}\right)\,.
\eeq

A commonly quoted extrapolation for the neutrino CC cross-section is
$\sigma_{31}=0.55$ at $10^{20}$~eV~\cite{GQRS}. Comparisons with the
``critical'' angles delimiting the various density boundaries in the
Earth, given in Table~(\ref{table:critical}) in the Appendix, then
reveals that over ocean, $\sig \gsim 4 \times 10^{-32}{\rm cm}^2$ 
gives rise to events whose trajectories were dominantly in only water;
and over land, $\sig \gsim 10^{-33}{\rm cm}^2$ gives rise to events
whose trajectories were dominantly in only surface rock (as opposed to
mantle or core). For these events, the Earth density is approximately a
constant. We also study smaller cross-sections, for which the density
is not constant along the path-integral. In the Appendix we present
our general calculation of $\dnu$.

The bound $\wth(\thn)$ on the depth of $\wint$ integration in
Eq.~\rf{Pnutotau} is determined by the requirement that the tau emerge 
from the Earth with sufficient energy, $\Eth$, to produce air showers
which trigger the detector apparatus. In general, $\wth$ is
angle-dependent because the density in the Earth is
angle-dependent. The mean energy of the tau emerging from the Earth is
obtained by integrating Eq.~\rf{dEdx}. The result is~\footnote{The
  $\alpha\rightarrow 0$ limit of the $\ln$ of the denominator in
  Eq.~\rf{Etau} is easily seen to be $\eye$, and so the
  $\alpha\rightarrow 0$ limit of $E_\tau$ is $E_0\,e^{-\eye}$.}
\beq{Etau}
E_{\tau}(\wint)=\frac{E_0}{\left[1+\alpha\,\eye (\wint )\,
\left(\frac{E_0}{10^{19}{\rm eV}}\right)^\alpha\right]^{1/\alpha}}\,,
\eeq
where $E_0=(1 - \langle y \rangle)\,E_{\nu}$ is the mean energy of a
tau created by an incoming neutrino with incident energy $E_{\nu}$,
and $\langle y \rangle$ is the average inelasticity parameter which we
will take as $\langle y \rangle = 0.2$~\cite{GQRS96,Aramo}.  Thus, we
take $E_0 = 0.8 \, E_\nu$. We define $\eye(\wint)$ as the
dimensionless tau ``opacity'' from point of production to the Earth's
surface, {\sl normalized to a tau with $E=10^{19}$~eV}, 
\beq{eye}
\eye(\wint )=\int_0^{\wint } dw\,\beta_{19}(z)\,\rhoearth(z)\,.
\eeq
This definition allows isolation of the energy-dependence of
$\beta_\tau (E)$  in a separate factor, evident in Eq.~\rf{Etau}. Note
that both $\beta_{19}$ and $\rhoearth$ in \rf{eye} depend on the 
Earth's composition (e.g., water versus rock), which in general
changes with depth $z$. For UAS rising from land, there is no
$z$-dependence: $\rhoearth=\rhosr$ and $\beta_{19}=1.0\times
10^{-6}{\rm cm}^2/{\rm g}$ are fixed, and the tau opacity is simply 
the path length in units of the tau energy-attenuation length at $E_0
= 10^{19}$ eV, $\wint/\lambda_{\tau}(10^{19}{\rm eV})$, with
$\lambda_{\tau} (10^{19}{\rm eV})=(\beta_{19}\,\rhosr)^{-1}$. On the
other hand, for UAS rising from the oceans, there is a discontinuity
at the ocean's bottom: $\rhoearth$ comprises two contributions, one
from ocean water and the other from the underlying rock. In the
Appendix, we show the calculation of $\eye$ for this case. We will
take the depth of the ocean $\zw$ to be 3.5~km.

Setting Eq.~\rf{Etau} equal to $\Eth$, one obtains
\beq{wth1}
\eye (\wth (\thn))=\frac{1}{\alpha}\,
\left[
\left(\frac{10^{19}{\rm eV}}{\Eth}\right)^\alpha
	-\left(\frac{10^{19}{\rm eV}}{E_0}\right)^\alpha
\right]
\eeq
as the equation defining the integration maximum
  $\wth$.~\footnote{Writing $(\Eth)^{-\alpha}$ as
  $e^{-\alpha\,\ln\Eth}$, and similarly for $E_0^{-\alpha}$, one 
  readily finds the $\alpha\rightarrow 0$ limit of Eq.~\rf{wth1} to be
  $\eye (\wth (\theta_z))=\ln\,(E_0/\Eth)$.} For UAS rising from land,
$\beta_{19}\,\rhosr$ is constant and the integration and inversion of
Eq.~\rf{wth1} to get $\wth$ is trivial. For UAS rising from the ocean,
the integration and inversion of Eq.~\rf{wth1} to get $\wth$ is more 
complicated, as the path comprises a water and a rock component. Both
cases, land and ocean, are dealt with in the Appendix.

The decay probability is 
\beq{Pdecay}
P_{\rm dk}(\mathit{L})=\int_0^\infty
\frac{d\wdk}{\tau}\;e^{-\frac{\wdk}{\tau}}
\eeq
with the tau lifetime in the lab-frame given by
\beq{lifetime}
\tau=\frac{E_\tau}{m_\tau}\,\tau_{\rm RF} 
= \frac{392\,(\Enu/10^{19}{\rm eV})\,{\rm km}}
{\left[
1+\alpha\,\eye (\wint)\,(0.8\,\Enu/10^{19}{\rm eV})^\alpha
\right]^{\frac{1}{\alpha}}}\,,
\eeq
where $\tau_{\rm RF}$ is the rest-frame value of the tau lifetime.
The numerical expression in Eq.~\rf{lifetime} properly includes 
the 0.8 mean factor for energy transfer between the incident $\nutau$
and the $\tau$. We remind the reader that for the simple case of UAS
over rock, the opacity is just $\eye =\beta_{19}\,\rhosr\,\wint$. The
more complicated case for UAS over oceans is dealt with in the
Appendix.

When the tau decays, it has a 64\% branching probability to decay to 
$\nutau+hadrons$.  For an unpolarized tau, $\sim 2/3$ of the energy
goes into the hadrons, and therefore into the shower. Accordingly, for
this mode we take the relation between tau and shower energies to be 
$\frac{2}{3}E^\tau=E^{\rm sh}$.  We define $\Esh$ to be the 
minimum-energy trigger for the detector. Thus, we have the threshold
relation $\Eth=\frac{3}{2}\,\Esh$. The tau also has 18\% branching
probabilities each into $\nu+\nubar+e$ and $\nu+\nubar+\mu$. The
electronic mode immediately creates an electromagnetic shower with
$\sim 1/3$ of the tau energy, on average. So for the electronic mode
we take the relation between thresholds to be $\Eth=3\,\Esh$. The
muonic mode is ignorable, for the decay length of the muon exceeds the
distance to the ground. In our calculation of the UAS acceptance, we
will weight each tau decay with 64\% for the hadron mode where
$\Esh=\frac{2}{3}\,\Eth$, and 18\% for the electron mode where
$\Esh=\frac{1}{3}\,\Eth$; the remaining 18\% is the unobservable
muon mode.~\footnote{The tau is 100\% polarized (to order
  $m_\tau/E_\tau$) at production in the Earth. It is possible that
  even after multi-scattering in the Earth (mainly due to
  photo-nuclear interactions), the tau retains some of its initial
  polarization. If so, then the decay particle with helicity opposite
  to that of the tau is softer on average. The net result is slightly
  more energy transferred to the electromagnetic shower, and slightly
  less energy transferred to the hadronic shower~\cite{taupolzn}.}

The two integrals in Eqs.~\rf{Pnutotau} and~\rf{Pdecay} are coupled
via the $\wint$-dependent lifetime of the tau. When we later introduce
constraints due to cloud covering, we will see further coupling among
the integration variables. The exponential in Eq.~\rf{Pdecay}
describes the survival probability of the upgoing tau lepton to reach
the decay point $\wdk$. There is some probability for the tau to decay
inside the Earth in the interval $[0,\wint]$, but as we mentioned
above it is negligibly small in the energy range of interest. There is
regeneration of tau neutrinos over the whole Earth due to the tau
production and decay chain, but the regenerated taus with their lower
energy contribute negligibly to the high-energy sample discussed here
and so are not included.

In practice, a sufficient column density of air beyond the tau decay
point $\wdk$ is required such that the decay products fully develop
into a shower. This requirement will cutoff the integration in
Eq.~\rf{Pdecay}, and provide an $L$-dependence (or
$\cos\thn$-dependence)  to $P_{\rm dk}$. In the original
study~\cite{KW}, a simple analytic result for the decay
integral~\rf{Pdecay} was obtained by invoking certain
approximations. The tau lifetime was taken to be a constant over the
energy-range of interest, and the integral was cutoff at the scale
height of the atmosphere, $h=8$~km. With these approximations, one
obtains for the decay integral $P_{\rm dk}=1-e^{-h/(\tau\cos\thn)} =
1-e^{-2R_\oplus\,h/L\tau}$. Also, the air shower rate per incident
$\nu_{\tau}$ was computed analytically for the case where the angle
above the horizon satisfies $\thhor \gg
(10^{17}\rm{eV}/E_{\tau})$~degrees so that $P_{\rm dk}\approx
{2R_\oplus\,h/L\tau}$. In this work, we do not adopt these
approximations. Here, the implicit energy-dependence of the UAS rate
in Eq.~\rf{UASrate} arises from the energy-dependences of $\wth$,
$\lambda_\nu$, and $\tau$, as well as from the differential flux
$F_{\nutau}$. We present results for the full nested integrals
of Eq.~\rf{UASrate}.

\subsection{Horizontal Air-Showers (HAS)} 
\label{sec:HASrate}

We now turn to the derivation of the HAS event rate. Neutrino-induced
air-showers come in several topologies~\cite{flavorID}. All three
neutrino flavors contribute equally to the neutral current (NC)
events, but these transfer on average only 20\% of the incident energy
to the shower. Furthermore, the NC interaction rate is smaller, about
44\%, than the charged-current (CC) rate. Among the CC events, the
leading muon and tau from incident $\numu$ and $\nutau$, respectively,
are not visible in the air (unless the tau decays in a ``double-bang''
event). In the CC process, only 20\% of the incident energy is
transferred to the visible shower. For a $\nu_e$-initiated CC event,
the produced electron contributes electromagnetically to the shower,
so the full incident energy converts to shower energy. In summary,
about one event in four (the $\nu_e$~CC interaction) will transfer
100\% of the incident energy to the shower, while three events in four
will transfer $\sim 20\%$ of the energy.~\footnote{If the incident
  neutrino spectrum is falling as a power, then {\sl at fixed energy}
  the  $\nu_e$~CC events dominate the total rate.}

\begin{figure}[t]
\includegraphics[width=.56\textwidth]{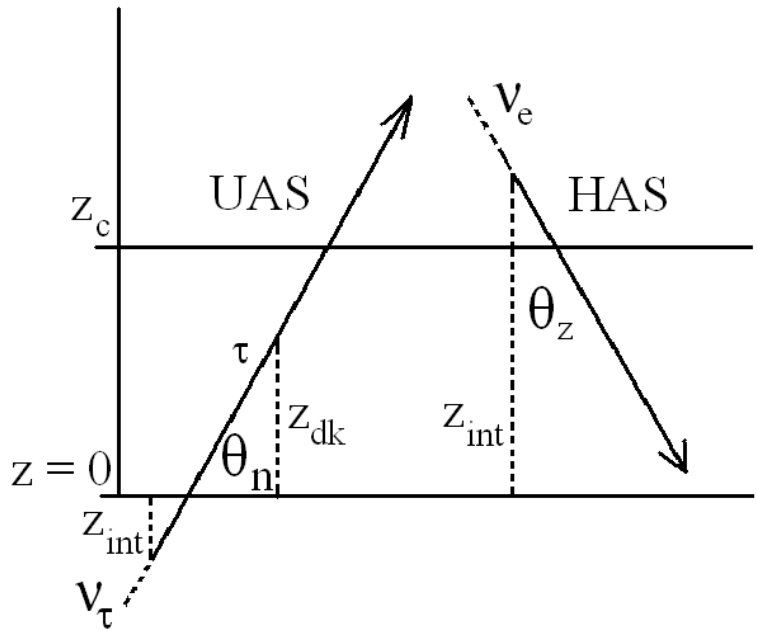}
\caption{
\label{fig:schem2} 
  Lateral snapshot of UAS and HAS; $z$ labels vertical altitudes and
  depths.
} 
\end{figure}

To be definite, we assume a $\nu_e$ CC interaction in what follows. 
We label the spatial axes as $z$ for vertical upward and $x$ and $y$
for the directions tangent to the Earth's surface; curvature of the
Earth's surface may be neglected for HAS. It is useful to consider
first a parallel neutrino flux perpendicular to ${\hat x}$, incident
with a zenith angle $\thz$, as illustrated in the projections of
Figs.~\ref{fig:schem2} and \ref{fig:schem4}. The horizontal air shower
probability is 
%
\beq{HASrate1}
R_{\nue} ({\rm HAS}) =
F_{\nue}\,\int d\Omega \int \sig \,\rhoatm({\vec \rint})\, d^3
\rint = F_{\nue}\,2 \pi A\,\sig \,\rhoatm (0)
\int d\cos\thz \int e^{-\zint/h} d\zint\,,
\eeq
%
where ${\vec \rint}$ is the point of interaction, $A=\int dx\int dy$
again, and the second expression follows from the first when the
atmospheric density function 
\beq{rhoatm}
\rhoatm (z)=\rhoatm (0)\,e^{-z/h}
\eeq
is inserted. We set the scale-height $h$ = 8~km. The factor of
$\int\,d\Omega =2\,\pi\,\int d\cos\thz$ in Eq.~\rf{HASrate1} rotates
the incident flux, initially assumed to be parallel and now assumed to
be isotropic, over the full sky.

In principle, we should include the curvature of the Earth's surface
in assessing the vertical height $z$ in $\rho(z)$ along the developing
shower. In practice, this is unnecessary as long as the shower length
is a small fraction of the Earth's radius, as is the case (we will
return to the curvature issue later in the discussion of UAS events). 

The HAS event rate scales linearly with the cross-section $\sig$.
This is because the absorption probability of the neutrino in the
atmosphere is negligibly small. The natural scales of atmospheric
column density are the vertical density
\beq{dvert}
\dvert\equiv\int^\infty_0 dz\,\rhoatm(z)=h\,\rhoatm(0) =1030\,{\rm
g/cm}^2\,,
\eeq
and the horizontal density
%
\beq{dhor}
\dhor=\int^\infty_0 dx\,\rhoatm(z=\sqrt{\Rearth^2+x^2}-\Rearth)\approx
\dvert\int_0^\infty du\,e^{-u^2h/2\Rearth}=
\sqrt{\pi\Rearth/2\,h}\,\dvert =36\,\dvert\,.
\eeq
%
In terms of the latter, the neutrino absorption probability in the
atmosphere is
%
\beq{HASnuabs}
P(\nu-{\rm air\;absorption}) =
\sig \,N_A\,\dhor\,\left(\frac{d}{\dhor}\right)
	= 2\times 10^{-3}\,\sigstd\,\left(\frac{d}{\dhor}\right)
\eeq
%
where $d\le\dhor$ is the column density of the neutrino's trajectory
in the atmosphere. Thus, for $\sig \lsim 10^{-29}{\rm cm}^2$,
atmospheric absorption is negligible even for horizontal neutrinos,
and so the neutrino interaction rate scales linearly with $\sig$.

Further restrictions on the integration variables result from further
assumptions for detector efficiencies. Let us assume that the air
shower must originate in the detector FOV of area $A$. Then the
straightforward integration of Eq.~\rf{HASrate1} gives 
\beq{HASrate2}
R_{\nue} (HAS) = 2\pi\,A\,F_{\nue}\,h\,\sig \,\rhoatm(0)\,.
\eeq
The value $h\,\sig \,\rhoatm (0) = 0.62\times 10^{-4}\,\sigstd$ sets
the scale for the interaction probability in the atmosphere per
incident neutrino. The resulting value of the acceptance~\footnote{One
  may also write the acceptance as $2\pi\,A\,(h/\lambda_\nu)$, where
  ${\lambda_\nu}^{-1}=\sig\,\rhoatm (0)$ is the neutrino MFP. This
  expression is the $\lambda_\nu\gg h$ limit of
  $\acc=2\pi\,A\,(1-e^{-h/\lambda_\nu})$. In this latter form, one sees
  the acceptance saturating its geometric value of $2\pi\,A$ in the
  strong cross-section limit.} is $\acc\equiv R_{\nue}(HAS)/F_{\nue}=
2\pi\,A\,h\,\sig \,\rhoatm(0) =
3.9\,\sigstd\,\left(\frac{A}{10^4\,{\rm km}^2}\right)\,{\rm 
km}^2\,{\rm sr}$. This value suggests that wide-angle, large-area 
detectors exceeding $10^4\,{\rm km}^2\,{\rm sr}$, and cosmic neutrino
fluxes exceeding  $1/{\rm km}^2\,{\rm sr}\,{\rm yr}$, are needed for
event collection. Put another way, full sky coverage of an air mass of 
$\sim 10^5\,{\rm km}^2\times h\,\rho(0)\sim$~teraton is required.

\section{Constraints from development and identification of showers}
\label{sec:visibleAS}

In this section we address conditions for the showers to be observable.
First of all, shower detection will require that within the FOV, the
length of the shower track projected on a plane tangent to the Earth's
surface (as would be seen from far above or far below) exceeds some
minimum length, $\lmin$. Space-based observatories are far above the
Earth, and so view the atmosphere as a two-dimensional plane. For
ground-based observatories, tangential projections may not be the
optimum way to describe the FOV constraint, but we use it as a guide.

In addition to the projected length constraint, there are three
``shower-development'' constraints to be applied to the events. A
minimum column density, $\dmin$, beyond the point of shower initiation
is required for the shower to develop in brightness. On the other
hand, after a maximum column density, $\dmax$, the shower particles are
below threshold for further excitation of the $N_2$ molecules which
provide the observable fluorescence signal. We therefore terminate
showers at $\dmax$, which implies a finite length for the visible
shower.

While the requirements of minimum projected length, and minimum and
maximum shower column-densities are correlated, no two of them implies
the third. Some reflection on the $\theta$ and $z$ dependences of the
varying densities and projected lengths reveals that this is so.

Finally, the fluorescent emission per unit length of the shower will
decline exponentially with the air density at altitude. At $z=2h$, the
fluorescent emission is down to $e^{-2}=14\%$ of that at sea level.
At $z=3h\,(4h)$, it is down to $5\%\,(2\%)$ of that at sea level.
Atmospheric absorption of the emitted fluorescence also affects the
signal. This absorption is thought to scale roughly as the atmospheric
density, up to about 20~km~\cite{EUSO}. Thus, it turns out that the
fluorescence signal could roughly be taken as constant between zero and
20~km. Accordingly, we will take $\zthin=3h=24$~km as the \toothin
altitude beyond which the signal becomes imperceptible.

Thus, there are four constraints that render the shower observable.
These are the $\lmin$, $\dmin$, $\dmax$, and \toothin (or
$\zthin$) conditions. The values which we choose for these parameters 
are listed in Table~\ref{table:parameters}. The choice for the
$\zthin$ value was discussed and motivated above. The $\dmin$ and
$\dmax$ choices are inferred from the observed longitudinal
development profiles of ultrahigh-energy cosmic ray showers (the
famous Fly's Eye event at energy $3\times 10^{20}$~eV provides a
splendid example~\cite{FlysEye}). Showers at 300-400 g/cm$^2$ of
column density (also called ``atmospheric depth'' or ``slant depth'')
comprise tens of billions of electrons, with a brightness roughly 10\%
of shower maximum. The electrons in showers at $\gsim 1200$~g/cm$^2$
are ranging out, reducing significantly the shower brightness.
We assign a relatively small value to $\lmin$ to maximize the
observable event rate.  For the EUSO experiment, each pixel is a map
of a square kilometer of the Earth's surface~\cite{EUSO}. Thus, an
$\lmin$ of 10~km corresponds to a signal in ten contiguous pixels. The
background for ten contiguous pixels should be small.  With ten
pixels, the angular reconstruction of the event direction is roughly
1/10 radian ($\sim 5$~degrees). With a cloud layer a smaller $\lmin$
value for event triggering may be needed. With an $\lmin$ of 5~km,
the signal/noise should still be acceptable. For $\lmin=5$~km, the
angular reconstruction is reduced to 1/5~radian ($\sim
10$~degrees), though. Table~\ref{table:parameters} contains a summary
of the values which we choose for the four shower-development
parameters $\zthin$, $\dmin$, $\dmax$, and $\lmin$.

\subsection{Effective area}

\begin{figure}[t]
\includegraphics[width=.56\textwidth]{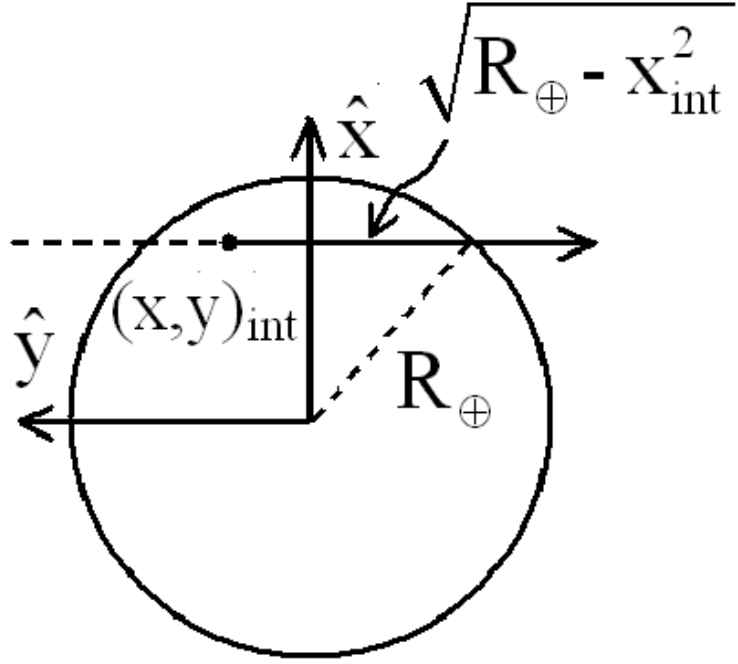}
\caption{\label{fig:schem4} 
  Overhead snapshot of the event projected onto the field of view
  (FOV).
}
\end{figure}

Let us describe the projected length $\lmin$ constraint, and the
$\dmin$ and $\dmax$ constraints in the general case. Consider a
detector with a FOV characterized by a radius $R_{\rm FOV}$, i.e., (see
Fig.~\ref{fig:schem4})  
\beq{circle}
x^2 + y^2 \le \rfov^2\,.
\eeq
A shower produced with initial coordinates $(x,y,z)$ cannot have a
visible projected length larger than the chord length $y+\sqrt{\rfov^2
  - x^2}$ in the FOV. However, the projected length may be smaller,
for three reasons. The first is that the shower must develop and
brighten before becoming visible. This requires traversing the column
density $\dmin$. The second reason is that the shower may hit the
ground (attain the \toothin\ altitude) before reaching the far
boundary of the FOV in the case of HAS (UAS). And thirdly, the shower
may extinguish before reaching the far boundary of the FOV. Extinction
occurs when the traversed column density attains the value
$\dmax$. For HAS (UAS), we label the upper altitude where the shower
becomes visible (extinguishes or strikes the  \toothin\ altitude) as
$z_U$, and the lower altitude where the shower extinguishes or strikes
the ground (becomes visible) as $z_L$. We will use the HAS or UAS
label on $z_U$ and $z_L$ to distinguish between these altitudes in
both cases. The altitudes $z_B$ and $z_E$, to be defined shortly, will
also carry a HAS or UAS label. For brevity, we will sometimes omit the
HAS and UAS labels when it is clear from the context which label
applies.

From the above shower-development considerations, the visible shower
length projected on the Earth's surface is $(\zu-\zl)\,\tan\theta$. 
Collecting the remarks above, the projected air-shower length within
the FOV of the detector is then
\beq{lFOV}
l_{\rm FOV}=\min\,\left\{ [\zu-\zl]\,\tan\theta,\, y+\sqrt{\rfov^2 -
x^2}\;\,\right\} \,. 
\eeq 
\noindent
We will discuss the maximum and minimum altitude values $z_U$ and
$z_L$ for each type of shower in the following subsections.

For the projected length in the FOV to exceed some minimum length, 
$\lmin$, we infer from Eq.~\rf{lFOV} the set of conditions
\beq{lmin1}
y+\sqrt{\rfov^2 - x^2} \geq \lmin\,,
\eeq
\beq{lmin2}
[z_U-z_L]\,\tan\theta \geq \lmin\,.
\eeq
After algebraic manipulation of constraints \rf{circle} and \rf{lmin1},
the area integral $\int dx \int dy$ is easily done, yielding
\beq{area}
A=\pi\,\rfov^2\;\left(\frac{\arcsin \eta -
\eta\,\sqrt{1-\eta^2}}{\pi/2}\right)\,,
\eeq
where
\beq{eta}
\eta \equiv \sqrt{1 - \frac{\lmin^2}{4\rfov^2}}
\eeq
is nearly one in a large-area experiment. This is then the
constraint-modified meaning of $A$. Since the arguments leading to it 
apply equally to the HAS and UAS geometry, the result in Eq.~\rf{area}
applies in both rates, Eqs.~\rf{UASrate} and \rf{HASrate1}. We note
that for $\lmin \ll 2\rfov$, the expression in parenthesis in
\rf{area} is nearly unity, with an expansion  
\beq{areaseries}
1-\frac{3}{\pi}\left(\frac{\lmin}{2\rfov}\right)
+ {\cal O}\left(\frac{\lmin}{2\rfov}\right)^3\,.
\eeq
Thus, with $\lmin \ll 2\rfov$, the constrained area is just the
geometric area.

\subsection{Four constraints for HAS}
\label{sec:visibleHAS}
\noindent
In this subsection we develop the
$\lmin$, $\dmin$, $\dmax$, and \toothin (or $\zthin$) 
constraints for HAS.

The \toothin constraint at high altitude is simple. It effectively
implies  $(\wint=\zint/\cos\thz$) 
\beq{toothinHAS}
\zg \le \zint \le \zthin\,,
\eeq
where $\zg=0$ labels the altitude of the ground. We will retain the
symbol $\zg$ even though it is zero in the present context. Retaining
$\zg$ will be useful for substitutions in later sections where we
include a cloud layer.

The projected track length condition, given in Eq.~\rf{lmin2}, is
\beq{HASlmin2}
\left[\zuhas-\zlhas \right]\,\tan\thz \geq \lmin\,.
\eeq
Now we consider the calculations of the maximum and minimum
visible-shower altitudes, $\zuhas$ and $\zlhas$, respectively, which
enter this formula. The column densities of the HAS showers evolved
from initial altitude $\zint$ to the lower altitude $z$ are given by
%
\beq{HAScolumn}
d({\rm HAS};\,z) =
\int^{\zint}_{z} \frac{dz}{\cos\thz}\,\rhoatm (z) =
\frac{\dvert}{\cos\thz}\,\left(e^{-z/h}-e^{-\zint/h}\right)\,.
\eeq
%
The evaluated integral uses the exponential decrease of the
atmospheric density with increasing altitude and the definition $\dvert
= h \, \rhoatm(0)$. Equating $d({\rm HAS};\,z)$ to $\dmin$, we get the
altitude where the shower first becomes visibly bright.  We call this
altitude $\zbhas$. Positivity of the integrand ensures that
$\zb<\zint$ for any atmospheric density profile. Implicitly, $\zbhas$
is given by
\beq{zB1}
e^{-\zbhas/h}-e^{-\zint/h}=\frac{\dmin}{\dvert}\cos\thz\,.
\eeq
Explicit solutions for $\zbhas$ as a function of $\zint$ and vice
versa are 
\beq{zB2}
\zbhas= -h\ln\left[e^{-\zint/h}+\frac{\dmin}{\dvert}\cos\thz \right]\,;
\quad\quad
\zint = -h\ln \left[e^{-\zbhas/h}-\frac{\dmin}{\dvert}\cos\thz
  \right]\,. 
\eeq
A shower becomes visible above the ground level only if $\zb>\zg$.
From Eq.~\rf{zB1}, the condition $z_B>\zg$ implies
\beq{HASdmin}
\zint > -h\ln
\left[e^{-\zg/h}-\frac{\dmin}{\dvert}\cos\thz \right]\,.
\eeq
Although this condition ensures a visible shower above the ground,
it allows the visible length to be arbitrarily small. The requirement
of a visible projected length in excess of $\lmin$ will lead to a
stronger constraint, presented below in Eq.~\rf{HASlmin4}. However,
Eq.~\rf{HASdmin} is useful in that it implies an absolute limit on the
shower direction. Since $\zint$ is less than $\zthin$ by construction
(Eq.~\rf{toothinHAS}), the limit is
\beq{abscos}
\cos\thz \le
\frac{\dvert}{\dmin}\,\left(e^{-\zg/h}-e^{-\zthin/h}\right) 
\approx \frac{\dvert}{\dmin}\,.
\eeq
There is no absolute restriction on angle from the $\dmin$ constraint
if $\dmin$ is less than $\dvert=1030\,{\rm g/cm}^2$, since then even
a vertical shower traverses enough column density to brighten. We will
take $400\,{\rm g/cm}^2$ as our standard value for $\dmin$. 

The calculation of $\zlhas$ proceeds analogously to the calculation of
$\zuhas$. Setting the column density $d({\rm HAS};\,z)$ equal to
$\dmax$, we get the altitude where the shower extinguishes. We call
this altitude $\zehas$. Implicitly, $\zehas$ is given by
\beq{zE1}
e^{-\zehas/h}-e^{-\zint/h}=\frac{\dmax}{\dvert}\cos\thz \,.
\eeq
The LHS of Eq.~\rf{zE1} is maximized by setting $\zehas$ equal to $\zg$
and $\zint$ to $\zthin$.  If the RHS exceeds this maximum value, then
for any $\zint$ the total column density remains less than $\dmax$,
the shower does not extinguish, and there is effectively no $\dmax$
constraint. So, for more-vertical showers obeying
\beq{coshas}
\cos\theta_z \ge \frac{\dvert}{\dmax}\,\left(
  e^{-\zg/h}-e^{-\zthin/h}\right)\,\equiv\coshathas\,,
\eeq
the shower hits the ground before extinction and thus, we have
$\zlhas=\zg$. 

On the other hand, when $\cos\thz<\coshathas$, then whether or not
the shower extinguishes before striking the ground depends on the
height in the atmosphere at which the shower originated, i.e., on
$\zint$. Solving Eq.~\rf{zE1} explicitly for $\zehas$ 
as a function of $\zint$ and vice versa,
one gets
\beq{zE2}
\zehas=
-h\ln\left[e^{-\zint/h}+\frac{\dmax}{\dvert}\cos\thz\right]\,;
\quad\quad
\zint = -h\ln \left[e^{-\zehas/h}-\frac{\dmax}{\dvert}\cos\thz\right]
	\equiv \zhath\,.
\eeq
The shower strikes the ground if $\ze\le\zg$, and extinguishes if
$\ze>\zg$. The critical value is $z_E=\zg$. At this critical value,
\rf{zE2} gives 
\beq{HASdmax}
\zint	= -h\ln
\left[e^{-\zg/h}-\frac{\dmax}{\dvert}\cos\thz\right]
	\equiv \zhath\,.
\eeq
Thus we have two cases: for $\zint\le\zhath$, the shower strikes the
ground and the minimum altitude is $\zl=\zg$; while for $\zint>\zhath$
(which implies $\cos\theta_z< \coshathas$ because $\zint<\zthin$ by
construction), the shower extinguishes above the ground and
$\zl=\zehas$. The latter case corresponds to Eq.~\rf{zE2} having a
real-valued solution in the physical interval $[\zg,\zthin]$, whereas
the former case corresponds to no such solution for Eq.~\rf{zE2}. The
high altitude $\zuhas$ is where the shower begins its visible track
length.  Accordingly, we set $\zuhas =\zbhas$ in the $\lmin$
constraint for both cases, where $\zbhas$ is given in Eq.~\rf{zB2}.

For $\zint> \zhath$, the shower extinguishes and we substitute
$\zuhas=\zbhas$ and $\zlhas=\zehas$, given in Eqs~\rf{zB2} and
\rf{zE2}, respectively, into Eq.~\rf{HASlmin2}. After a bit of
algebra, one finds that the resulting $\lmin$ constraint can be
expressed as
\beq{HASlmin3}
\zint>-h\ln
\left[\frac{\cos\thz}{\dvert}\,
\left(
\frac{\dmax-\dmin\,e^{\frac{\lmin}{h\tan\thz}}}{e^{\frac{\lmin}{h
      \tan\thz}}-1} \right)\right]\,.
\eeq
Real-values of $\zint$ in the interval $[\zhath,\zthin]$ which satisfy
this equation, if any, satisfy all four constraints for $\cos\thz <
\coshathas$, and so contribute to the integral for the observable
event rate.

For the other case, where $\zint<\zhath$, the shower strikes the
ground.  We have $\zuhas=\zbhas$, the latter given in Eq.~\rf{zB2},
and $\zlhas=\zg$. Inputing these expressions into Eq.~\rf{HASlmin2},
one finds an explicit expression for the $\lmin$ constraint, 
\beq{HASlmin4}
\zint> -h\ln
\left[
e^{-(\frac{\lmin}{\tan\thz}+\zg)/h}-\frac{\dmin}{\dvert}
\cos\thz \right]\,.
\eeq
This constraint ensures a visible projected length exceeding $\lmin$.
It replaces the constraint of Eq.~\rf{HASdmin}, which ensured only a
nonzero visible track length. Of course, in the limit $\lmin=0$, the
two constraints are identical. Real-values of $\zint$ in the interval
$[\zg,\zthin]$ for $\cos\thz \ge \coshathas$, or in the interval
$[\zg,\zhath]$ for $\cos\thz < \coshathas$, which satisfy this
equation, if any, satisfy all four constraints and so contribute to
the integral for the observable event rate.

To summarize the HAS rate formulas in the absence of cloud cover, we
have the general rate equation, Eq.~\rf{HASrate1}, with the area given
in Eq.~\rf{area}, and the \toothin constraint in
Eq.~\rf{toothinHAS}. There are two alternate ways to express the 
$\dmin$, $\dmax$, and $\lmin$ constraints.  The first way is to define
the boundaries of $(\zint, \thz)$-integration physically but
implicitly. This is done with Eq.~\rf{HASlmin2} implementing the
$\lmin$ constraint, where $\zuhas=\zbhas$ implements the $\dmin$
constraint with $\zbhas$ given in Eq.~\rf{zB2}, and $\zlhas=\max\{\zg,
\zehas\}$ implements the $\dmax$ constraint with $\zehas$ given in
Eq.~\rf{zE2}. If Eq.~\rf{zE2} has no real-valued solutions in the
interval $[\zg,\zthin]$, then $\zlhas=\zg$. The value of $\zg$ is zero
(in the absence of clouds).

The alternative way to express the $\dmin$, $\dmax$, and $\lmin$ 
constraints is to solve the constraints of the first approach for
explicit boundaries on the $(\zint, \thz)$-integration. The
results of this approach bifurcate, depending on whether the shower
extinguishes, or the shower strikes the ground. For the case where the
shower extinguishes, the boundaries are  given by $\zint >\zhath$,
with $\zhath$ defined in Eq.~\rf{HASdmax}, and by
Eq.~\rf{HASlmin3}. For the case where the shower strikes the ground,
the boundaries are given by $\zint<\zhath$, and by Eq.~\rf{HASlmin4}.
{\sl A priori}, there is no guarantee that Eqs.~\rf{HASlmin3} and
\rf{HASlmin4} have real-valued solutions in the physical region of
$\zint$.

\subsection{Four Constraints for UAS}
\label{sec:visibleUAS}

The calculation of the $\lmin$, $\dmin$, $\dmax$, and \toothin (or
$\zthin$) constraints for UAS events proceeds analogously to the
calculation for HAS events. The \toothin altitude constraint is again
\beq{toothinUAS}
\zg \le \zdk \le \zthin\,.
\eeq
At sea level, $\zg=0$.
The projected track length condition for UAS events, analogous to
Eq.~\rf{HASlmin2} for HAS events, is 
\beq{UASlmin2}
\left[\zuuas-\zluas \right]\,\tan\thn \ge \lmin\,. 
\eeq
The values of $\zuuas$ and $\zluas$ differ from $\zuhas$ and $\zlhas$.
To calculate them, we turn to calculations of UAS column densities,
given by 
%
\beq{UAScolumn}
d({\rm UAS};z)=
\int^{z}_{\zdk} \frac{dz}{\cos\thn}\,\rhoatm (z) = 
	\frac{\dvert}{\cos\thn}\,\left(e^{-\zdk/h}-e^{-z/h}\right)
\eeq
%
Setting this equal to $\dmin$ defines implicitly the brightness
altitude $\zbuas$: 
\beq{UASdmin1}
e^{-\zdk/h}-e^{-\zbuas/h}=\frac{\dmin}{\dvert}\cos\thn \,.
\eeq
Solving this equation explicitly for $\zbuas$ as a function of $\zdk$
and vice versa, one gets
\beq{UASdmin2}
\zbuas=-h\ln\left[
e^{-\zdk/h}-\frac{\dmin}{\dvert}\cos\thn
\right]\,;
\quad\quad
\zdk=-h\ln\left[\frac{\dmin}{\dvert}\cos\thn + e^{-\zbuas/h}\right]\,.
\eeq
We require that $\zbuas <\zthin$; otherwise, the shower invisibly
disappears into thin air. From Eq.~\rf{UASdmin2}, the
condition $\zbuas <\zthin$ can be written 
\beq{UASdmin}
\zdk\le
-h\ln\left[\frac{\dmin}{\dvert}\cos\thn + e^{-\zthin/h}\right]\,.
\eeq
This condition ensures visibility of the shower, but with a visible
length arbitrarily small. The requirement of a visible projected
length in excess of $\lmin$ will lead to a stronger constraint,
presented in Eq.~\rf{UASlmin4} below. Positivity of $\zdk$ and
Eq.~\rf{UASdmin} lead to the same angular constraint for
$\cos\theta_z$ as was found for HAS's $\cos\theta_n$ in
Eq.~\rf{abscos}.

Setting $d({\rm UAS};z)$ equal to $\dmax$, we get the high altitude
$\zeuas$ where the UAS shower extinguishes. Implicitly, this highest
visible altitude is given by 
\beq{UASzE1}
e^{-\zdk/h}-e^{-\zeuas/h}=\frac{\dmax}{\dvert}\cos\thn \,,
\eeq
The LHS of Eq.~\rf{UASzE1} is maximized by setting $\zdk$ equal to
$\zg$ and $\zeuas$ to $\zthin$. If the RHS exceeds this maximum value,
then for any $\zdk$ the total column density remains less than $\dmax$,
the shower does not extinguish, and there is effectively no $\dmax$
constraint. So, for more-vertical showers obeying
\beq{cosuas}
\cos\thn \ge \frac{\dvert}{\dmax}\,\left(
  e^{-\zg/h}-e^{-\zthin/h}\right)\,\equiv\coshatuas\,,
\eeq
we have $\zuuas=\zthin$. 

On the other hand, when $\cos\thn < \coshatuas$, then whether or not
the shower extinguishes before reaching the \toothin boundary
depends on the height in the atmosphere at which the shower
originated, i.e., on $\zdk$. Solving Eq.~\rf{UASzE1} explicitly for
$\zeuas$ as a function of $\zdk$ and vice versa,
one gets
\beq{UASzE2}
\zeuas=-h\ln\left[e^{-\zdk/h}-\frac{\dmax}{\dvert}\cos\thn \right]
\quad ; \quad
\zdk = -h\ln
\left[\frac{\dmax}{\dvert}\cos\thn + e^{-\zeuas/h}\right]\,.
\eeq
The shower extinguishes if $\zeuas < \zthin$, and hits the \toothin
boundary if $\zeuas \ge \zthin$. The critical value is
$\zeuas=\zthin$. Inputting this critical $\ze$ into Eq.~\rf{UASzE2},
one finds a critical value for the decay altitude
\beq{UASdmax}
\zdk =  -h\ln
\left[\frac{\dmax}{\dvert}\cos\thn + e^{-\zthin/h}\right]
	\equiv \zhatu \,.
\eeq
For $\zdk<\zhatu$, the shower attains the ``length'' $\dmax$ and
extinguishes, whereas for $\zdk>\zhatu$, the shower reaches the
\toothin boundary $\zthin$ without extinction.

For $\zdk<\zhatu$, the shower extinguishes and so $\zuuas=\zeuas$.
Substituting this (Eq.~\rf{UASzE2}) and $\zluas=\zbuas$ from
Eq.~\rf{UASdmin2} into the $\lmin$ constraint Eq.\rf{UASlmin2} leads
to an explicit expression for the $\lmin$ constraint:
%
\beq{UASlmin3} 
\zdk\ge -h\ln\left[
\frac{\cos\thn}{\dvert} \, \left(
\frac{\dmax\,e^{\frac{\lmin}{h\tan\thn}}-\dmin}
{e^{\frac{\lmin}{h\tan\thn}}-1} \right)\right] \equiv z_< \, ,
\eeq
%
This limit, like the analogous one for HAS in Eq.~\rf{HASlmin3},
forces the shower initiation to occur at a higher altitude where the
air is thinner, and therefore, for fixed $\dmax$, the shower and its
projection are longer. There is no $\dmin$ constraint for a shower
that saturates $\dmax$. 

For $\zdk\ge\zhatu$, the shower reaches $\zthin$ without
extinction. Therefore, there is no $\dmax$ constraint and we set 
$\zuuas=\zthin$. Substituting this and $\zluas=\zbuas$ into the
$\lmin$ constraint Eq.~\rf{UASlmin2}, we find the following explicit
expression for the $\lmin$ constraint:
%
\beq{UASlmin4}
\zdk < -h\ln\left[
e^{(\frac{\lmin}{\tan\thn}-\zthin)/h} 
+ \frac{\dmin}{\dvert}\cos\thn \right] \equiv z_>\, .
\eeq
%
For this class of showers which reach the \toothin boundary, this
upper bound on $\zdk$ ensures a visible projected length exceeding
$\lmin$. It supersedes the constraint of Eq.~\rf{UASdmin}, which
ensured only a visible track of non-zero length. Of course, in the
limit $\lmin=0$, the two conditions are identical. 

The essence of the $\lmin$ constraint is that the shower must have
sufficient normal angle to attain a minimum horizontal
projection. Thus, there is a critical angle $\thn^{\rm crit}$ for 
which the conditions of Eqs.~\rf{UASlmin3} and \rf{UASlmin4} collapse
to $z_<=\zhatu = z_>$. For normal angles smaller than  $\thn^{\rm
  crit}$, i.e., for $\cos\thn > \cos\thn^{\rm crit}$, there are no
observable events. Setting $z_<=\zhatu = z_>$, one finds that this
critical angle is given implicitly by
\beq{thncrit}
\zthin = -h \ln\left[ \frac{\dmax - \dmin}{\dvert} \,
  \frac{\cos\thn^{\rm crit}}{e^{\frac{\lmin}{h \tan\thn^{\rm crit}}}
    -1} \quad
  \right] \,.
\eeq
This critical angle encapsulates a relatively weak constraint
in the cloudless case.  However, it will become a strong constraint 
when we consider cloudy skies.

To summarize the UAS rate formulas in the absence of cloud cover, we
have the general rate equation, Eq.~\rf{UASrate}, with inputs from 
Eqs.~\rf{Pnutotau}, \rf{dnu}, \rf{Etau}-\rf{lifetime} and \rf{area},
and the \toothin constraint in Eq.~\rf{toothinUAS}. As was the
case for the HAS events, there are two alternative ways to express the
$\dmin$, $\dmax$, and $\lmin$ constraints for UAS events. The physical
but implicit approach bounds the $(\zdk, \thn)$-integration with 
Eq.~\rf{UASlmin2} implementing the $\lmin$ constraint, where
$\zuuas=\min\{\zthin, \zeuas\}$ implements the $\dmax$ constraint with
$\zeuas$ given by Eq.~\rf{UASzE2}, and where $\zluas=\zbuas$
implements the $\dmin$ constraint with $\zbuas$ given in
Eq.~\rf{UASdmin2}.

Alternatively, one can solve the constraints of the first approach
explicitly for the $(\zdk, \thn)$-integration boundaries. As with
the HAS events, the results again bifurcate, depending on whether 
$\zeuas<\zthin$ (the shower extinguishes), or $\zeuas>\zthin$ (the
shower runs out of air). For the case where the shower extinguishes,
the boundaries are given by $\zdk <\zhatu$ with the latter quantity
defined in Eq.~\rf{UASdmax}, and by Eq.~\rf{UASlmin3}. For the case
where the shower strikes the \toothin altitude, the boundaries are
given by $\zdk > \zhath$ and by Eq.~\rf{UASlmin4}.

\subsection{Four Constraints for UAS, Including Earth's Curvature}
\label{sec:curvature}

So far we have treated $z$ as the vertical height above a flat Earth.
As remarked in the section on HAS rates, this is a valid approximation
as long as the trajectory in the atmosphere is small relative to the 
Earth's radius $\Rearth$.  Such is the case with HAS events. However,
at $10^{20}$~eV, the decay MFP for a tau is nearly 5000~km,
comparable to $\Rearth$ (6371~km). At extreme energies, curvature
effects cannot be neglected for UAS events. Specifically, the error
made in the vertical height, as a function of the atmospheric path
length $w$ and angle $\thhor$, is (neglecting terms of order ${\cal
  O}(w^4/\Rearth^3)$): 
\beq{zerror}
\delta z =\frac{w^2}{2\Rearth}\,
	\frac{\cos^2\thhor}{1+\frac{w}{\Rearth}\sin\thhor}
	\le \frac{w^2}{2\Rearth}\,.
\eeq
The far RHS expression, $w^2/2\Rearth=78\,(w/10^3{\rm km})^2$~km, is
saturated for (near) horizontal events. The error $\delta z$ in the
height of the decaying tau can be considerable: $\sim 2000$~km for
nearly horizontal events with $E_\tau\sim 10^{20}$~eV, and $\sim
20$~km for nearly horizontal events with $E_\tau\sim 10^{19}$~eV. So
our neglect of Earth's curvature means that we underestimate the
height of the shower, and so overestimate the air-density for shower
development. The height underestimate will erroneously reduce
(increase) the event rate as viewed from space (ground) when we
introduce clouds. The density overestimate will erroneously enhance
shower development.

\begin{figure}[t]
\includegraphics[width=.56\textwidth]{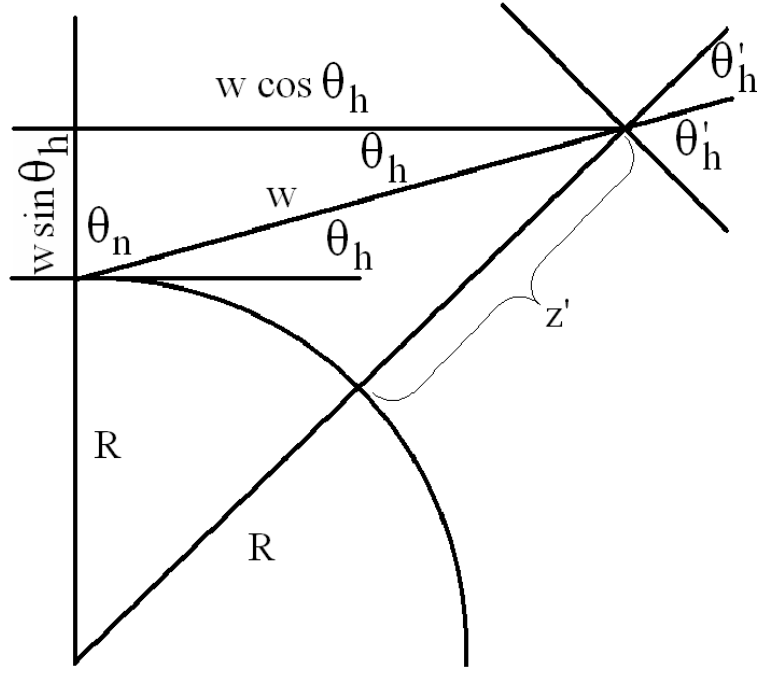}
\caption{\label{fig:curvature} 
  Illustration of the dependence of shower variables on the Earth's
  curvature. For clarity, the various parts are not drawn to
  proportion.
} 
\end{figure}

The curved geometry is shown in Fig.~\ref{fig:curvature}. The net
effects of curvature are twofold. First, for a given pathlength $w$
and trajectory angle $\thn$ at emergence from the Earth, the
curvature-corrected altitude, which we label as $z'$, is increased.
Second, the angle of the shower with respect to a plane tangent to the
Earth directly below, which we label as $\thhorp$ and $\thnp$ for the
horizontal and nadir angles, respectively, are rotated relative to the
comparable emergence angles (again, for a given $w$ and $\thn$). Note
that the primed variables are the altitude $(\zp)$ and angle $(\thp)$
seen by a detector. The unprimed variables describe the shower's
prehistory.

We now list the geometric relations which we need. From applying the
Pythagorean theorem to the right triangle in Fig.~\ref{fig:curvature},
we get $\zp$ in terms of $w$ and $\thhor$:
\beq{zprime}
1+\frac{\zp}{\Rearth} =
\left(
1+\left(\frac{w}{\Rearth}\right)^2+2\left(\frac{w}{\Rearth}\right)
\sin\thhor \right)^{1/2}\,,
\eeq
Applying the Law of Sines to the same right triangle, we get
\beq{sinprime}
\sin\thnp = \cos\thhorp 
= \frac{\sin\thn}{1+\frac{\zp}{\Rearth}} 
= \frac{\cos\thhor}{1+\frac{\zp}{\Rearth}}\,,
\eeq
with $\zp(w,\thhor)$ given in Eq.~\rf{zprime}. From this comes
\beq{cosprime}
\cos\thnp=\sin\thhorp=\frac{\sin\thhor+\frac{w}{\Rearth}}
			{1+\frac{\zp}{\Rearth}}\,,
\eeq
and
\beq{tanprime}
\cot\thnp=\tan\thhorp=\tan\thhor +\frac{w}{\Rearth\,\cos\thhor}\,.
\eeq
This last expression shows clearly that for nearly tangent
(``Earth-skimming'') events, the observed angle ($\thhorp$) is
increased from the emergent angle by the term $\approx w/\Rearth$.

In particular, the altitude and the angle at the decay point of the tau
are obtained by setting $w$ equal to $\wdk$ in
Eqs.~\rf{zprime}--\rf{tanprime}. For example, the decay altitude 
$\zpdk$ is given by 
\beq{zpdk}
\frac{\zpdk}{\Rearth}= \left(
	1+\left(\frac{\wdk}{\Rearth}\right)^2
	+2\left(\frac{\wdk}{\Rearth}\right)\sin\thhor\right)^\half
-1\,.
\eeq
After the tau decays, the resulting shower has a length which is short
on the scale of $\Rearth$. Consequently, we may ignore the Earth's
curvature from $\wdk$ onward, and use $\thhorp$ and $\thnp$ as {\sl
the} shower angles.

The development of the shower constraints with curvature parallels
that for UAS events without curvature, but with $(\zp,\thnp)$ rather
than $(z,\thn)$ parameterizing the shower altitude and angle.

The \toothin shower-constraint now becomes
\beq{zmax_constraint1}
\zpdk(\wdk,\,\thhor) \le \zthin \,.
\eeq
Using Eq.~\rf{zpdk}, this constraint can be cast as a constraint on
the integration variables $\wdk$ and $\thhor$ (equivalently, $L$). The
result is 
\beq{zmax_constraint2}
\wdk \le \wdkmax(\thhor,\zthin)\,,
\eeq
with
\beq{wdkmax}
\frac{\wdkmax}{\Rearth}=
	\sqrt{\sin^2\thhor+\left(\frac{\zthin}{\Rearth}\right)
	\left(2+\frac{\zthin}{\Rearth}\right)}
	-\sin\thhor\,.
\eeq
This constraint requires the decay to occur at an altitude below
$\zthin$, but does not require a minimum of shower development or a
minimum of projected length. The shower development and $\lmin$
constraints to come will be stronger. However, one useful feature of
this constraint is that it defines the maximum decay distance
available to a tau in our curved atmosphere. Taking $\thhor=0$ to
maximize the available decay length, one finds 
%
\beq{wdkmax2}
\wdkmax\le\wdkmax(\thhor=0) = \sqrt{\zthin\,(2\Rearth+\zthin)} 
       \simeq 550\,\sqrt{\frac{\zthin}{3\,h}}\;{\rm km}\,.
\eeq
%
The mean decay length for a tau is already 490~km at $10^{19}$~eV, and
grows linearly with its energy. Thus, it is clear from Eq.~\rf{wdkmax2}
that the Earth's curvature significantly reduces the UAS rate at
energies at and above $10^{19}$~eV. It is also clear from \rf{wdkmax2}
that since $\wdkmax\ll\Rearth$, leading order expansions in the ratio
$w/\Rearth$ are valid. For example, to very good approximations,
Eqs.~\rf{cosprime} and \rf{zpdk} can be written as
\beq{cosprime2}
\sin\thhorp=\sin\thhor+\frac{\wdk}{\Rearth}\,,
\eeq
and
\beq{zpdk2}
\zpdk\approx \wdk\,\sin\thhor +\frac{\wdk^2}{2\Rearth}\,.
\eeq
The first term on the RHS is just $\zdk$ for a ``flat Earth''. So we
learn that the replacement of $\zdk$ with $\zpdk$ in Eq.~\rf{zpdk2},
like that of $\thhor$ (at small angle) with $\thhorp$ in
Eq.~\rf{tanprime}, is a simple translation.

The constraints from shower development require explicit expressions
for the shower column density $d(\wdk,\thn)$ and shower length. The
column density in the observer's primed coordinates is
%
\beq{coldensN}
d({\rm UAS};\zp) = \int_{\wdk} dw\,\rho(\zp) =
	\int^{\zp}_{\zpdk} \frac{d\zp}{\cos\thnp}\,\rho(\zp) =
\frac{\dvert}{\cos\thnp}\left(e^{-\zpdk/h} - e^{-\zp/h}\right)\,,
\eeq
%
with the variables $\thnp$ and $\zpdk$ depending on just $\wdk$ and
$\thn$. Changing $z\rightarrow\zp$ and $\theta\rightarrow\thp$ in
Eqs.~\rf{UASdmin1} and \rf{UASzE1} defines $\zpbuas$ and $\zpeuas$,
respectively. The explicit expressions for $\zpbuas$ and $\zpeuas$ are
given by applying $z\rightarrow\zp$ and $\thn \rightarrow \thnp$ to
Eqs.~\rf{UASdmin2} and \rf{UASzE2}, respectively. The results are
\beq{UASdmin2curv}
\zpbuas=-h\ln\left[
e^{-\zpdk/h}-\frac{\dmin}{\dvert}\cos\thnp
\right]\,,
\eeq
and
\beq{UASzE2curv}
\zpeuas=-h\ln\left[e^{-\zpdk/h}-\frac{\dmax}{\dvert}\cos\thnp\right]\,.
\eeq
The $\lmin$ constraint is
\beq{lminprime}
\left[\zpu-\zpl \right]\,\tan\thnp > \lmin\,,
\eeq
with 
\beq{minmaxprime}
\zpu=\min\{\zpeuas, \zthin\}\,,\quad {\rm and}\; \zpl=\zpbuas\,.
\eeq
An algorithm has emerged for including Earth-curvature in our prior
calculations. We simply replace the unprimed variables $\zdk$, $\thn$,
$\zbuas$, $\zeuas$ with $\zpdk$, $\thp$, $\zpbuas$, $\zpeuas$, where
$\zpdk(\zdk,\thn)$ and $\thnp(\zdk,\thn)$ are given in
Eqs.~\rf{zprime}-\rf{tanprime}, and $\zpbuas$ and $\zpeuas$ are given 
in Eqs.~\rf{UASdmin2curv} and \rf{UASzE2curv}. Note that the
parameters $\zthin,\zg$, and soon-to-be-introduced $\zc$ are never
primed, for they define layers concentric with the spherical Earth. 

At this point we can summarize very easily the UAS constraints
including the Earth's curvature (in the absence of cloud cover). We
have again the general rate equation, Eq.~\rf{UASrate}, with inputs
from Eqs.~\rf{Pnutotau}, \rf{dnu}, \rf{Etau}-\rf{lifetime} and
\rf{area}. The \toothin constraint is given by
Eq.~\rf{zmax_constraint1}, or equivalently, by
Eqs.~\rf{zmax_constraint2} and \rf{wdkmax}. The $\lmin$ constraint is
given by Eqs.~\rf{lminprime} and \rf{minmaxprime}, with $\zpbuas$ in
Eq.~\rf{UASdmin2curv} implementing the $\dmin$ constraint and
$\zpeuas$ in Eq.~\rf{UASzE2curv} implementing the $\dmax$
constraint. Unprimed variables are obtained from primed variables via
Eqs.~\rf{zprime}-\rf{tanprime}, or to a very good approximation via
the simplified Eqs.~\rf{cosprime2}-\rf{zpdk2}. In any case, we have
used the latter for our numerical computations.

As in the ``flat Earth'' calculation, more algebra can be done when
Eq.~\rf{minmaxprime} is substituted into Eq.~\rf{lminprime}. The
result is Eq.~\rf{UASlmin3} for the $\lmin$ constraint if
$\zpuuas<\zphatu$, and Eq.~\rf{UASlmin4} if $\zpuuas>\zphatu$, where
$(\zdk,\theta_n)\rightarrow (\zpdk,\thnp)$ in Eqs.~\rf{UASlmin3}
and~\rf{UASlmin4}, and $\zphatu$ is defined by priming
Eq.~\rf{UASdmax}:
\beq{zhatprime}
\zphatu\equiv  -h\ln
\left[\frac{\dmax}{\dvert}\cos\thnp+e^{-\zthin/h}\right]\,.
\eeq
However, when these explicit constraints are expressed in terms of the 
unprimed variables, no easy separation of $\zdk$ and $\thn$ appears
to be possible.  Consequently, it seems best to treat the constraints
as non-linear relations in the $(\zdk, \theta)$-integration space.
This is exactly what we do.

As a check on our work, we have regained the ``flat-Earth'' results 
for UAS from the curved-Earth formalism, by taking the Earth's radius
to be very large in the constraint equations (but keeping $\Rearth$
physical in Eq.~\rf{UASrate} and the $L=2\,\Rearth\,\cos\theta_n$
relation).

\section{Shower rates with clouds}
\label{sec:clouds}

The presence of clouds, their distribution, altitude and optical depth
would obviously affect the observed event rates. Hence, when we
consider the effect of a cloud layer on the observable event rate, the
resulting constraints become more complicated. In addition, the
constraints come to depend on whether the detector is above or below
the clouds, i.e., on whether the detector is space-based or
ground-based. We will model the cloud layer in a very simplified way
as an infinitely thin layer, but with infinite optical depth and we
will assign $\zc$ to be the height of the relevant cloud
boundary. However, we will not take into account how the cloud
presence could affect the reconstructed shower geometry and
energy~\cite{AJLclouds}. For a space-based detector, the only visible
air showers are those at $z > \zc$, while for a ground-based detector
only showers at $z < \zc$ can be seen. Let us name the four possible
event-detection types, UAS and HAS as seen from space (S) or from the
ground (G), as \usp, \ugd, \hsp, and \hgd, in obvious notation. Our
calculation is partitioned into four parts, corresponding to these
four event types. We will see that \ugd\ and \hsp\ event types are
easily calculated with only simple modifications of our prior,
cloudless formulas. However, the \usp\ and \hgd\ event types require
more care, since the observed shower may have its origination above or
below the cloud boundary for these cases.

In our simplified model, the actual development of the shower does not
depend on the presence or absence of clouds.  Thus, the expressions
for $\zb$ and $\ze$, determined by the $\dmin$ and $\dmax$
constraints, are unchanged. However, the visible projected length of
the shower certainly depends on the presence or absence of clouds. 

In this section we do not include the effect of the Earth's curvature.
We have seen that Earth curvature does not affect the HAS calculation.
In the next section, \S\ref{sec:cloudsncurvature}, we include the
Earth's curvature along with clouds in the calculations of the UAS
rates.

\subsection{UAS in Ground-Based Detectors}
\label{sec:UASG}

Looking upward with a \ugd\ detector, the observable atmosphere is
bounded from above by the cloud layer. In the absence of clouds, the
observable atmosphere was bounded from above by $\zthin$. Thus, the
prior, cloudless calculation applies to the cloudy atmosphere if we
just reset the \toothin height $\zthin$ to the cloud boundary
$\zc$. 

One feature of this replacement is that the angular constraint on UAS
events, traceable back to Eq.~\rf{abscos}, becomes
\beq{UASGcos}
\cos\thn \le 
\frac{\dvert}{\dmin}\,\left(1-e^{-\zc/h}\right)\nonumber\\
	\equiv \cosuasg 	
\eeq
If $\zc\le -h\ln[(\dvert-\dmin)/\dvert]$, then $\cosuasg$ is less than
1, presenting a real constraint on the shower direction. With
$\dmin=400\,{\rm g/cm}^2$, $\cosuasg \le 1$ occurs for $\zc \le
3.9$~km.  Thus, for cloud boundaries below this value, near-vertical 
showers do not satisfy the $\dmin$ condition. For example, with a
cumulus cloud layer at $\zc =2$~km, only shower angles $\thn>55^\circ$
are allowed by the $\dmin$ constraint. For common extrapolations of
the neutrino-nucleon cross-section to very high energies, only
``Earth-skimming'' neutrinos are expected to emerge from the Earth at
very high energies. For these ``Earth-skimming'' neutrinos, clouds
must be very low to affect the rate.

We rewrite the relevant UAS constraint equations with
$\zthin\rightarrow\zc$ to include the cloud boundary. The \toothin
constraint in Eq.~\rf{toothinUAS} becomes
\beq{toothinUASG}
0\le\zdk < \zc \,.
\eeq
The $\lmin$ constraint Eq.~\rf{UASlmin2} is replaced with
\beq{UASGlmin2}
\left[\min\{\zeuas, \zc\}-\zbuas\right] \tan\thn \ge \lmin\,,
\eeq
with $\zeuas$ and $\zbuas$ as before, given in Eqs.~\rf{UASzE2} and
\rf{UASdmin2}, respectively. This concludes the implicit calculation
of the integration boundaries.

Explicit boundaries are obtained by substituting Eqs.~\rf{UASzE2} and
\rf{UASdmin2} into \rf{UASGlmin2}. The result is that the $\lmin$
constraint is given by Eq.~\rf{UASlmin3} when $\zdk<\zhatugd$, 
and by 
(derivative from Eq.~\rf{UASlmin4} via $\zthin\rightarrow\zc$) 
%
\beq{UASGlmin4}
\zdk < -h\ln\left[ \frac{\dmin}{\dvert}\cos\thn
+e^{(\frac{\lmin}{\tan\thn}-\zc)/h} \right]  \,,
\eeq
%
when $\zdk>\zhatugd$;
the critical altitude $\zhatugd$ (derivative from Eq.~\rf{UASdmax} via
$\zthin\rightarrow\zc$) is given by 
\beq{UASGdmax}
 \zhatugd \equiv  -h\ln\left[
e^{-\zc/h}+\frac{\dmax}{\dvert}\cos\thn \right] \,.
\eeq

As with the no-clouds case, these $\lmin$ constraints require a
sufficiently large normal angle so that the horizontal projection of
the shower is visible. The critical angle is obtained from
Eq.~\rf{thncrit} with the substitution $\zthin$ by $\zc$. The
resulting equation is
\beq{UASGBclouds}
\zc = -h\ln\left[\frac{\dmax-\dmin}{\dvert} \, 
  \frac{\cos\thn^{\rm crit}}{e^{\frac{\lmin}{h \tan\thn^{\rm crit}}}
    -1} \right] \,. 
\eeq
The meaning is that given a cloud layer at $\zc$, there are no visible
events for $\cos\thn > \cos\thn^{\rm crit}$. Since $\zc$ is
a monotonically increasing function of $\cos\thn^{\rm crit}$, this
result may be stated in a different way: there are no visible events
at $\cos\thn > \cos\thn^{\rm crit}$ if there is a cloud layer lower
than that of Eq.~\rf{UASGBclouds}. There are no visible events at all
if $\cos\thn^{\rm crit}=0$. From Eq.~\rf{UASGBclouds}, this occurs for
the critical cloud altitude 
\beq{UASGBclouds2}
\zcc = -h\ln\left[\frac{\dmax-\dmin}{\dvert} \,
  \frac{h}{\lmin} \right] \,.  
\eeq
Thus, clouds completely obscure the detector if $\zc < \zcc$.

To summarize the formulas giving the \ugd\ events even in the presence
of cloud cover, the general rate equation is Eq.~\rf{UASrate}, with
inputs from Eqs.~\rf{Pnutotau}, \rf{dnu}, \rf{Etau}-\rf{lifetime} and
\rf{area}, the \toothin constraint is given by Eq.~\rf{toothinUASG},
the $\lmin$ constraint by Eq.~\rf{UASGlmin2}, with $\zeuas$ and
$\zbuas$ unchanged from their cloudless expressions, Eqs.~\rf{UASzE2}
and \rf{UASdmin2}, respectively. Alternatively, explicit $\lmin$
constraints are available in Eq.~\rf{UASlmin3} for $\zdk<\zhatugd$,
and in \rf{UASGlmin4} for $\zdk>\zhatugd$, with $\zhatugd$ defined in
Eq.~\rf{UASGdmax}. These twin constraints reflect the two possible
outcomes of $\min\{\zeuas, \zc\}$ in Eq.~\rf{UASGlmin2}.

\subsection{HAS in Space-Based Detectors}
\label{sec:HASS}

Looking downward with a \hsp\ detector, the visible atmosphere is
bounded below by the cloud layer. In the absence of clouds, it is
bounded below by $\zg$. Thus, the cloudless calculation applies when
$\zg$ is reset to $\zc$. With this type of substitution in mind, we
retained the symbol $\zg$ in our prior cloudless formulas, even though
its value was zero there.
For example, for the \hsp\ events, 
the angular constraint of Eq.~\rf{abscos} becomes
\beq{HASScos}
\cos\thz \le
\frac{\dvert}{\dmin}\,\left(e^{-\zc/h}-e^{-\zthin/h}\right)
\equiv\coshass \,.
\eeq
This constraint ensures that the shower brightens sufficiently above
the clouds to become visible. If $\zc\ge -h\ln(\dmin/\dvert +
e^{-\zthin/h})$, then $\coshass < 1$, being a real constraint on the
shower direction. With $\dmin=400\,{\rm g/cm}^2$, $\coshass \le 1$ for
$\zc \ge 6.6$~km. Cumulus cloud layers rarely rise to this height, and
so there is no angular constraint on HAS resulting from cumulus
clouds. However, cirrus clouds populate the high atmosphere, and
therefore do constrain the HAS angle $\thz$.

We rewrite the other relevant HAS constraint equations with
$\zg\rightarrow\zc$ to include the cloud boundary. The \toothin
constraint becomes 
\beq{toothinHASS}
\zc\le\zint\le \zthin \,.
\eeq
The $\lmin$ constraint Eq.~\rf{HASlmin2} becomes
\beq{HASSlmin2}
\left[(\zbhas-\max\{\zehas,\zc\})\right]\,\tan\thz \geq \lmin\,,
\eeq
with $\zbhas$ and $\zehas$ as before, given in Eqs.~\rf{zB2} and
\rf{zE2}. This concludes the implicit calculation of the integration
boundaries.

Explicit boundaries are obtained by substituting Eqs.~\rf{zB2} and
\rf{zE2} into \rf{HASSlmin2}. The result is that the $\lmin$
constraint is given by Eq.~\rf{HASlmin3}, for $\zint>\zhathsp$, and by
the following (derived from Eq.~\rf{HASlmin4} via $\zg\rightarrow\zc$) 
for $\zint<\zhathsp$:
\beq{HASSlmin4}
\zint> -h\ln
\left[
e^{-(\frac{\lmin}{\tan\thz}+\zc)/h}-\frac{\dmin}{\dvert}
\cos\thz \right]\,;
\eeq
the critical altitude $\zhathsp$, derivative from Eq.~\rf{HASdmax} via
$\zg\rightarrow\zc$, is
\beq{HASSdmax}
 \zhathsp \equiv  -h\ln\left[
e^{-\zc/h}-\frac{\dmax}{\dvert}\cos\thz \right] \,.
\eeq
To summarize the formulas giving the \hsp\ events even in the presence
of a cloud layer, the general rate equation is Eq.~\rf{HASrate1}, with
the area given in Eq.~\rf{area}, the \toothin constraint is given
by Eq.~\rf{toothinHASS}, and the $\lmin$ constraint by
Eq.~\rf{HASSlmin2}, with $\zbhas$ and $\zehas$ unchanged from their
cloudless expressions. Alternatively, explicit expressions for the
$\lmin$ constraint are available in Eq.~\rf{HASSlmin4} for $\zint
<\zhathsp$, and in Eq.~\rf{HASlmin3} for $\zint>\zhathsp$, with
$\zhathsp$ defined in Eq.~\rf{HASSdmax}. These twin constraints
reflect the two possible outcomes of $\max\{\zehas,\zc\}$ in
Eq.~\rf{HASSlmin2}.

\subsection{UAS in Space-Based Detectors} 
\label{sec:UASS}

Looking downward with a \usp\ detector, the visible atmosphere is
bounded below by the cloud layer, but the visible UAS may have begun
its development above or below the clouds. Thus, the \toothin
constraint remains Eq.~\rf{toothinUAS} as in the cloudless
calculation. The new $\lmin$ constraint is
%
\beq{UASSlmin2}
\left[\min\{\zeuas,\zthin\}-\max\{\zbuas,\zc\}\right]\,\tan \thn \ge
\lmin 
\eeq
%
with $\zeuas$ and $\zbuas$ as before, given in Eqs.~\rf{UASzE2} and
\rf{UASdmin2}, respectively. This concludes the implicit calculation
of the integration boundaries. The left-hand side of
Eq.~\rf{UASSlmin2} makes it clear that generation of a visible UAS
requires $\zeuas>\zc$ and $\zbuas<\zthin$, and that the clouds are
irrelevant when $\zbuas>\zc$.

Explicit boundaries, if desired, are obtained by substituting
Eqs.~\rf{UASzE2} and \rf{UASdmin2} into \rf{UASSlmin2}. There are
$4!=24$ {\sl a priori} orderings of the four parameters in the $\lmin$
constraint.  However, the orderings $\zeuas>\zbuas$ and $\zthin>\zc$
are fixed. This leaves $4!/(2\cdot2)=6$ possible orderings of the
parameters. Of these six orderings, one has $\zbuas>\zthin$ and
another has $\zeuas<\zc$. These orderings do not produce an observable
shower, the former showering too late and the latter showering too
early. We are left with four relevant orderings:\\
(a) $\zthin>\ze>\zb>\zc$;\\
(b) $\ze>\zthin>\zb>\zc$;\\
(c) $\zthin>\ze>\zc>\zb$;\\
(d) $\ze>\zthin>\zc>\zb$.\\
The first two, (a) and (b), are characterized by $\zb>\zc$, which
restricts $\zdk$ according to
\beq{new1}
\zdk > -h \ln\left[ e^{-\zc/h}+\frac{\dmin}{\dvert}\,\cos\thn
  \right]\,.
\eeq
Ordering (a) characterizes the shower that extinguishes (i.e.\
$\zdk<\zhatu$), while (b) characterizes the shower that reaches the
\toothin air boundary (i.e.\ $\zdk>\zhatu$). For the two orderings~(a)
and~(b), the clouds do not obscure any part of the visible shower, 
and the $\lmin$ formulas of the cloudless section
\S\ref{sec:visibleUAS} apply. The next two orderings, (c) and (d), are
characterized by $\zc>\zb$. For these two orderings, the clouds do
obscure part of the visible shower.

However, it may be that for low values of $\zc$, there are no events
satisfying the topologies specified in (c) and (d) {\sl regardless of
  whether the cloud layer is actually present at $\zc$.} For example,
with our canonical value $\zthin=3 h$, the conditions in~(d) require
that the shower remain visible over a vertical length at least as long
as ($3 h - \zc$). For a small value of $\zc$, such a shower will never
happen, as showers cannot both begin below $\zc$ and survive beyond
$3h$. Thus, in order to see if there is some critical altitude
below which clouds would not suppress the rates in \usp\ detectors, we
seek the conditions for which categories~(c) and (d) do not contribute
events {\sl even in the absence of clouds}. Under such conditions, the
acceptance is just that of the cloudless case.

For ordering (d), i.e.\ $\zeuas>\zthin$ and $\zc>\zbuas$, the $\lmin$
constraint presents a restriction on $\thn$ alone,
\beq{UASSlmin3}
\left[\zthin-\zc\right]\,\tan\thn \geq \lmin\,, \quad \rm{or} \quad
\zc \le \zthin - \lmin/\tan\thn \,.
\eeq
For ordering (c), i.e., $\zthin>\zeuas$ and $\zc>\zbuas$, the $\lmin$
constraint is
\beq{UASSlmin4}
\left[\zeuas-\zc\right]\,\tan\thn \geq \lmin \,.
\eeq
Inputting $\zeuas$ from Eq.~\rf{UASzE2} leads to a restatement of this
latter constraint as 
\beq{UASSlmin5}
\zdk>-h\ln\left[
\frac{\dmax}{\dvert}\cos\thn
+e^{-\left(\frac{\lmin}{\tan\thn}+\zc\right)/h} \right] \,.
\eeq
Hence, the altitude on the RHS expresses the minimum altitude above
which tau decays contribute events to category~(c). Since we have
argued that the minimum altitude for shower development in
category~(d) is higher than in~(c), the RHS also expresses the
minimum altitude above which tau decays contribute events to
categories~(c) and~(d). Thus, the RHS is the minimum altitude above
which tau decays produce showers partially obscured by clouds.

This constraint on $\zdk$ is analogous to the one in
Eq.~\rf{UASlmin3} which sets the minimum value for $\zdk$ in the
absence of clouds. Since clouds must be more restrictive than no
clouds, the RHS of Eq.~\rf{UASSlmin5} must be larger than $z_<$ (RHS
of Eq.~\rf{UASlmin3}) if categories~(c) and~(d) are to have events. 
This happens if  
\beq{UASSBclouds}
\zc > -h\ln\left[\frac{\dmax-\dmin}{\dvert} \, \frac{\cos\thn \,
    e^{\frac{\lmin}{h \tan\thn}}}{e^{\frac{\lmin}{h \tan \thn}}-1}
  \right] 
\eeq
This is a necessary condition for events to fall into categories~(c)
and (d). Consequently, the inequality of opposite sign is the {\sl
sufficient} condition for clouds to {\sl not} obscure the showers. The
RHS of Eq.~\rf{UASSBclouds} is a critical cloud altitude.

For the typical parameter choices which we consider, the RHS
of Eq.~\rf{UASSBclouds}, when positive, has a very weak dependence on 
$\cos\thn$.  In particular, it does not differ much from its value 
evaluated at $\cos\thn = 0$, which is
\beq{UASSBcritz}
\zcc = -h\ln\left[\frac{\dmax-\dmin}{\dvert} \,
  \frac{h}{\lmin} \right] \,.  
\eeq
Thus, we may use Eq.~\rf{UASSBcritz} as a very good approximation to 
the RHS of~\rf{UASSBclouds}. The approximation becomes even better as
the cross-section becomes larger, for then events come from more
horizontal neutrino trajectories. Conveniently, the definition of
$\zcc$ in Eq.~\rf{UASSBcritz} is identical to that in
Eq.~\rf{UASGBclouds2}. Thus, for all practical purposes we can use the 
same $\zcc$ as the critical altitude for space-based and ground-based
detectors.~\footnote{We have checked numerically the accuracy of this
  statement.} Thus we have the complementary situation that clouds
below $\zcc$ completely obscure \ugd, but do not affect \usp\ at all. 
Of course, clouds above $\zcc$ will partially obscure both \ugd\ and
\usp. 

This concludes the explicit construction of the \usp\ constraints for
all four allowed orderings of the parameters. The final event rate is
the sum of the contributions from the four allowed orderings. As a
check, we note that in the limit $\zc\rightarrow 0$, orderings (c) and
(d) no longer contribute, since $\zb > 0$. Thus, we are left with just
the two orderings (a) and (b) of the cloudless limit.

To summarize the formulas giving the \usp\ events even in the presence
of a cloud layer, the general rate equation is Eq.~\rf{UASrate}, with
inputs from Eqs.~\rf{Pnutotau}, \rf{dnu}, \rf{Etau}-\rf{lifetime}, and
\rf{area}. The \toothin constraint is given by \rf{toothinUAS},
the $\lmin$ constraint by Eq.~\rf{UASSlmin2}, with $\zeuas$ and
$\zbuas$ unchanged from their cloudless expressions~\rf{UASzE2} and
\rf{UASdmin2}. Explicit solutions for the $\lmin$ constraint, if
desired, are given above for the four allowed orderings of the
parameters $\{\ze,\zthin,\zb,\zc\}$. The total event rate is the sum
of the four contributions.

\subsection{HAS in Ground-Based Detectors}
\label{sec:HASG}

Looking upward with a \hgd\ detector, the visible atmosphere is
bounded above by the cloud layer, but the visible HAS may have begun
its development above or below the clouds. Thus, the \toothin
constraint remains Eq.~\rf{toothinHAS} as in the cloudless
calculation. The $\lmin$ constraint is
\beq{HASGlmin2}
\left[\min\{\zbhas,\zc\}-\max\{\zehas,0\}\right]\,\tan\thz\ge\lmin
\eeq
with $\zbhas$ and $\zehas$ as before, given in Eqs.~\rf{zB2} and
\rf{zE2}, respectively. This concludes the implicit calculation of the
integration boundaries. Note that the left-hand side of
Eq.~\rf{HASGlmin2} makes it clear that generation of a visible shower
requires $\zehas<\zc$, and that the clouds are irrelevant when
$\zbhas<\zc$.

As was the case with \usp\ events, there are four orderings that
contribute to the \hgd\ events:\\
(a) $\zc>\zb>\ze> 0 $;\\
(b) $\zc>\zb> 0 >\ze$;\\
(c) $\zb>\zc> 0 >\ze$;\\
(d) $\zb>\zc>\ze> 0 $.\\
The first two, (a) and (b), are characterized by $\zc>\zb$. For these
two, the clouds do not obscure any part of the visible shower, 
and the $\lmin$ formulas of the
cloudless section \S\ref{sec:visibleHAS} apply (ordering (a)
characterizes the shower that extinguishes, while (b) characterizes
the shower that reaches the ground). 
The next two orderings, (c) and (d), are characterized by $\zb>\zc$. 
Here, the clouds do obscure part of the visible shower. 
For ordering (c), i.e., $\zbhas>\zc$ and $0>\zehas$,
the $\lmin$ constraint presents a restriction on $\theta_n$ alone.
It is 
\beq{HASGlmin3}
\zc\tan\thz \geq \lmin \, .
\eeq
For ordering (d), i.e.\ $\zbhas>\zc$ and $\zehas>0$,
\beq{HASGlmin4}
\left[\zc-\zehas\right]\,\tan\thz \geq \lmin \, .
\eeq
A short calculation leads to a restatement of this latter constraint as
\beq{HASGlmin5}
\zint < -h\ln\left[
e^{\left(\frac{\lmin}{\tan\thz}-\zc\right)/h}-
\frac{\dmax}{\dvert}\cos\thz
\right]
\eeq
This concludes the explicit construction of the constraints for all
four allowed orderings of the parameters. The final event rate is the
sum of the contributions from the four allowed orderings. As a check,
we note that in the limit $\zc\rightarrow\zthin$, orderings (c) and
(d) no longer contribute, since $\zb <\zthin$. Thus, we are left with
just the two orderings (a) and (b) of the cloudless limit.  

To summarize the formulas giving the \hgd\ events even with a cloud
layer present, the general rate equation is Eq.~\rf{HASrate1}, with the
area given in Eq.~\rf{area}. The \toothin constraint is given by
Eq.~\rf{toothinHAS}, the $\lmin$ constraint by Eq.~\rf{HASGlmin2}, with
$\zehas$ and $\zbhas$ unchanged from their cloudless
expressions Eqs.~\rf{zE2} and \rf{zB2}. Explicit solutions for the
$\lmin$ constraint, if desired, are given above for the four allowed
orderings of the parameters $\zbhas,\zc,\zehas$, and $\zg=0$. The
total event rate is the sum of the four contributions.

\subsection{Remark on UAS with clouds}
\label{subsec:UASremark}

The RHS of Eq.~\rf{UASGBclouds} gives the critical altitude below
which clouds would completely obscure ground-based detection. The RHS
of Eq.~\rf{UASSBcritz} gives the critical altitude below which clouds
would not affect space-based detection.~\footnote{To be accurate, it is
  the minimum of the RHS of~\rf{UASSBclouds} which gives the
  critical altitude, as we explained in Section~\ref{sec:UASS}.}
These two equations are inverse to each other in meaning, but
numerically the critical altitudes on the RHS's in these two equations
are identical:
\beq{zcritcloud}
\zcc \equiv
-h\ln\left[\frac{\dmax-\dmin}{\lmin}\,\,\frac{h}{\dvert}\right] \,. 
\eeq
Here we have factored the argument of the logarithm into a ratio
$(\dmax-\dmin)/\lmin$ which is determined by the experimental triggers 
(i.e., by humans and their optics), and the ratio
$h/\dvert=1/\rho(0)$, which is Nature's gift of our atmospheric
density at sea level (cf.\ Eq.~\rf{dvert}).  Numerically, the latter
term is  $8\,{\rm  km}/1030\,{\rm g}\,{\rm cm}^{-2}= (129\,{\rm
  g}\,{\rm cm}^{-2}/{\rm km})^{-1}$. When the argument of the
logarithm is $<1$, this equation has a positive solution, and so
sufficiently low-lying clouds will completely obscure \ugd, but not
affect at all \usp. There will always be a positive solution $\zc$
whenever $(\dmax-\dmin)/\lmin$ is less than
$\dvert/h=\rho(0)=129\,{\rm g}\,{\rm cm}^{-2}/{\rm km}$.

The range of a visible shower at or near $10^{20}$~eV is comparable to
$\dvert$, and so $(\dmax-\dmin)/\dvert$ is of order unity. Typically,
the visible length required for shower identification is of order of
$h=8$~km, so $h/\lmin$ is also of order unity. Thus, the argument of
the logarithm is of order unity. Consequently, whether there can be
significant cloud obscuration in \ugd\ or no effect in \usp\ will
depend critically on an experiment's choice of shower parameters,
$\dmin$, $\dmax$, and $\lmin$. As relevant examples, with the choices
$\dmin=400\,(300)\,{\rm g/cm}^2$ and $\dmax=1200\,(1500)\,{\rm
  g/cm}^2$, for $\lmin = 10$~km the argument of the log is $<1$,
$\zcc$ is 3.8~(0.56)~km, and so a cloud layer at a lower altitude will
completely suppress observation in \ugd\ detectors and have no effect
in \usp\ detectors. In contrast, for $\lmin = 5$~km the argument of
the logarithm exceeds unity, there is no $\zc$,  and so there is
partial rate suppression due to clouds at any altitude for both \ugd\
and \usp. 

\begin{figure}[t]
\includegraphics[height=3.25in]{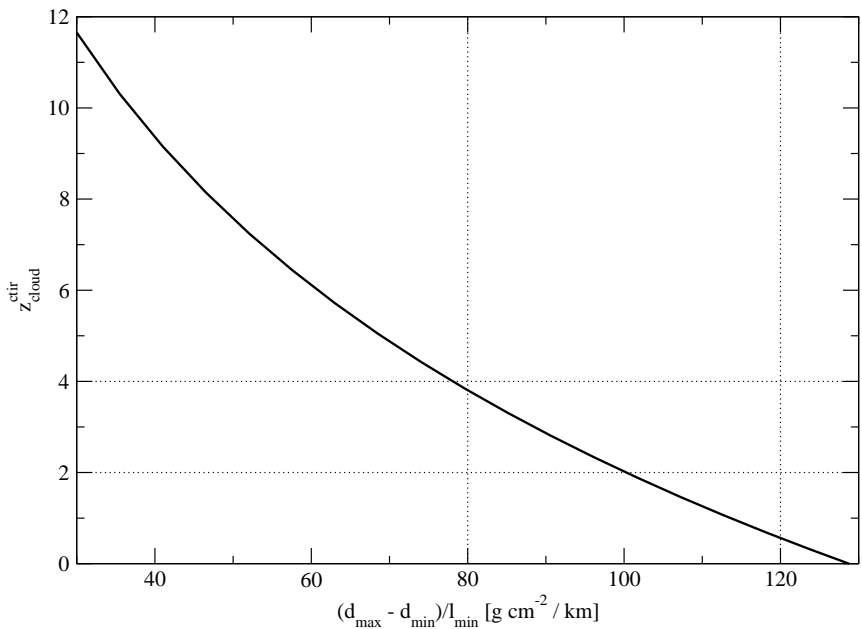}
\caption{\label{fig:zcrit} 
  Critical altitude below which clouds would obscure the detector for 
  \ugd\ and have virtually no effect for \usp. Above the critical
  altitude, clouds would partially obscure both  \ugd\ and \usp.
  The label on the abscissa can be thought of as an experimental
  sensitivity to showers. We have taken $\dvert/h=1030\,{\rm
  g\,cm}^{-2}/8\,{\rm km}$.
} 
\end{figure}

In Fig.~\ref{fig:zcrit} we plot $\zcc$ versus the trigger-parameter
combination $(\dmax-\dmin)/{\lmin}$, over the range
30-130~km/g\,cm$^{-2}$. As foretold, for values above
$\dvert/h=\rho(0)=129\,{\rm km}/{\rm g}\,{\rm cm}^{-2}$, the solution
to Eq.~\rf{zcritcloud} is negative and partial cloud suppression
occurs in both \ugd\ and \usp. But for $(\dmax-\dmin)/{\lmin}$ less
than $129\,{\rm km}/{\rm g}\,{\rm cm}^{-2}$, there is a positive
critical altitude delineating total cloud suppression as seen from
ground and no cloud suppression as seen from space, from partial
suppression of both. We infer from the figure that an experimental
trigger $(\dmax-\dmin)/{\lmin}$ exceeding $(50, 80, 100)\,{\rm
  km}/{\rm g}\,{\rm cm}^{-2}$ is required for (i) \ugd\ to avoid
complete rate suppression from clouds below $(7.6,3.8,2.0)$~km,
while suffering partial suppression from clouds above
$(7.6,3.8,2.0)$~km; (ii) \usp\ to have partial rate suppression from 
clouds above $(7.6,3.8,2.0)$~km, while suffering no suppression from
clouds below $(7.6,3.8,2.0)$~km.

The ground-based result is intuitive, in that as the trigger
sensitivity $(\dmax-\dmin)/{\lmin}$ is increased, the experiment may
tolerate clouds ever closer to the ground. The space-based result is
less intuitive. In this case, as the trigger sensitivity is increased,
the critical cloud altitude above which UAS are partially obscured is
{\sl again lowered} (after all, the same $\zcc$ is common to \ugd\ and
\usp). The reason is that with better triggering, a space-based
experiment is sensitive to more effective atmospheric volume, and so
to the presence of lower cloud layers. Put another way, with better
sensitivity a space-based experiment may see deeper into the
atmosphere where the air is denser, but not if there are low-lying
clouds.

Notice that we have used our flat-Earth formulas to derive $\zc$, and
the conclusions that follow from it. In particular, we have expressed
the condition for cloud suppression of \ugd\ rates and (to very good
approximation) cloud non-suppression of \usp\ rates analytically
without regard to the angles of UAS trajectories. One may ask whether
Earth's curvature alters our discussion. Unfortunately, inclusion of
curvature leads to transcendental equations, rather than to an
improved simple analytic expression. However, from numerical studies
we can attest that curvature does not alter our qualitative
conclusions. In fact, when only small horizontal-angle events
contribute (which holds for most of the cross-section range, cf.\
Eq.~{\rf{thetahor}), then our quantitative conclusions are accurate,
  too.

Finally, we remark that there is no analogue of $\zc$ for cloud
suppression of HAS rates. The HAS constraints are different from the
UAS, and HAS trajectories are not restricted to small
horizontal-angles.

\section{UAS with Clouds and Earth-Curvature}
\label{sec:cloudsncurvature}

The distance between interaction point and shower extinction for HAS
events is sufficiently small that Earth-curvature can be
neglected. However, we have seen that Earth-curvature cannot be 
neglected for UAS events, since the tau decay path at high energy 
provides a length large on the scale of the atmospheric height
$\zthin$.

We consider again the two UAS possibilities, viewed from space and
viewed from ground. The results of sections~\ref{sec:UASG} and
\ref{sec:UASS} for UAS rates with clouds are extended to include also
Earth's curvature by priming appropriate variables.

The algorithm for priming was presented in \S\ref{sec:curvature}.
According to the algorithm, the Earth's curvature is added to our
prior ``flat-Earth'' calculations by simply replacing the unprimed
variables $\zdk,\thn,\zbuas,\zeuas$ with primed variables
$\zpdk,\thnp,\zpbuas,\zpeuas$, where $\zpdk(\zdk,\thn)$ and
$\thnp(\zdk,\thn)$ are given in Eqs.~\rf{zprime}-\rf{tanprime}, and
$\zpbuas$ and $\zpeuas$ are given by priming $\zbuas,\zeuas,\zdk,\thn$
in Eqs.~\rf{UASdmin2} and \rf{UASzE2} to get Eqs.~\rf{UASdmin2curv}
and \rf{UASzE2curv}. The parameters $\zthin,\zg,\zc$ are never primed,
for they define layers concentric with the spherical Earth. In what
follows, we apply this algorithm explicitly to the ground-based and
space-based UAS rates.

\subsection{UAS Viewed from Ground, with Clouds and Curvature}
\label{sec:UASGcurvature}

Priming the appropriate variables of \S\ref{sec:UASG}, the \toothin
constraint in Eq.~\rf{toothinUASG} becomes
\beq{toothinUASGcurv}
0\le\zpdk < \zc \,.
\eeq
The $\lmin$ constraint Eq.~\rf{UASGlmin2} becomes
\beq{UASGlmin2curv}
\left[\min\{\zpeuas, \zc\}-\zpbuas\right] \tan\thnp \ge \lmin\,.
\eeq
This concludes the inclusion of curvature in the integration boundaries
of the \ugd\ rate with clouds. 

To summarize the formulas giving the \ugd\ events, the general rate
equation is Eq.~\rf{UASrate}, with inputs from Eqs.~\rf{Pnutotau},
\rf{dnu}, \rf{Etau}-\rf{lifetime} and \rf{area}, the \toothin
constraint is given by \rf{toothinUASGcurv}, and the $\lmin$
constraint by Eq.~\rf{UASGlmin2curv}, with $\zpeuas$ and $\zpbuas$
given in Eqs.~\rf{UASzE2curv} and \rf{UASdmin2curv}, respectively.

\subsection{UAS Viewed from Space, with Clouds and Curvature}
\label{sec:UASScurvature}

Priming appropriately, the \toothin constraint in Eq.~\rf{toothinUAS}
becomes 
\beq{toothinUASScurv}
0\le\zpdk < \zthin \,,
\eeq
and the $\lmin$ constraint in Eq.~\rf{UASSlmin2} becomes
\beq{UASSlmin2curv}
\left[\min\{\zpeuas,\zthin\}-\max\{\zpbuas,\zc\}\right]\,
\tan\thnp\ge\lmin
\eeq
This concludes the inclusion of curvature in the integration boundaries
of the \usp\ rate with clouds. 

To summarize the formulas giving the \usp\ events, the general rate
equation is Eq.~\rf{UASrate}, with inputs from Eqs.~\rf{Pnutotau},
\rf{dnu}, \rf{Etau}-\rf{lifetime}, and \rf{area}. The \toothin
constraint is given by Eq.~\rf{toothinUASScurv}, and the $\lmin$ 
constraint by Eq.~\rf{UASSlmin2curv}, with $\zpeuas$ and $\zpbuas$
given in Eqs.~\rf{UASzE2curv} and \rf{UASdmin2curv}, respectively.

\section{Results}
\label{sec:results}

In this section, we present the results of our semi-analytical
approach. For ground- and space-based detectors, we show the 
dependence of the acceptance for HAS and UAS events on neutrino
energy, threshold energy, shower length, and shower column density, as 
a function of the neutrino-nucleon cross-section. For incident neutrino
energies, we choose $E_\nu=10^{20}$ and $10^{21}$ for illustration, and
demonstrate that the ratio of HAS-to-UAS events resulting from these
energies would be of great help in determining the neutrino-nucleon
cross-section at these very high energies. For the UAS sample, we
compute the acceptance for taus emerging over land from pure rock, and
separately for taus emerging over the ocean from a water layer
overlaying a rock layer; we take the water layer to have a uniform
depth of 3.5~km. We also consider the deleterious effects of low or
high cloud layers in the atmosphere, as viewed from space and from the
ground.

In all figures, we take the FOV and solid angle entering the
acceptance calculations to be that of the EUSO design
report~\cite{EUSO}. This FOV area, entering Eq.~\rf{area}, is
$\pi\times(400/\sqrt{3})^2\,{\rm km}^2$. The solid angle is $2\,\pi$ 
for either the HAS or the UAS events. The product of area and solid
angle is then, very nearly $10^6\,{\rm km}^2\;{\rm sr}$.~\footnote{A
  simple estimate of the instantaneous EUSO acceptance for HAS
  cosmic-ray events is readily obtained by multiplying this 
  $A\times\,2\,\pi$ value by $\half$ to account for the mean
  projection of the FOV normal to the source. The result is a
  na\"{\i}ve HAS acceptance of $\sim 5\times 10^5\,{\rm km}^2\;{\rm
    sr}$ for cosmic-rays. For neutrinos, the detection efficiency is 
  less than unity by the factor $\sim 2\,h\,\rho(0)\,\sig$. The factor
  of two arises because the mean path length in the atmosphere of a
  neutrino is twice the vertical value. Put another way, the increased
  interaction probability for oblique trajectories compensates the
  $\half$ coming from projecting the FOV normal to the mean neutrino
  direction (cosines cancel). These simple HAS acceptances assume
  100\% detection efficiencies. Incidentally, the discriminator
  between cosmic-ray initiated HAS and neutrino initiated HAS is the
  depth of origin of the shower in the atmosphere.} The OWL
proposal~\cite{OWL} (two (or more) free-flying satellites) has a
larger FOV, and stereo eyes.

Before proceeding with a comparison of the various acceptance curves,
it is worthwhile to reflect on what kind of event rates might arise in
very-large EUSO/OWL-scale neutrino experiments. The event rate is
obtained by simply multiplying the acceptance by Nature's cosmic
neutrino flux per appropriate flavor. As discussed in earlier
sections, the appropriate flavor for HAS is $\nue$ (and $\nuebar$),
since $\numu$ and $\nutau$ interactions ``lose'' 80\% of their energy
to the escaping charged muon or tau. For UAS, the appropriate flavor is
$\nutau$, since among the charged leptons only the tau has a radiation
length long enough to allow a significant fraction of taus to escape
from the Earth. For the UAS case, we weight the $\nutau$ flux by tau
branching fractions and $\tau\rarr shower$ energy-transfers (2/3 for
hadronic showers, 1/3 for electronic showers, and zero for the muonic
mode).

The cosmic neutrino flux is a matter for pure speculation at present. 
A collection of theoretical fluxes is shown in Ref.~\cite{EEfluxes}.
We will choose as our benchmark a neutrino flux which is ten times the
integrated flux of cosmic-rays at $\egzk$, just below the GZK
suppression. Our benchmark (BM) value is
\beq{benchmark}
 \frac{d{\cal F}_{\rm BM}}{dA\,d\Omega\,dt}\equiv 
   10\times \frac{d{\cal F}_{\rm CR}(>\egzk)}{dA\,d\Omega\,dt}
   = \frac{1}{{\rm km}^2\,{\rm sr}\,{\rm yr}}\,.
\eeq
The factor of ten is included to give a simple number for the
benchmark flux.  

A popular alternative benchmark neutrino flux is that of Waxman and
Bahcall~\cite{WB}, who offered arguments relating the high-energy
neutrino flux to the observed high-energy cosmic-ray flux. They
obtained 
\beq{WBflux}
\frac{d{\cal F}_{\rm WB}}{dA\,d\Omega\,dt} 
= \frac{6\times 10^{-2}\,(10^{20}{\rm eV}/E_\nu)}
    {{\rm km}^2\,{\rm sr}\,{\rm yr}}\,. 
\eeq
Subsequent discussion has shown that their arguments, while sensible,
are not compelling. Predictions of the cosmogenic neutrino
flux~\cite{BZflux}, resulting from charged-pion production in the GZK
process and subsequent pion decay, gives fluxes of order of the WB
benchmark. Proposed sources of a more exotic nature give larger fluxes.

In reality, only Nature knows the value of the real flux. It could be
larger than these benchmarks, it could be smaller, or it could even be
zero. For our benchmark flux, an acceptance of one ${\rm km}^2$-sr is
required to yield one event per year. For Nature's flux, the event
rate is $\frac{d{\cal F}_\nu(E_\nu>E_*)}{dA\,d\Omega\,dt}/
\frac{d{\cal F}_{\rm BM}}{dA\,d\Omega\,dt}$, where $E_*$ is the
minimum neutrino energy producing observable events in the
detector with efficiency of order unity. 
We use units of (${\rm km}^2$-sr) when we plot
acceptances. The fluxes and rates discussed here give a real physical
meaning to these acceptance units. 

\begin{figure}
\includegraphics[width=1.\textwidth]{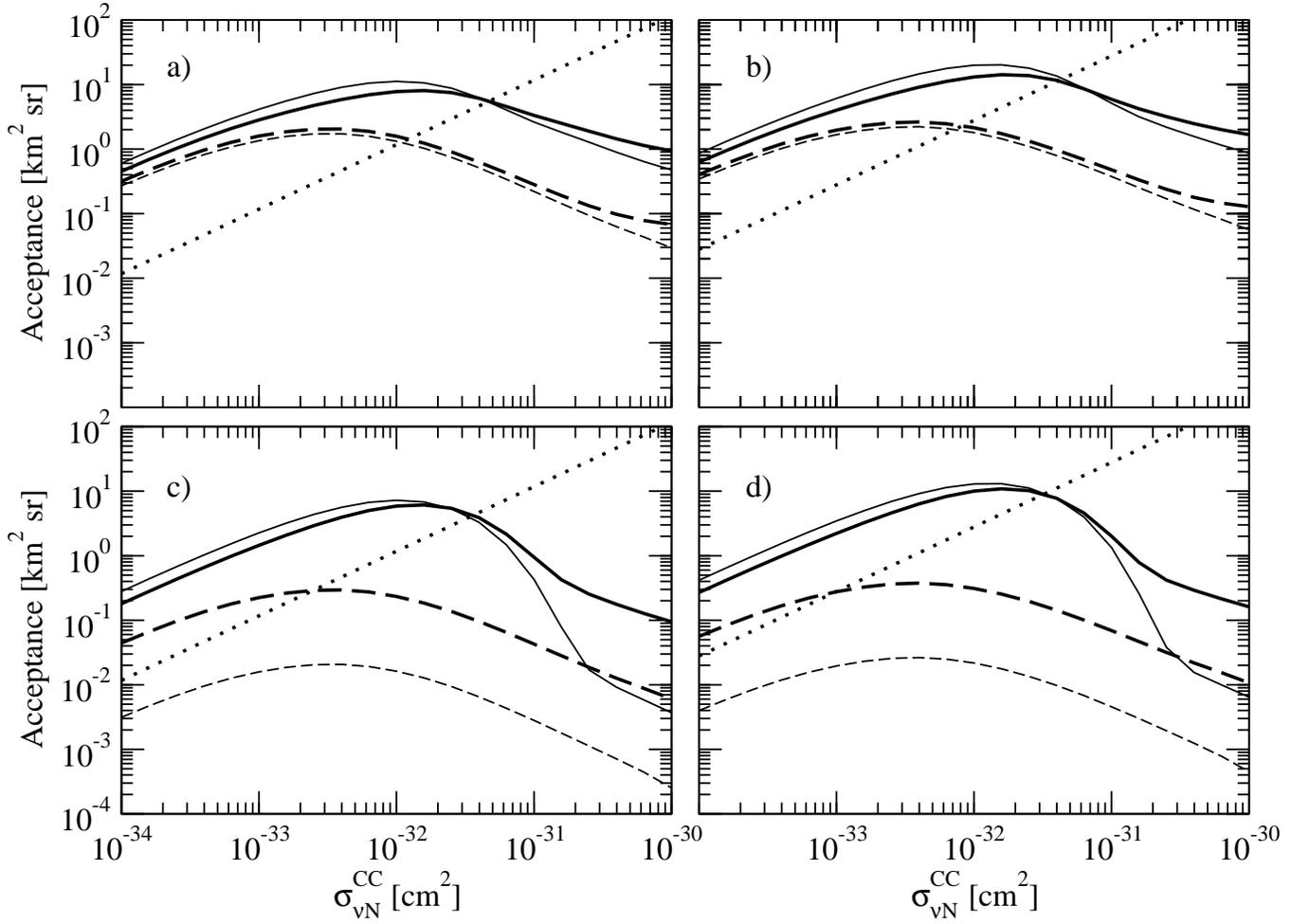}%
\caption{\label{noclouds} 
  Acceptances for space-based (or ground-based) detectors in the
  absence of clouds. Values of $\dmin$ and $\dmax$ are fixed at 400
  and 1200~${\rm g}/{\rm cm}^2$, respectively. The curves correspond
  to HAS (dotted line), which are independent of $E_\nu$ except
  through $\sig (E_\nu)$; and UAS over ocean with $E_\nu=10^{21}$~eV
  (thick solid line), ocean with $E_\nu=10^{20}$~eV (thin solid line),
  land with $E_\nu=10^{21}$~eV (thick dashed line), and land with
  $E_\nu=10^{20}$~eV (thin dashed line). Panels are for 
(a) $\Esh=10^{19}$~eV and $l_{\rm{min}} = 10$ km;
(b) $\Esh=10^{19}$~eV and $l_{\rm{min}} = 5$ km;
(c) $\Esh=5 \times 10^{19}$~eV and $l_{\rm{min}} = 10$ km;
(d) $\Esh=5 \times 10^{19}$~eV and $l_{\rm{min}} = 5$ km.
  For reference, a popular QCD-extrapolation of the neutrino-nucleon
  cross-section~\cite{GQRS} gives 0.54 and 1.2 times $10^{-31}\,{\rm 
  cm}^2$ at $E_\nu=10^{20}$ and $10^{21}$~eV, respectively; the known
  CC cross-section is $2\times 10^{-34}\,{\rm cm}^2$ at an equivalent
  fixed-target energy of $5\times 10^{13}$~eV, the highest energy for
  which measurement has been made (at HERA).
}
\end{figure}

In Fig.~\ref{noclouds} are plotted UAS (solid and dashed) and HAS
(dotted) acceptances in our standard units of (${\rm km}^2$-sr),
versus fixed values of $\sig$, for the ideal case of a cloudless sky. 
Five separate dependences are illustrated in this figure: UAS vs\ HAS; 
$E_\nu=10^{20}$~eV vs.\ $10^{21}$~eV; over ocean vs.\ over land;
shower threshold energy $\Esh=10^{19}$~eV vs.\ $5\times 10^{19}$~eV;
and minimum shower length $\lmin=10$~km vs.\ 5~km. Shower-evolution
parameters are set to $\dmin=400 {\rm g}/{\rm cm}^2$ and $\dmax=1200
{\rm g}/{\rm cm}^2$. A sixth possible dependence is whether the shower
is viewed from above by a space-based observatory, or from the below
by a ground-based observatory. Within the approximations of this
paper, there is no difference between the acceptances for ground-based
and space-based detectors {\sl in the cloudless case}. However, there
are significant up-down differences when the sky includes clouds.

The HAS acceptances depend on neutrino energy only via $\sig (E_\nu)$,
and rise linearly with $\sig$. Plotted against fixed $\sig$, then, the
straight-line HAS curves (dotted) are universal curves valid for any
$E_\nu$ exceeding the trigger threshold $\Esh$. The UAS acceptances
have a complicated dependence on $E_\nu$; it arises from the energy
dependences of $\nu$ propagation in the Earth, tau propagation in the
Earth, and path-length of the tau in the atmosphere before it decays,
the latter also affecting the visible shower characteristics. In each
panel, we show UAS acceptances for two different incident neutrino
energies, $10^{21}$~eV (thick lines) and $10^{20}$~eV (thin
lines). The solid lines show the UAS acceptances for trajectories 
emerging from the ocean, and dashed lines show the UAS acceptances for
trajectories emerging from land, having traveled through only rock. 

Two different shower threshold energies are shown, $\Esh=10^{19}$~eV
in the upper two panels and $\Esh=5\times 10^{19}$~eV in the lower two
panels. The EUSO experiment is working to lower its threshold trigger
from $\Esh=5\times 10^{19}$~eV to $10^{19}$~eV, in order to better
overlap events from the Auger experiment (the Auger threshold is $\sim
10^{18}$~eV). Also explored in the different panels of
Fig.~\ref{noclouds} is the dependence of the acceptance on the minimum
shower-length required for experimental identification. In the two left
panels we have taken $\lmin=10$~km, while in the two right panels we
took $\lmin$ equal to half of that, 5~km.

Several trends are evident in Fig.~\ref{noclouds}. We can clearly see
that the UAS acceptance (and so also the rate) is typically an order
of magnitude larger when neutrinos traverse a layer of ocean water,
compared to a trajectory where they only cross rock. Thus, the UAS
event rate is enhanced over the ocean relative to over
land~\cite{waterenhanced}. The value of this enhancement depends on
the shower threshold-energy $\Esh$ of the detector (upper versus lower
panels) and on the neutrino-nucleon cross-section (the abscissa) in a
non-trivial way. One sees general trends that (i) the larger the
cross-section and  threshold energy are, the larger is the relative
enhancement; (ii) the lower the threshold energy is, the closer
are the acceptances for different initial neutrino energies (thick
vs.\ thin lines); and for high $\Esh$ approaching $E_\nu$, there is a
significant suppression of events over land, and over water for larger
cross-sections. Lower threshold energies are of course also
advantageous in that they necessarily imply larger total event rates.

The sensitivity to $\Eth$ is partly due to the various energy
transfers from the tau to the shower in the different tau-decay
modes. We have remarked that for the hadronic/electronic/muonic decay
modes, $\frac{2}{3}/\frac{1}{3}/0$ of the tau energy goes into the
shower. This means that a neutrino with an incident energy of
$10^{20}$~eV characteristically produces a tau with energy $0.8\times
10^{20}$~eV, which then produces hadronic/electronic showers with mean
energies {\sl at most} (after allowing for the tau's $dE/dx$ in the
Earth) $5.3/2.7\times 10^{19}$~eV. Clearly, the electronic mode is 
below the $\Eth=5\times 10^{19}$~eV threshold, and the hadronic mode
is barely above. Both modes are above the $\Eth=10^{19}$~eV threshold.

We obtain benchmark event rates by multiplying our calculated
acceptances with the benchmark integrated flux, Eq.~\rf{benchmark}, of
one neutrino per $({\rm km}^2\,{\rm sr}\,{\rm yr})$. The result is a
signal exceeding an event per year for an acceptance exceeding a
(${\rm km}^2$-sr).  Thus we see that the benchmark flux gives a HAS
rate exceeding 1/yr if $\sig$ exceeds $10^{-32}\,{\rm cm}^2$; and an
UAS rate exceeding 1/yr over water for the whole cross-section range
with $\Esh=10^{19}$~eV, and over land if $\sig \lsim 10^{-31}\,{\rm
  cm}^2$. When $\Esh$ is raised to $5\times 10^{19}$~eV, however, the
UAS signal over land is seriously compromised, while  UAS rates over
the ocean are little changed. HAS rates are unchanged, as long as
$\Eth$ exceeds $E_\nu$.

It is interesting that in the UAS case over the ocean, the acceptances
at $E_\nu=10^{20}$ and $10^{21}$~eV as a function of $\sig$ are seen
to cross. For lower values of the cross-section, the
acceptance is larger when the initial neutrino energy is smaller,
unlike what might na\"{\i}vely be expected. This is due to the 
combined effect of 
larger nadir angles $\thn$ contributing at lower neutrino energy,
the nature of energy losses in water vs.\ rock, and
the trigger constraints imposed on the showers. For larger
cross-sections, only small-angle Earth-skimming neutrinos contribute,
so the propagation of the tau lepton happens mostly in water;
complications are mainly absent and hence larger initial neutrino 
energies give larger acceptances, in agreement with intuition.

We call attention to the fact that for UAS over both ocean and land,
there is a maximum in the UAS acceptance at cross-section values $\sig
\sim (1-2) \times 10^{-32}~\rm{cm}^2$ and $\sig \sim (0.3-0.5) \times
10^{-32}~\rm{cm}^2$, respectively. For cross-sections similar or
smaller than those at the maximum, the acceptance for UAS is larger 
than that for HAS; conversely, for cross-sections above those at the
maximum, HAS events will dominate UAS events. The cross-section value
at the maximum lies just below the extrapolation of the Standard Model
cross-section, which for the two initial neutrino energies considered,
$10^{20}$~eV and $10^{21}$~eV, is $0.54 \times 10^{-31} \rm{cm}^2$ and
$1.2 \times 10^{-31} \rm{cm}^2$, respectively. If this extrapolation
is valid, then one would expect comparable acceptances (and event
rates) for UAS over water and for HAS; the acceptance for UAS over
land is down from these by an order of magnitude. If the true
cross-section exceeds the extrapolation, then HAS events will dominate
UAS events; if the true cross-section is suppressed compared to the
extrapolation, then UAS events will dominate HAS events. Importantly,
the very different dependences on the cross-section of the HAS
(linear) and UAS acceptances offers a practical method to measure
$\sig$. One has simply to exploit the ratio of UAS-to-HAS event rates.

Furthermore, the shape of the UAS acceptance with respect to $\sig$
establishes the ``can't lose theorem''~\cite{KW}, which states that
although a large cross-section is desirable to enhance the HAS rate, a
smaller cross-section still provides a robust event sample due to the
contribution of UAS.  The latter is especially true over ocean.

Finally, from the comparison of left ($\lmin=10$~km) and right
($\lmin=5$~km) panels, one infers the sensitivity of acceptance to the 
experimental trigger for visible shower length. We see that reducing
the minimum shower length by a factor of two here, increases the
acceptance by roughly a factor of three for HAS, and slightly less for
UAS. So for a cloudless sky, not too much is lost by choosing longer
showers for event reconstruction.  This is fortunate, for, as remarked
early in Section~\ref{sec:visibleAS}, the signal/noise and angular
reconstruction are greater for longer showers. We forewarn that the
sensitivity to the $\lmin$ trigger will become extreme when we
consider a sky with clouds, which we address next.

\begin{figure}
\includegraphics[width=1.\textwidth]{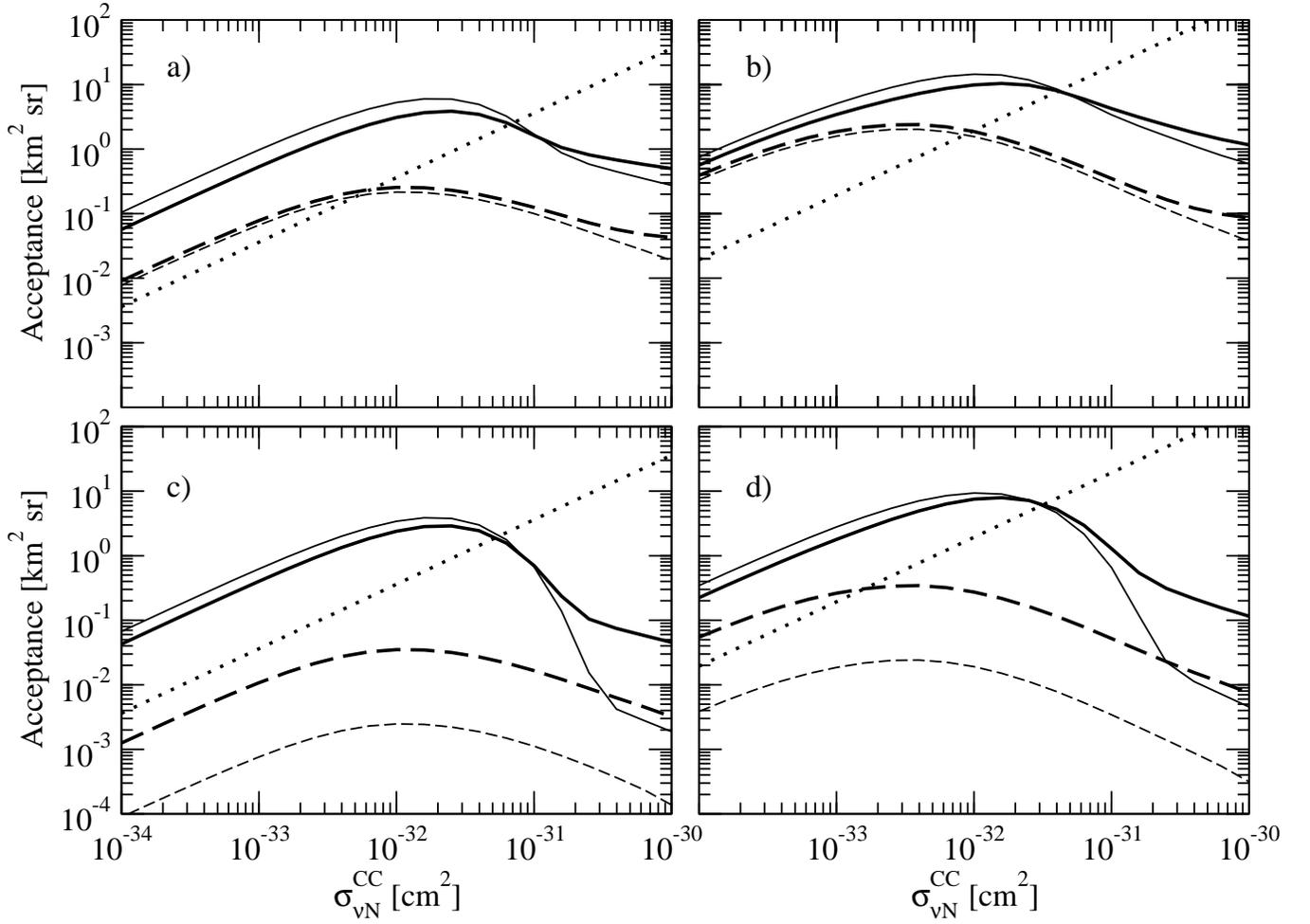}%
\caption{\label{clouds} 
  Acceptances in the presence of a cloud layer at $z_{\rm{cloud}} = 2$
  km; with $l_{\rm{min}}$ fixed at 5~km, $\dmin$ at $400 {\rm g}/{\rm
  cm}^2$, and $\dmax$ at $1200 {\rm g}/{\rm cm}^2$. The curves
  correspond to HAS (dotted line), which are independent of $E_\nu$
  except through $\sig (E_\nu)$; and UAS over ocean with
  $E_\nu=10^{21}$~eV (thick solid line), ocean with $E_\nu=10^{20}$~eV
  (thin solid line), land with $E_\nu=10^{21}$~eV (thick dashed line),
  and land with $E_\nu=10^{20}$~eV (thin dashed line). Panels are for
(a) ground-based detectors with $\Esh=10^{19}$~eV;
(b) spaced-based detectors with $\Esh=10^{19}$~eV;
(c) ground-based detectors with $\Esh=5 \times 10^{19}$~eV;
(d) spaced-based detectors with $\Esh=5 \times 10^{19}$~eV;
} 
\end{figure}

In Fig.~\ref{clouds} are shown the acceptances in the presence of a
cumulus cloud layer at 2 km, again as a function of $\sig$, and again
with $\dmin=400 {\rm g}/{\rm cm}^2$ and $\dmax=1200 {\rm g}/{\rm
  cm}^2$. We model the cloud layer as infinitely thin with altitude
$\zc$, but with an infinite optical depth so that showers are
completely hidden on the far side of the cloud layer. Details of this
modeling were given in Sections~\ref{sec:clouds} and
\ref{sec:cloudsncurvature}. We call low-lying cloud layers
``cumulus'', and high-lying layers ``cirrus'', for obvious reasons.

For a sky with clouds, we show acceptances for $\lmin$ set to 5~km.
We do not show the case with $\lmin=10$~km, because with low-lying
clouds, the UAS acceptances for ground-based detectors are essentially
zero with $\lmin=10$~km (and $\dmin=400\,{\rm g/cm}^2$ and
$\dmax=1200\,{\rm g/cm}^2$). On the other hand, the space-based rates
are virtually unaffected by the low clouds. For the UAS as seen from
the ground, there simply is not enough space below the cloud layer for
the Earth-skimming tau to decay and for the subsequent shower to
develop (see Eqs.~\rf{UASGBclouds} and \rf{UASSBclouds}). The smaller
$\lmin=5$~km that we do show allows enough UAS events to develop into
an observable shower below the cloud layer to establish a meaningful
acceptance for ground-based detectors. Recall, however, that the UAS
rate suppression due to clouds depends sensitively on the value of
$(\dmax-\dmin)/\lmin$. If the value of this is larger than $\dvert/h$,
the suppression is aggravated for space-based UAS and alleviated for
ground-based UAS, and vice versa (as discussed below
Eqs.~\rf{UASGBclouds} and \rf{UASSBclouds}, and more extensively in  
Section~\ref{subsec:UASremark}). 

Since we show one value of $\lmin$, not two, in Fig.~\ref{clouds},
the number of panels is half that in Fig.~\ref{noclouds}. On the
other hand, the symmetry between upward-looking ground-based
detectors, and downward-looking space-based detectors is broken by the
cloud layer, so we must now show separate panels for the space-based
and ground-based detectors. This brings the number of panels back to
four. The HAS and UAS curves are represented in the same way in
Fig.~\ref{clouds} as for Fig.~\ref{noclouds}. The thick and thin lines 
bear the same meanings for the initial neutrino energies. The left
panels show acceptances for ground-based detectors, whereas the right
panels show those of space-based detectors. As with the previous
figure, the upper panels show results for a threshold energy of $\Esh
= 10^{19}$ eV, and the lower panels for $\Esh = 5 \times 10^{19}$ eV.

We see from this figure that qualitative features learned for the
cloudless case apply also in this cloudy case. One difference is that
the UAS acceptances over land  in panel (c) are smaller than the HAS
acceptance over the entire range of $\sig$. One may judge the effect
of clouds by comparing Fig.~\ref{clouds} against the $\lmin=5$~km
panels (b) and (d) of the clear-sky Fig.~\ref{noclouds}.
Quantitatively, the ground-based acceptances (left panels in
Fig.~{noclouds}) are quite reduced by the low-lying clouds, whereas
the space-based acceptances (right panels) are not, as one would
expect. The suppression of the ground-based acceptance is most severe
for small cross-sections, for which the tau leptons emerge more
vertically and disappear into the clouds before their eventual shower
occurs and develops. Ground-based UAS acceptances are reduced by up to
an order of magnitude over water, and even more over land.
Ground-based HAS acceptances, still linear in $\sig$, are reduced by
an order of magnitude. For space-based detectors, the UAS acceptance
is reduced little by clouds at 2~km. Larger neutrino cross-sections
lead to more tangential tau-showers which may hide below a low-lying
cloud layer. We see that UAS reductions are a factor of 2 for the
larger cross-sections shown, and less for the smaller values of cross
section.

The dramatic reduction of ground-based acceptances by low-lying
cumulus clouds begs the question, ``what are the effects of
higher-altitude clouds on space-based detectors?'' In
Fig.~\ref{curvaturedep}, we continue the study of the dependence of 
space-based acceptances on cloud altitude. We also examine the
suppressing effect of the Earth's curvature. 

\begin{figure}
\includegraphics[width=1.\textwidth]{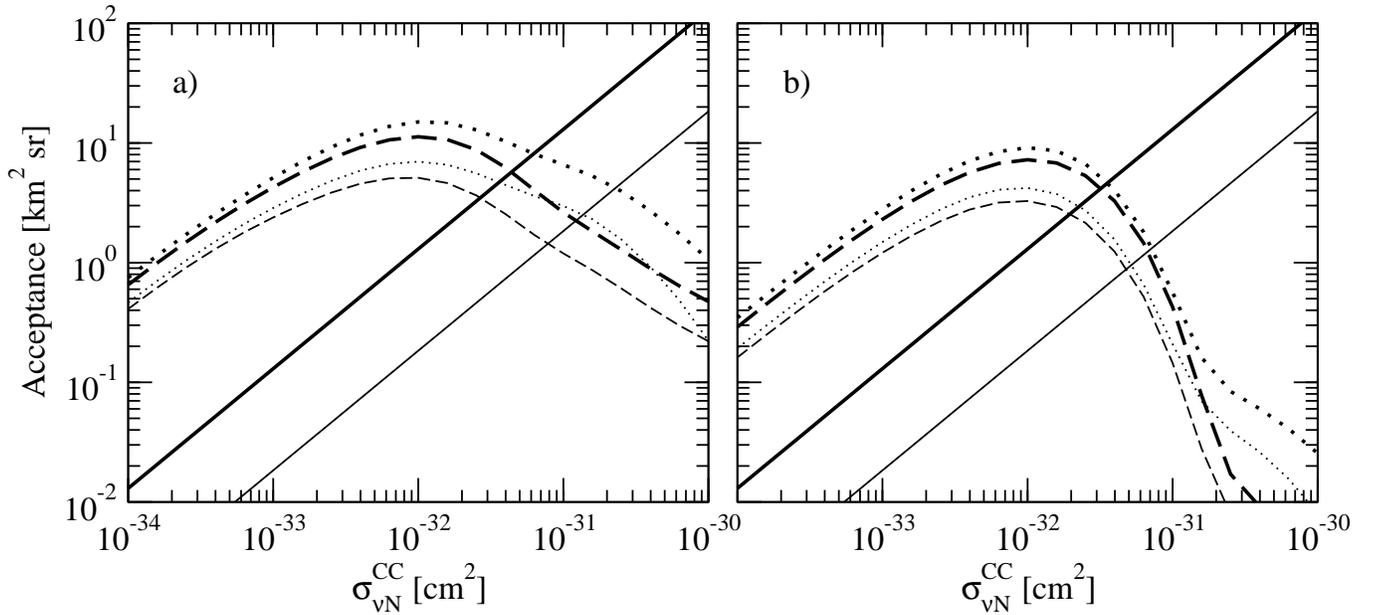}
\caption{\label{curvaturedep} 
  Dependence of acceptance on cloud altitudes for space-based
  fluorescence detectors. Fixed values are $\lmin = 5$~km, and
  threshold energies of $\Esh = 10^{19}$~eV in the left panel and $5
  \times 10^{19}$~eV in the right panel. All curves representing UAS
  assume trajectories over water and an initial neutrino energy of
  $10^{20}$~eV; curves for HAS are valid for any energy exceeding
  $\Esh$. Solid lines show HAS, while dashed and dotted lines show UAS
  with and without Earth-curvature effects, respectively. Thick lines
  are for a cloud layer at 4 km and thin lines for a cloud layer at 12
  km.
}
\end{figure}

Fixed values in Fig.~\ref{curvaturedep} are $\lmin = 5$~km, and
threshold energies of $\Esh = 10^{19}$~eV in the left panel and $5
\times 10^{19}$~eV in the right panel. All curves representing UAS
assume trajectories over water and an initial neutrino energy of
$10^{20}$~eV; curves for HAS are valid for any energy exceeding 
$\Esh$. Acceptances for two cloud altitudes, $\zc = 4$~km (thick 
curves) and 12 km (thin curves), are shown for HAS (solid curves) and 
UAS (dashed curves). Results are to be compared with the thin solid
(UAS) and thin dotted (HAS) lines in panels (b) and (d) of
Figs.~\ref{noclouds} and~\ref{clouds}. We infer from comparing these
three figures that the effect on a space-based detector of higher
cumulus clouds, and even higher cirrus clouds, is more dramatic for
downgoing HAS than for upcoming UAS. 
The HAS acceptance is reduced by factors of $\sim$ 1.5,
3, and 10 when the cloud layer lies at $\zc=$ 2~km, 4~km, and 12~km,
respectively. In contrast, the UAS acceptance is reduced by factors of
$\sim$ 1.5, 2 and 3 when the cloud layer lies at $\zc=$ 2~km, 4~km and
12~km, respectively. Since cloud layers are common, they will
compromise the acceptance of space-based detectors.

Also shown in both panels of Fig.~\ref{curvaturedep} are the UAS
acceptances (dotted lines) for a flat Earth. One sees that correct
inclusion of the Earth's curvature lowers the acceptance, since it
puts the tau decay and the subsequent onset of shower evolution into
the thinner air of higher altitudes. Curvature does little harm for
smaller cross-sections, but reduces the acceptance for $\sig\gsim
0.5\times 10^{-31}\,{\rm cm}^2$. Coincidentally, $0.5\times
10^{-31}\,{\rm cm}^2$ is the popular value for the QCD-extrapolated
cross-section. The reduction of acceptance for for larger
cross-sections is understandable, because the for larger $\sig$ the
taus emerge from the Earth more horizontally, and hence travel more
lateral distance before they decay. The Earth ``falls away'' from the
taus as (lateral displacement)$^2 /2\,\Rearth$. Beyond $\sim
10^{-31}\,{\rm cm}^2$, the reduction factor is about 2.5 for cloud
layers at either 4 or 12~km (and quite different for $E_\nu$ near
$\Eth$).

Since curvature raises the altitude of the tau-shower (and rotates it
toward the vertical by $\theta\sim$~(lateral displacement)/$\Rearth$),
the net effect is to remove the bottom layer of atmosphere from the
UAS shower development. Clouds remove the bottom layer from view for
\usp. Thus, one expects the reduction in acceptance due to Earth's
curvature to be largest in the cloudless case. We have checked
numerically that for the parameters of Fig.~\ref{clouds}, a reduction
factor of $\sim$ 3.5 is obtained for the cloudless case.

However, it is dangerous to generalize that Earth's curvature causes 
event suppression. One sees in the left panel of
Fig.~\ref{curvaturedep} that for high clouds, curvature effects may
even {\it increase} the event rates. This is because more events that
would not have been visible above the cloud altitude are now
``lifted'' to visibility, compared to the number of events that would
have been visible but are not  ``lifted'' to invisibility. Curvature
does not increase rates in the right panel, which points again to the
dangers of generalization. 

\begin{figure}
\includegraphics[width=1.\textwidth]{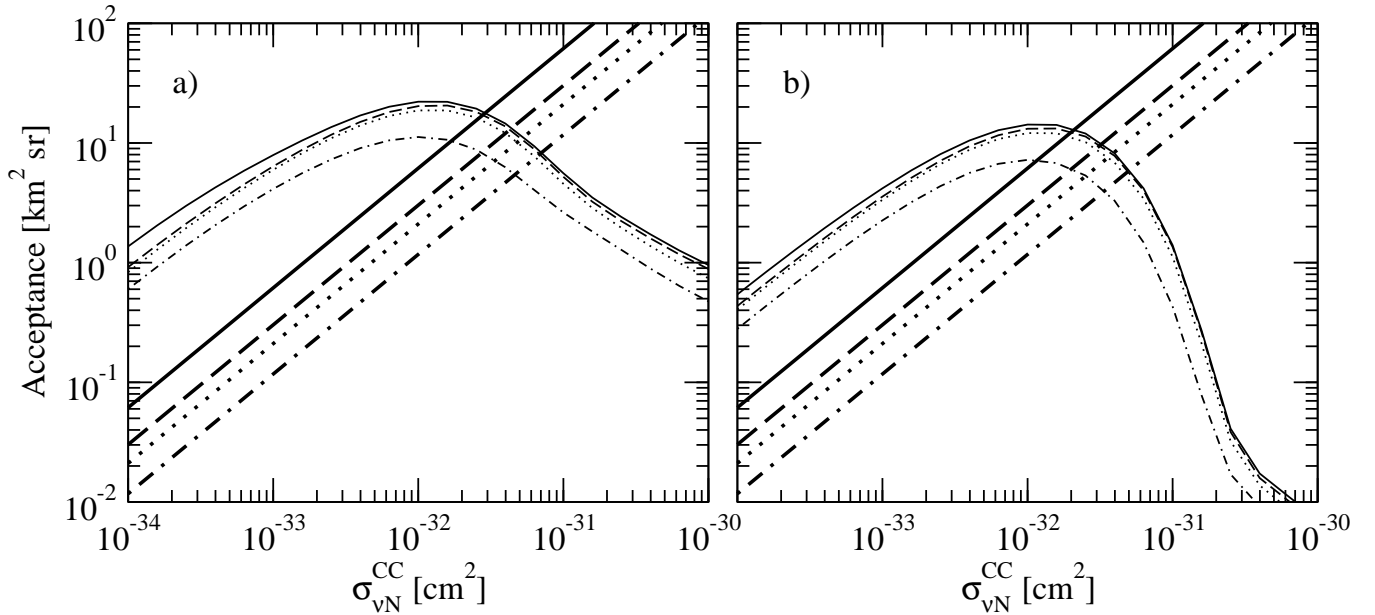}
\caption{\label{showerdep} 
  Dependences on $\dmin$, $\dmax$, and $\lmin$ without clouds of
  acceptances for space-based fluorescence detectors. Threshold
  energies are $\Esh = 10^{19}$~eV in the left panel, and $5 \times
  10^{19}$~eV in the right panel. All curves representing UAS assume
  trajectories over water, and an initial neutrino energy of
  $10^{20}$~eV; curves for HAS are valid for any energy exceeding
  $\Esh$. Thick lines represent HAS, and thin lines UAS. Solid lines
  correspond to ($\dmin,\dmax,\lmin)= (0\,{\rm g/cm}^2,10^5\,{\rm
  g/cm}^2,0$~km), dashed lines to ($300\,{\rm g/cm}^2, 1500\,{\rm
  g/cm}^2, 5$~km), dotted lines to ($300\,{\rm g/cm}^2, 1500\,{\rm
  g/cm}^2, 10$~km), and dash-dotted lines to ($400\,{\rm g/cm}^2,
  1200\,{\rm g/cm}^2, 10$~km).
}
\end{figure}

In Fig.~\ref{showerdep}, we show dependences of the acceptances on the
parameters $\lmin$ and $\dmin$ describing the shower triggers, and the
parameter $\dmax$ characterizing shower extinction. In this figure we
assume a cloudless sky. Threshold energies are $\Esh = 10^{19}$~eV in
the left panel and $5 \times 10^{19}$~eV in the right panel. Thick
curves present HAS acceptances and thin curves present UAS
acceptances. All curves representing UAS assume trajectories over
water, and an initial neutrino energy of $10^{20}$~eV; curves for HAS
are valid for any energy exceeding $\Esh$. We depict four different
sets of shower parameters: ($\dmin,\dmax,\lmin)= (0\,{\rm
  g/cm}^2,10^5\,{\rm g/cm}^2,0$~km) shown in solid curves, ($300\,{\rm
  g/cm}^2, 1500\,{\rm g/cm}^2, 5$~km) in dashed curves, ($300\,{\rm
  g/cm}^2, 1500\,{\rm g/cm}^2, 10$~km) in dotted curves, and
($400\,{\rm g/cm}^2, 1200\,{\rm g/cm}^2, 10$~km) in dash-dotted
curves. The value $\dmax =10^5\,{\rm g/cm}^2$ is to be interpreted as
an effectively infinitely long shower persistence, i.e.\ an
illustration of shower development without extinction. The acceptances
in the solid curves correspond to the most liberal shower-trigger
requirements, with basically all showers declared observable. 
Those in he dashed, dotted,and dashed-dotted curves correspond to a 
realistic set of
choices for the shower-development parameters $\dmin$, $\dmax$ and for
the shower length $\lmin$. 

One sees that even in the cloudless case shown here, the dependences
on the shower parameters is considerable. The UAS acceptance gets
reduced by $\sim$ 2 when $\lmin$ is increased from 5~km to 10~km and
the visible column density ($\dmax-\dmin$) is reduced from 1200 to
800~g/cm$^2$. Changes in the HAS acceptance are a bit more
dramatic. As the shower triggers are tightened according to our 
examples, the HAS acceptance falls by a factor $\sim$ 6. 

The acceptances in Fig.~\ref{showerdep} can also be compared to the
thin solid (UAS) and dotted (HAS) curves in panels (b) and (d) of the
no-cloud Fig.~\ref{noclouds}, where ($\dmin,\dmax,\lmin)= (400\,{\rm
  g/cm}^2,1200\,{\rm g/cm}^2,5$~km). As can be seen from this figures,
the effect of reducing the constraint on the minimum shower length
from 10~km to 5~km increases the acceptance by roughly a factor of
$\sim$ 3 for HAS, and $\sim$ 2 for UAS. 

With clouds, the sensitivity to shower-development parameters is more
acute. The dotted case corresponds to a good trigger sensitivity
(discussed in Section~\ref{subsec:UASremark}) of
$(\dmax-\dmin)/\lmin=120\,{\rm g}\,{\rm cm}^{-2}/{\rm km}$, and so to
a critical cloud-altitude of $\zcc=0.56$~km (refer to
Fig.~\ref{fig:zcrit}). The dashed-dotted case corresponds to a less 
good $80\,{\rm g}\,{\rm cm}^{-2}/{\rm km}$ sensitivity, and a less
pleasing (for ground-based detectors) critical cloud-altitude
$\zcc=3.8$~km. Thus one expects the dashed-dotted case to be quite
sensitive to cloud layers, and the other three cases to be relatively
insensitive to cloud layers.

\section{Discussion}
\label{sec:discussion}

In this paper we have presented a mostly analytic calculation of the
acceptances of space-based and ground-based fluorescence detectors of
air-showers at extreme-energies. Included in the calculation are the
dependences of the acceptances on initial neutrino energy,
trigger-threshold for the shower energy, composition of Earth (surface 
rock or ocean water), and several shower parameters (the minimum and
maximum column densities for shower visibility, and the tangent length
of the shower). Also included in the calculation are suppression of the
acceptances by cloud layers of arbitrary altitude, and in the UAS
case, by the Earth's curvature. And most importantly, included in the
calculations are the dependences on the unknown neutrino cross
section. The dependence is trivial and linear for HAS, but nontrivial
and nonlinear for UAS.

The merits of the analytic construction are two-fold: it offers an
intuitive understanding of each ingredient entering the total
calculation; and it allows one to easily re-compute when parameters
governing the shower or the atmosphere are varied. While a Monte Carlo
approach may be simpler to implement, it sacrifices some insight and
efficiency.

The differing dependences of HAS and UAS on $\sig$ enable two very
positive conclusions: (1) the ``no-lose theorem'' is valid, namely,
that acceptances are robust for the combined HAS plus UAS signal
regardless of the cross-section value; and (2) an inference of the
cross-section at $10^{20}$~eV is possible if HAS and UAS are both
measured.

Our formulas are valid for the energy range $\sim 10^{18}-10^{21}$~eV.
The lower limit is necessary to validate our assumption that the
$\tau$ decay length is much larger than its radiation length. Below
$10^{18}$~eV, the $\tau$'s boost factor is insufficient to provide a
$\tau$ lifetime in compliance with this assumption. The upper limit
arises from the fact that above $10^{21}$~eV, the weak charged-current
losses of the $\tau$'s, not included in our calculation, exceed their
electromagnetic losses, which we have included. Although the energy 
range of validity is limited, it covers the range of interest for
extreme-energy cosmic neutrino studies.

\subsection{A tale of two media (ocean and land), 
  and two lengths ($\lambda_\nu$ and $\lambda_\tau$)}
\label{sec:twolengths}

One may sensibly ask, ``what difference does it make whether the 
UAS interaction takes place in water or in rock?'' After all, one has
only to look at a different value of $L$ or $\thn$ to compensate
for any change in target density. Nevertheless, we have shown that the
UAS rate over ocean is an order of magnitude larger than over land.
Let us explore this a bit. 

The rate for producing upgoing taus which exit the Earth's surface 
has its peak near the chord length $L_{\rm
  peak}\sim\lambda_\nu+\lambda_\tau$. Increasing the solid angle  
optimizes the rate. This is accomplished by making $L_{\rm
  peak}$ as large as possible. Equivalently then, we ask that
$\lambda_\nu$ and $\lambda_\tau$ be as large as possible. Explicitly,
we seek to maximize
\beq{Lopt}
L_{\rm peak}\sim \lambda_\nu+\lambda_\tau
= \frac{1}{\rhoearth}\left(\frac{1}{\sig\,N_A}+\frac{1}
  {\beta_\tau}\right) \,.
\eeq
Apparently, Nature may optimize $L_{\rm peak}$ in one of three ways:
either reducing the cross-section, reducing the tau's radiation loss,
or reducing the density. Thus, the oceans, less dense than surface
rock, will serve to increase the emerging tau rate by the factor
$\rhosr/\rhow=2.65$ (for fixed $\sig$ and $\beta_\tau$). For
generation of the visible UAS, there is a further
enhancement. Trajectories through water exit the Earth nearly
horizontally. Horizontally-emergent taus have more time and more
atmosphere in which to decay and evolve into a visible shower. This
further increases the UAS acceptance. And there is a third effect:
$\beta_\tau$ in water is nearly half of $\beta_\tau$ in rock, leading
to a larger $\lambda_\tau$ in Eq.~\rf{Lopt}. So several effects
conspire to enhance the UAS acceptance over water vs.\ over land. In
our numerical work presented in Figs.~\ref{noclouds} and \ref{clouds},
we found the enhancement to be typically a factor of 10. This
enhancement over water is fortunate, in that 70\% of the Earth's
surface is ocean, and ocean nights are not polluted with man-made
lights.

Note that 
\beq{lambdaratio}
\frac{\lambda_\tau}{\lambda_\nu}=\sigstd\,\left(\frac{\beta_{19}}
     {\beta_\tau (E)}\right) \times \left\{ \barr{ll}
0.06 & {\rm for\ \ rock} \\
0.10 & {\rm for\ \ water} 
\earr
\right.
\eeq
Furthermore, the energy dependence of $\beta_\tau (E)\propto E^{0.2}$
is mild, changing $\beta_\tau$ by just 1.6 per decade. Thus, we have
that  $\lambda_\nu > \lambda_\tau$ for $\sig \lsim 10^{-30}\,{\rm
  cm}^2$, i.e.\ for the entire range of cross-sections which we
consider. So the story of the UAS enhancement over water is really the
story of the neutrino's longer MFP in water than in rock, followed by
the tau's greater probability to decay and shower following its nearly
tangential emergence from water.

For very weak cross-sections, the MFP becomes large compared to the
size of the critical chord in the ocean
$\lwsr=422\,\sqrt{\zw/3.5\,{\rm km}}$~km. Then the neutrino is forced  
to spend part of its trajectory in rock, mitigating the difference
between UAS over water and over land. This phenomenon is included in
our calculations. We have seen that the neutrino MFP exceeds the
critical chord for $\sig < 4 \times 10^{-32}{\rm cm}^2$. Far below
this cross-section value, the ocean portion of the neutrino's chord is
too small to affect the UAS rate. In Fig.~\ref{noclouds} one can see
that the two UAS acceptances, over water and over land, tend toward
each other below $\sig\sim 10^{-32}\,{\rm cm}^2$, and to a ten-fold 
enhancement for water above $\sig\sim 10^{-32}\,{\rm cm}^2$.

\subsection{Remark on Observations over Elevated Land}
\label{subsec:elevationremark}

We will present results for events over land and ocean, both taken to
have zero, i.e.\ ``sea level'', elevation. However, much of the
Earth's land surface is at higher elevation (fortunately, for land
animals). Furthermore, some ground-based observatories are sited at
high elevations to reduce various backgrounds. The Auger observatory
in Argentina, for example, is at 1400 meters, above $\sim 15\%$ of the
atmosphere. The HiRes siting in Utah is at a similar elevation. For
HAS viewed from these sites (in cloudless skies), the acceptance is
the same as that for space-based viewing with a cloud layer at 1.4~km. 
For UAS viewed from elevated sites, one cannot proceed by simple
analogy. 

However, inclusion of elevation into our formalism, for ground-based
or space-based, HAS or UAS, with clouds and without, is simple. One
merely replaces the sea-level atmospheric density, $\rho(0)$, with the
ground-level density.  For an elevation of $\zell$, the replacement
value is $e^{-\zell/h}\,\rho(0)$. This leads to the further
replacement  $\dvert\rarr e^{-\zell/h}\,\dvert$. With these 
substitutions, all previous formulas may be used, with altitudes of
clouds and other ``z-parameters'' now understood to be with respect to
ground-level, not sea-level. Strictly speaking, $\zthin$ should be
replaced by $\zthin-\zell$. This would allow accurate comparisons of
acceptances from elevation with those from sea-level. In practice,
keeping or not keeping the additional kilometer or so of thin
atmosphere makes little difference. And for comparing ground-based and
space-based detectors at a common elevation, the difference is
irrelevant.

In summary, ground elevation reduces the amount of atmospheric volume
available, sometimes substantially. This in turn reduces the target
mass for HAS, the decay volume for UAS, and the grammage available for
shower development. Consequently, elevation disadvantages observations
over land, compared to space-based observations over the
zero-elevation ocean. The disparity between the two becomes more acute 
in the presence of clouds.

\subsection{Comparisons with prior work}
\label{sec:compare}

Among the aims of this paper is the detailed extension of the idea
introduced in Ref.~\cite{KW}, that $\sig$ at $10^{20}$~eV can be
inferred from a measurement of the UAS/HAS ratio. So let us first
compare our calculation with the more qualitative one in
Ref.~\cite{KW}. In Ref.~\cite{KW}, several energy-dependences were
purposely frozen, for simplicity. The tau energy loss was set
constant $\beta_\tau (E)=\beta_{19}$ (i.e., $\alpha = 0$), and the
energy for the produced tau lepton was assumed to be that of the
incoming neutrino (i.e., $\langle y \rangle = 0$). The decay length
for the taus exiting the Earth was fixed to a single value, ignoring
the dependence on the tau's initial energy, production point in the
Earth, and zenith angle.  Furthermore, all taus were considered to
decay within an atmospheric height of 10 km. These approximations do
not affect the main conclusions of~\cite{KW}, but do impact the
results quantitatively. For example, they lead to unphysical behavior
in the UAS rate for high values of the cross-section; for large $\sig$
the incident neutrino must interact with the Earth's surface with
probability one, and the UAS acceptance should asymptote to a
$\sig$-independent constant  determined solely by tau physics. Also,
the case where the trajectory's chord length in the Earth is smaller
than the distance for which the tau lepton energy is reduced to the
threshold value $\Eth$, is incorrectly calculated. The impact of this
is small, for it only affects very Earth-skimming neutrinos ($\thhor <
0.1^\circ$). Finally, Ref.~\cite{KW} only considered neutrinos
traveling through rock, not water, and did not include suppression
effects from realistic shower formation, clouds, or curvature of the
Earth. Our work considerably improves upon the original work of
Ref.~\cite{KW}.

The prior work in Refs.~\cite{Feng1,Aramo,DHR05} (and related work in
Ref.~\cite{HEtaus}) is semi-analytic, like our own. A main difference
between them and us is the manner in which the tau energy loss is
parametrized. The parameter $\beta_\tau$ is taken to be constant in
Ref.~\cite{Feng1} (where $\langle y \rangle = 0$ is also assumed), to
depend linearly on the tau energy in Ref.~\cite{Aramo}, and assigned a
logarithmic dependence on the energy in Ref.~\cite{DHR05}. Also,
Refs.~\cite{Feng1,Aramo} assume some maximum tau decay distance, and
do not implement any constraints from the subsequent shower formation.
Furthermore, they do not consider events over water, or the presence
of clouds. Ref.~\cite{DHR05} computes the flux of tau leptons exiting
the Earth, but doesn't consider the important process of tau decay and
shower formation. On the other hand, \cite{DHR05} does include the
possibility of the tau decaying inside the Earth. However, as we
mentioned earlier in our paper, including tau decay within the Earth 
reduces the UAS acceptance by less than a few per cent for the
energies of interest here. Also in Ref.~\cite{DHR05}, a Monte Carlo
calculation is performed and shown to agree with the semi-analytical
approach to very good accuracy. Where comparisons are possible, our
results agree qualitatively with~\cite{DHR05}. Let us note that
Ref.~\cite{MPPAuger} offers an improved and more detailed evaluation
of the effective acceptance for fluorescence detection at the Pierre
Auger Observatory. Taken into account is the real elevation profile of
the surrounding mountains. They find a significant increase in the
event rate due to the nearby mountains, compared to the
semi-analytical, mountain-less calculations of
Refs.~\cite{Feng1,Aramo} and us. Unfortunately, even with the
enhancement, the predicted event rates are $\lsim 0.5$~events/yr for
neutrino fluxes motivated by the GZK process and topological defect
decay models. This small rate points again to one of the major
benefits of space-based detectors: the much larger FOV.

In Ref.~\cite{Zas}, a detailed semi-analytical computation of UAS and
HAS was performed.  This work considered a larger range of energies
than we do. Hence, it was necessary for Ref.~\cite{Zas} to include the
possibility of tau decay inside the Earth. This work assumed
$\beta_\tau$ to be constant, but otherwise the energy-loss of the
produced tau was calculated accurately. In addition, constraints on
the shower formation were included in a semi-analytical approach, in
order to calculate event rates for an air shower array. The main
differences between this work and our calculation is that we consider
a water layer as well as rock, and we include the possibility of
clouds.

Detailed Monte Carlo simulations of acceptances are presented in
Refs.~\cite{Bertou1,Bottai1}. Neutrino scattering inelasticities and
tau energy losses are accurately included. We have checked that we get
very good agreement with the results of these
papers. Ref.~\cite{Bertou1} is specific ground-based Auger 
detector. It includes realistic shower formation and detector
response, but it does not consider clouds. It also considers only
neutrino and tau propagation in rock. While rock is the dominant
material in the vicinity of Auger, there are trajectories reaching
Auger from the West which will travel in the Pacific
Ocean. Ref.~\cite{Bottai1}, undertaken mainly with EUSO in mind, does
study acceptances over both water and land.

In Ref.~\cite{FargionRate}, the analytic approach is different, and
more optimistic rates are obtained. However, there are several
questionable approximations. How the energy threshold constraint is
implemented is obscure. The treatment of the Earth's atmosphere is too
simplistic. And although the calculation is meant to be valid for an
arbitrary Earth density profile, the derived expression for the event
number is only valid for a constant Earth density. Our
acceptances for UAS events do not support the optimistic UAS rates of
Ref.~\cite{FargionRate}. We do however support the results found by
the more detailed analysis, e.g.\ that in Ref.~\cite{Bottai1}.  

In summary, the main advances we present in our study are the
inclusion of a new analytic and accurate power-law parametrization for
the tau energy loss pre-staging UAS events, the analytical
implementation of shower constraints, cloud boundaries, and Earth
curvature, and the consideration of UAS events over the ocean as well
as over land. We note that the tau energy loss parametrization we
implement in the present study was already used in Ref.~\cite{Bertou1}
for the case of taus propagating in rock. Here we present also the
parametrization when taus cross a water layer.

\subsection{Odds and Ends}
\label{sec:oddend}

In this subsection we offer remarks on issues possibly relevant to
this paper.

\subsubsection{Incident neutrino flavor ratios}
\label{sec:flavorratios}

One of the main points of this paper is to hone the argument that the
CC neutrino cross-section can be inferred from a comparison of UAS and
HAS rates. The ratio of these rates is the product of a flux ratio
times acceptance ratio. We have focused on the electron-neutrino as
the primary particle for HAS initiation, and the tau-neutrino as the
primary particle for UAS initiation, for the good reasons given in the
text. The relevant flux ratio, therefore, is the ratio of the $\nue$
flux to the $\nutau$ flux.

We have calculated acceptances. From these, one may simply form the
acceptance ratio. What is not known at present is the relevant $\nue$
to $\nutau$ flux ratio at $\sim 10^{20}$~eV. A general theorem for
neutrino flavor-mixing states that if the atmospheric mixing angle is
nearly maximal (it is!), and if the short-baseline angle $\theta_{13}$
is nearly zero (it is!), then $\numu$ and $\nutau$ equilibrate over
cosmic distances. A corollary to the theorem then, is that if cosmic
neutrinos originate from the complete pion decay chain,
$\pi\rarr\mu+\numu\rarr e+\nue+2\numu$, then after equilibration the
neutrinos will arrive at Earth with the democratic flavor ratio of
1:1:1. However, dynamics at the source, or new physics enroute from
the source, could alter this favorable ratio. {\sl Caveat emptor}!

\subsubsection{Ratio of neutral- and charged-current cross-sections}
\label{sec:NCtoCC}

It is an implicit assumption in this work that the ratio of the
neutral to charged-current cross-section is small, $\sim 0.44$
according to the Standard Model of particle physics. However, it is
possible that above $10^{15}$~eV and below $10^{20}$~eV a threshold is
passed at which the NC interaction becomes strong and the CC
interaction does not. Such would be the case, for example, in models
of low-scale gravity unification.

Crossing such a hypothetical threshold would change the physics in
this paper dramatically. First of all, even though the NC interaction
typically puts $\sim 5$~times less energy into the shower than does
the $\nue$~CC interaction, with a much larger NC cross-section, even
at fixed $E^{\rm sh}$ the NC events would dominate the $\nue$~CC
events. Secondly, UAS acceptances would be reduced because the energy
losses of neutrinos passing through the Earth would be larger.

\subsubsection{Weakly-interacting non-neutrino primaries}
\label{sec:WIMP}

The range of cross-sections we consider in this work spans the cross
sections of weakly-interacting massive particles (WIMP), a popular
candidate for dark matter. We believe, therefore, that our figures 
may be useful in assessing the qualitative features of acceptances 
for WIMP detection. However, we caution that there are substantial
differences between WIMP-initiation of showers and
neutrino-initiation. The WIMP carries considerable inertial mass, and
so transfers less energy to its shower. Also, the UAS generated by a
WIMP flux would not proceed through the tau production and decay chain.

\section{Conclusions}
\label{sub:conclude}

We have presented analytic formulas for the acceptances of
fluorescence detectors, both space-based and ground-based, for
neutrino-initiated events, as a function of the unknown extreme-energy 
neutrino cross-section. For the downgoing HAS events, the dependence
of acceptance on cross-section is linear, but for upcoming UAS events
the acceptance is quite complicated. It turns out to be somewhat flat
and relatively large, which validates the ``can't lose'' theorem which
says that if the HAS rate is suppressed by a small $\sig$, then the
UAS rate compensates to establish a robust signal. 

We have studied the dependence of acceptances on the incident neutrino
energy, the trigger-energy $\Esh$ for the shower, shower-development
parameters $\dmin$ and $\dmax$, and observable (tangent) shower length
$\lmin$; and on the ``environmental'' conditions of cloud layers for
HAS and UAS, and events over ocean versus over land for UAS. UAS
showers typically originate at a considerable distance
($c\tau_\tau=4900\,(E_\tau/10^{20}$~eV)~km) from the point on the
Earth where the parent tau emerged. Therefore, due to the earth's
curvature, they originate at higher altitudes with thinner air. Thus,
it is necessary to include the Earth's curvature in the calculation of
UAS acceptances. We have done so. We find that inclusion of the Earth's
curvature reduces the UAS acceptance by a factor of a few when
$\sig\gsim 0.5\times 10^{-31}\,{\rm cm}^2$. The meaning of ``a few''
depends on the various parameters entering the calculation.

Clearly lower shower-trigger energies are better. This is especially
true when clouds are present. We have quantified the sensitivity to
$\Esh$ by comparing two realistic values, $10^{19}$~eV and $5\times
10^{19}$~eV in the face of incident neutrino energies of $10^{20}$~eV
and $10^{21}$~eV.

Cloud layers may severely suppress acceptances. For UAS acceptances,
there is a strong dependence on the combination of shower-trigger
parameters $(\dmax-\dmin)/\lmin$, especially with clouds
present. Maximizing this combination to a value of
$\dvert/h=\rho(0)=129\,{\rm g}\,{\rm cm}^{-2}/{\rm km}$ or greater,  
significantly minimizes the suppression from clouds. For
$(\dmax-\dmin)/\lmin \lsim \rho(0)$, there is a critical altitude 
$\zcc\equiv -h\ln\left[(\dmax-\dmin)/\lmin)\,(h/\dvert)\right]$ 
below which a cloud layer would totally obscure the acceptance
of a ground-based detector, but leave the acceptance of a space-based 
detector unaltered.  Clouds above the critical altitude would
partially obscure UAS events, and therefore suppress the acceptances, 
of both space-based and ground-based detectors.

Concerning UAS events over water versus over land, we find that
acceptances over water are larger, typically by an order of
magnitude. We have traced this enhancement over water to the increased
pathlength in water of both neutrinos and taus, and to the increased
pathlength in air for tau decay and increased column density in air
for shower development, when a tau emerges with small horizontal angle
from the relatively shallow ocean. We also noted the smaller
enhancement from the fact that the atmospheric grammage over water 
integrates from sea-level, whereas the grammage over land is often
15\% less. It is difficult to imagine a ground-based detector over the
ocean, so the ``water advantage'' clearly belongs to the orbiting
space-based detectors. Perhaps a ground based detector could be
positioned near an ocean to realize the ``water advantage'' for much
of its solid angle.

In the spirit with which we began this study, we are led to two
bottom-line conclusions:\\ 
(i) Inference of the neutrino cross-section at and above $10^{20}$~eV
from the ratio of UAS and HAS events appears feasible, assuming that a
neutrino flux exists at these energies.\\
(ii) Space-based detectors enjoy advantages over ground-based
detectors for enhancing the event rate. The advantages are a much 
higher UAS rate over water compared to land, and the obvious advantage
that space-based FOV's greatly exceed ground-based FOV's.

Our hope is that space-based fluorescence-detection becomes a reality,
so that the advantages of point (ii) can be used to discover/explore
the extreme-energy cosmic neutrino flux. According to point (i), part
of the discovery/exploration can be the inference of the neutrino
cross-section at $E_\nu\sim 10^{20}$~eV.

\begin{acknowledgments}

Encouragement from J. Adams, S. Bottai, O. Catalano, D. Cline,
D. Fargion, P. Lipari, A. Santangelo, L. Scarsi, Y. Takahashi, and the
EUSO community is acknowledged. Figure~\ref{fig:nuchord} contributed
by Liguo Song. SPR and TJW are supported by NASA Grant ATP02-0000-0151
for EUSO studies, SPR by the Spanish Grant FPA2002-00612 of the MCT, 
and  TJW by the U.S. Department of Energy Grant no. DE-FG05-85ER40226 
and by a Vanderbilt University Discovery Award.

\end{acknowledgments}

\appendix

\section{Density Integrations in the Earth}
\label{sec:app}

Two column densities in the Earth are of relevance for UAS
probabilities. The first is the path-integral of $\rhoearth$ for the 
incident neutrino from entrance to interaction in the Earth. This
column density controls the neutrino absorption probability, and
therefore, the neutrino survival probability to the point of
interaction $\wint$.  The column density $\dnu(\thn)$ is stated in
Eq.~\rf{dnu} as
\beq{nucolumn1}
\dnu(\thn)=\int^L_{\wint} dw\,\rhoearth(w)\,.
\eeq
For constant density, which applies only for Earth-skimming neutrinos
entirely in surface rock ($\rhoearth=\rhosr=2.65\,{\rm g/cm}^3$) or
entirely in ocean water ($\rhoearth=\rhow=1.0\,{\rm g/cm}^3$), the
result is simply $\dnu(\thn)=(L-\wint)\,\rhoearth\,$. This constant
density result holds in ocean for angles relative to the horizon
smaller than $1.90^\circ$, and it holds in rock for angles relative to
the horizon smaller than $22.17^\circ$, as we show below.

The second relevant column density in the Earth is that of the
emerging tau in the UAS event sequence. This column density
$\dtau(\thn)$ is the path-integral of $\rhoearth$ from the interaction
point in the earth to the earth's surface, 
\beq{taucolumn1}
\dtau (\theta_z)=\int^{\wint}_0 dw\,\rhoearth(w)\,.
\eeq
For constant density, which applies for taus emerging from rock, or
for taus emerging from water with horizon-angle less than
$1.90^\circ$, the result is simply
$\dtau(\theta_z)=\wint\,\rhoearth\,$. This column density, when  
suitably weighted with the tau energy-attenuation factor $\beta_{19}$,
controls the tau energy-loss probability, and therefore, the probably
for the tau energy and lifetime at emergence from the Earth as given
in Eqs.~\rf{Etau} and \rf{lifetime}. 

For our purposes, concentric shells of constant density provide a
sufficiently accurate approximation to the Earth's profile, and allow
for an analytic evaluation of the path integrals. We take a simple
model of this kind for the Earth density:
%
\beq{Earthdensity}
\rhoearth(r)=
\left\{
\begin{array}{lll}
\rhow=1.0\;{\rm g/cm}^3   & {\rm for\ } 
	& \rsr\equiv\Rearth-\zw<r\le\Rearth\,, \\
\rhosr=2.65\;{\rm g/cm}^3 & {\rm for\ } & \rmantle<r\le\rsr\,, \\
\rhom=4.0\;{\rm g/cm}^3   & {\rm for\ } & \rc<r\le\rmantle\,, \\
\rhoc=12.0\;{\rm g/cm}^3  & {\rm for\ } & 0<r\le\rc\,, 
\end{array}
\right.
\eeq
%
with the radii of the boundaries listed in
Table~(\ref{table:critical}). For UAS events over land, we replace the 
outermost $\zw=3.5$~km of water with surface rock. Thus, there are in
this Earth model four (three) concentric density zones for UAS events
over water (land).

The chord length $L$, nadir angle $\thn$, and horizon angle $\thhor$
are related to the sagitta $s$ by the formulas 
\ba{sagitta}
L(s)&=&2\sqrt{2R_\oplus\,s -s^2}\,,\\
\cos\thn(s)&=&\sin\thhor(s)
=\sqrt{2\frac{s}{R_\oplus}-\left(\frac{s}{R_\oplus}\right)^2}\,,\\ 
\sin\thn(s)&=&\cos\thhor(s)=1-\frac{s}{R_\oplus}\,;
\ea
and to the boundary radius $\rb$ by
\ba{bradius}
L(\rb)&=&2\sqrt{R_\oplus^2-\rb^2}\,,\\
\cos\thn(\rb)&=&\sin\thhor(\rb)
	=\sqrt{1-\left(\frac{\rb}{R_\oplus}\right)^2}\,,\\
\sin\thn(\rb)&=&\cos\thhor(\rb)=\frac{\rb}{R_\oplus}\,.
\ea
In Table~\ref{table:critical} we collect the critical values for $L$
and $\thn$ at the various boundary layers.

\begin{center}
\begin{table}
\label{table:critical}
\begin{tabular}{|l|l|c|c|c|} \hline
Boundary & Radius & Critical $L$ & Critical $\thn$ & Critical
$\thhor$ \\ \hline  \hline
Earth/atmosphere & $R_\oplus=6371$~km & 0 km & $90^\circ$ & $0^\circ$
\\ \hline
water/surface-rock & $\rsr=\Rearth-\zw$, $\zw=3.5$~km
   & $\lwsr\equiv 422\left(\frac{\zw}{3.5\,{\rm km}}\right)^\half$~km  
   & $88.10^\circ$ & $1.90^\circ\,\left(\frac{\zw}{3.5\,{\rm
    km}}\right)^\half$  \\ \hline
surface-rock/mantle& $\rmantle=5900$~km & $\lsrm\equiv 4808$~km &
$67.83^\circ$ & $22.17^\circ$ \\ \hline 
mantle/core & $\rc=3485.7$~km & $\lmc\equiv 10666$~km & $33.17^\circ$ &
$56.83^\circ$ \\ \hline 
\end{tabular}
\caption{Critical chord lengths and nadir and horizon angles for the
  indicated Earth boundaries.}
\end{table}
\end{center}

A further useful formula is the path-length $w(\rb;L)$ from the
Earth's surface to the boundary of radius $\rb$, for fixed $L$ or
$\thn$:
\ba{wrb}
w(\rb;L\;{\rm or}\;\thn)&=&\frac{L}{2}-
	\sqrt{\left(\frac{L}{2}\right)^2+\rb^2-\Rearth^2}\,,\\
	&=& \Rearth\,\cos\thn -\sqrt{\rb^2-\Rearth^2\,\sin^2\thn}\,.
\ea
Consider the calculation of $\dtot(L)$. Using the path-lengths defined
in Eq.~\rf{wrb}, the result is:
%
\beq{Stot2}
\dtot= 
\left\{
\begin{array}{llrl}
L\,\rhow\,,&{\rm for\ } & 0	&\le L\le\lwsr \\
2\,\{w(\rsr;L)\,\rhow+[\frac{L}{2}-w(\rsr;L)]\,\rhosr\}
	\,,&{\rm for\ } & \lwsr &<L\le\lsrm \\
2\,\{w(\rsr;L)\,\rhow+[w(\rmantle;L)-w(\rsr;L)]\,\rhosr+[\frac{L}{2}
  -w(\rmantle;L)]\,\rhom\}\,,   
	&{\rm for\ } & \lsrm &<L\le\lmc \\
2\,\{w(\rsr)\,\rhow+[w(\rmantle)-w(\rsr)]\,\rhosr+[w(\rc)-w(\rmantle)]
\,\rhom 
	+[\frac{L}{2}-w(\rc)]\,\rhoc\}\,, &{\rm for\ } & \lmc  &<L\le
	2\Rearth  
\end{array}
\right.
\eeq
%
where for compactness, we have suppressed the $L$-dependence of the
$w$-function in the final line. For events over land, $\rhow$ must be
replaced in Eq.~\rf{Stot2} with $\rhosr$.

Next we consider the calculation of $\dtau(L)$. The tau radiation
length is very short on the scale of $\rmantle$, and so the tau path
is confined to just surface rock and ocean. Over land, then, we have
simply $d_\tau({\rm land})=\wint\rhosr$. Over water, the calculation
has two contributions in general, from water and from surface-rock.
For $L<\lwsr$, the tau encounters just water, and so $d_\tau({\rm
ocean};L<\lwsr)=\wint\rhow$. For $L>\lwsr$, the tau encounters rock and
then water. However, the tau never encounters first water and then
rock and then water again, for this requires a tau trajectory
exceeding $1/2$ of the critical path-length $\lwsr$, which is
$211\,\left(\frac{\zw}{3.5{\rm km}}\right)^\half$~km, far exceeding
the tau radiation length. We summarize these results, again making use
of Eq.~\rf{wrb}: 
%
\beq{dtau}
\dtau= 
\left\{
\begin{array}{lll}
\wint\,\rhosr\,,&{\rm over\ land,} & {\rm for\ all\ }\wint \\
\wint\,\rhow\,,&{\rm over\ oceans,} & {\rm for\ }\wint<w(\rsr;L) \\
w(\rsr;L)\,\rhow+[\wint-w(\rsr;L)]\,\rhosr\,,&{\rm over\ oceans,}&
{\rm for\ }\wint>w(\rsr;L) 
\end{array}
\right.
\eeq
%
To obtain $\dnu$, we may use the simple relation $\dnu=\dtot-\dtau$.
Thus, we are finished with calculating column densities.

The tau opacity is easily obtained in the constant-density
concentric-shells approximation.  Weighting the segments in
Eq.~\rf{dtau} with the corresponding values of $\beta_{19}$, either
$\betasr=1.0\times 10^{-6}$~cm$^2$/g or $\betaw=0.55\times
10^{-6}$~cm$^2$/g, we have
\beq{opacity}
\eye(\wint)= 
\left\{
\begin{array}{lll}
\wint\,\betasr\,\rhosr\,,&{\rm over\ land,} & {\rm for\ all\ }\wint \\
\wint\,\betaw\,\rhow\,,&{\rm over\ oceans,} & {\rm for\
}\wint<w(\rsr;L) \\ 
w(\rsr;L)\,\betaw\,\rhow+[\wint-w(\rsr;L)]\,\betasr\,\rhosr\,,
	&{\rm over\ oceans,}& {\rm for\ }\wint>w(\rsr;L)
\end{array}
\right.
\eeq
We also need an explicit formula for $\wth(L)$, defined implicitly in
Eqs.~\rf{eye} and~\rf{wth1}. For notational brevity, let us recall the
notation in Eq.~\rf{wth1}: 
\beq{wth2}
\eye(\wth)\equiv\frac{1}{\alpha}\,
\left[
\left(\frac{10^{19}{\rm eV}}{\Eth}\right)^\alpha
	-\left(\frac{10^{19}{\rm eV}}{0.8\,\Enu}\right)^\alpha
\right]\,.
\eeq
Then, a calculation similar to the ones above leads to
%
\beq{wth3}
\wth(L)=
\left\{
\begin{array}{lll}
\frac{1}{\betasr\,\rhosr}\,\eye(\wth)\,,
	&{\rm over\ land,} & {\rm for\ all\ }L\,, \\

\frac{1}{\betaw\,\rhow}\,\eye(\wth)\,,
	&{\rm over\ oceans,} & {\rm for\ }L\le\lwsr\;
	{\rm or\ }\eye(\wth)<\betaw\,\rhow\,w(\rsr;L)\,, \\
w(\rsr;L)\,\left(1-\frac{\betaw\,\rhow}{\betasr\,\rhosr}\right)
	+\frac{1}{\betasr\,\rhosr}\,\eye(\wth)\,,
	&{\rm over\ oceans,} & {\rm for\ }L>\lwsr\;
	{\rm and\ }\eye(\wth)>\betaw\,\rhow\,w(\rsr;L)\,.
\end{array}
\right.
\eeq
%
Finally we note that, in all the above formulas for events over oceans,
the correct result over land may be found by simply setting $\rhow$
equal to $\rhosr$ and $\betaw$ equal to $\betasr$. 



\begin{thebibliography}{00}



\bibitem{KW}
  A.~Kusenko and T.~J.~Weiler,
  Phys.\ Rev.\ Lett.\  {\bf 88}, 161101 (2002).


\bibitem{G66}
  K.~Greisen,
  Phys.\ Rev.\ Lett.\  {\bf 16}, 748 (1966).


\bibitem{ZK66}
  G.~T.~Zatsepin and V.~A.~Kuzmin,
  JETP Lett.\  {\bf 4}, 78 (1966)
  [Pisma Zh.\ Eksp.\ Teor.\ Fiz.\  {\bf 4}, 114 (1966)].


\bibitem{Letessier1}
  A.~Letessier-Selvon,
  AIP Conf.\ Proc.\  {\bf 566}, 157 (2000),
  astro-ph/0009444.


\bibitem{Zayyad1}
  T.~Abu-Zayyad {\it et al.}  [High Resolution Fly's Eye
  Collaboration], 
  Astropart.\ Phys.\  {\bf 23}, 157 (2005).


\bibitem{FlysEye}
  D.~J.~Bird {\it et al.},
  Astrophys.\ J.\  {\bf 441}, 144 (1995).


\bibitem{flavorID}
  J.~F.~Beacom, N.~F.~Bell, D.~Hooper, S.~Pakvasa and T.~J.~Weiler,
  Phys.\ Rev.\ D {\bf 68}, 093005 (2003)
  [Erratum-ibid.\ D {\bf 72}, 019901 (2005)].


\bibitem{pimuchain} 
For some recent discussions of this assumption, see 

  L.~A.~Anchordoqui, H.~Goldberg, F.~Halzen and T.~J.~Weiler,
  Phys.\ Lett.\ B {\bf 621}, 18 (2005);

  J.~P.~Rachen and P.~Meszaros,
  Phys.\ Rev.\ D {\bf 58}, 123005 (1998);

  T.~Kashti and E.~Waxman,
  Phys.\ Rev.\ Lett.\  {\bf 95}, 181101 (2005).

\bibitem{nulifetime}
  J.~F.~Beacom, N.~F.~Bell, D.~Hooper, S.~Pakvasa and T.~J.~Weiler,
  Phys.\ Rev.\ Lett.\  {\bf 90}, 181301 (2003).


\bibitem{pDirac}
  J.~F.~Beacom, N.~F.~Bell, D.~Hooper, J.~G.~Learned, S.~Pakvasa and
  T.~J.~Weiler,
  Phys.\ Rev.\ Lett.\  {\bf 92}, 011101 (2004).


\bibitem{fluoresence} The use of fluorescence profiles to infer
  ultrahigh cosmic ray energies has recently been validated with
  electron bunches at SLAC: 
%
  J.~Belz {\it et al.},
  astro-ph/0510375.

\bibitem{GQRS}
  R.~Gandhi, C.~Quigg, M.~H.~Reno and I.~Sarcevic,
  Phys.\ Rev.\ D {\bf 58}, 093009 (1998).


\bibitem{sigma20}
  A.~Z.~Gazizov and S.~I.~Yanush,
  Phys.\ Rev.\ D {\bf 65}, 093003 (2002);

  M.~H.~Reno, I.~Sarcevic, G.~Sterman, M.~Stratmann and W.~Vogelsang,
  in {\it Proc. of the APS/DPF/DPB Summer Study on the Future of
  Particle Physics (Snowmass 2001) } ed. N.~Graf, eConf {\bf C010630},
  P508 (2001), 
  hep-ph/0110235;

  R.~Basu, D.~Choudhury and S.~Majhi,
  JHEP {\bf 0210}, 012 (2002);

  J.~Jalilian-Marian,
  Phys.\ Rev.\ D {\bf 68}, 054005 (2003),
  [Erratum-ibid.\ D {\bf 70}, 079903 (2004)];

  R.~Fiore, L.~L.~Jenkovszky, A.~Kotikov, F.~Paccanoni, A.~Papa and
  E.~Predazzi, 
  Phys.\ Rev.\ D {\bf 68}, 093010 (2003);

  M.~V.~T.~Machado,
  Phys.\ Rev.\ D {\bf 71}, 114009 (2005).


\bibitem{saturation}
  L.~V.~Gribov, E.~M.~Levin and M.~G.~Ryskin,
  Phys.\ Rept.\  {\bf 100}, 1 (1983);

  A.~H.~Mueller and J.~W.~Qiu,
  Nucl.\ Phys.\ B {\bf 268}, 427 (1986).


\bibitem{miniBH}
  L.~Anchordoqui and H.~Goldberg,
  Phys.\ Rev.\ D {\bf 65}, 047502 (2002);

  L.~A.~Anchordoqui, J.~L.~Feng, H.~Goldberg and A.~D.~Shapere,
  Phys.\ Rev.\ D {\bf 65}, 124027 (2002).


\bibitem{branewraps}
  E.~J.~Ahn, M.~Cavaglia and A.~V.~Olinto,
  Phys.\ Lett.\ B {\bf 551}, 1 (2003);

  P.~Jain, S.~Kar and S.~Panda,
  Int.\ J.\ Mod.\ Phys.\ D {\bf 12}, 1593 (2003);

  L.~A.~Anchordoqui, J.~L.~Feng and H.~Goldberg,
  Phys.\ Lett.\ B {\bf 535}, 302 (2002).


\bibitem{EWinstanton}
  H.~Aoyama and H.~Goldberg,
  Phys.\ Lett.\ B {\bf 188}, 506 (1987);

  A.~Ringwald,
  Nucl.\ Phys.\ B {\bf 330}, 1 (1990);

  O.~Espinosa,
  Nucl.\ Phys.\ B {\bf 343}, 310 (1990);

  L.~D.~McLerran, A.~I.~Vainshtein and M.~B.~Voloshin,
  Phys.\ Rev.\ D {\bf 42}, 171 (1990);

  P.~B.~Arnold and M.~P.~Mattis,
  Phys.\ Rev.\ D {\bf 42}, 1738 (1990);

  V.~V.~Khoze and A.~Ringwald,
  Phys.\ Lett.\ B {\bf 259}, 106 (1991);

  S.~Y.~Khlebnikov, V.~A.~Rubakov and P.~G.~Tinyakov,
  Nucl.\ Phys.\ B {\bf 350}, 441 (1991);

  A.~Ringwald,
  Phys.\ Lett.\ B {\bf 555}, 227 (2003);

  F.~Bezrukov, D.~Levkov, C.~Rebbi, V.~A.~Rubakov and P.~Tinyakov,
  Phys.\ Lett.\ B {\bf 574}, 75 (2003);

  A.~Ringwald,
  JHEP {\bf 0310}, 008 (2003);

  Z.~Fodor, S.~D.~Katz, A.~Ringwald and H.~Tu,
  Phys.\ Lett.\ B {\bf 561}, 191 (2003);

  T.~Han and D.~Hooper,
  Phys.\ Lett.\ B {\bf 582}, 21 (2004).


\bibitem{composite}
  G.~Domokos and S.~Nussinov,
  Phys.\ Lett.\ B {\bf 187}, 372 (1987);

  G.~Domokos and S.~Kovesi-Domokos,
  Phys.\ Rev.\ D {\bf 38}, 2833 (1988);

  J.~Bordes, H.~M.~Chan, J.~Faridani, J.~Pfaudler and S.~T.~Tsou,
  hep-ph/9705463;

  ibid.,
  Astropart.\ Phys.\  {\bf 8}, 135 (1998).


\bibitem{string}
  G.~Domokos and S.~Kovesi-Domokos,
  Phys.\ Rev.\ Lett.\  {\bf 82}, 1366 (1999);

  G.~Domokos, S.~Kovesi-Domokos, W.~S.~Burgett and J.~Wrinkle,
  JHEP {\bf 0107}, 017 (2001);


\bibitem{extradimensions}
  S.~Nussinov and R.~Shrock,
  Phys.\ Rev.\ D {\bf 59}, 105002 (1999);

  P.~Jain, D.~W.~McKay, S.~Panda and J.~P.~Ralston,
  Phys.\ Lett.\ B {\bf 484}, 267 (2000);

  M.~Kachelriess and M.~Plumacher,
  Phys.\ Rev.\ D {\bf 62}, 103006 (2000);

  L.~Anchordoqui, H.~Goldberg, T.~McCauley, T.~Paul, S.~Reucroft and
  J.~Swain,
  Phys.\ Rev.\ D {\bf 63}, 124009 (2001);

  A.~V.~Kisselev and V.~A.~Petrov,
  Eur.\ Phys.\ J.\ C {\bf 36}, 103 (2004).


\bibitem{disprelation}
  H.~Goldberg and T.~J.~Weiler,
  Phys.\ Rev.\ D {\bf 59}, 113005 (1999).


\bibitem{Icecubecs}
  D.~Hooper,
  Phys.\ Rev.\ D {\bf 65}, 097303 (2002);

  L.~A.~Anchordoqui, Z.~Fodor, S.~D.~Katz, A.~Ringwald and H.~Tu,
  JCAP {\bf 0506}, 013 (2005);

  L.~Anchordoqui and F.~Halzen,
  hep-ph/0510389.


\bibitem{Augercs}
  L.~A.~Anchordoqui, J.~L.~Feng, H.~Goldberg and A.~D.~Shapere,
  Phys.\ Rev.\ D {\bf 66}, 103002 (2002);

  L.~Anchordoqui, T.~Han, D.~Hooper and S.~Sarkar,
  hep-ph/0508312;
  
  and the last reference in~\cite{Icecubecs}.

\bibitem{HERAcs}
  I.~Abt {\it et al.}  [H1 Collaboration],
  Nucl.\ Phys.\ B {\bf 407}, 515 (1993);

  T.~Ahmed {\it et al.}  [H1 Collaboration],
  Nucl.\ Phys.\ B {\bf 439}, 471 (1995);

  S.~Aid {\it et al.}  [H1 Collaboration],
  Phys.\ Lett.\ B {\bf 354}, 494 (1995);

  M.~Derrick {\it et al.}  [ZEUS Collaboration],
  Phys.\ Lett.\ B {\bf 316}, 412 (1993);

  M.~Derrick {\it et al.}  [ZEUS Collaboration],
  Z.\ Phys.\ C {\bf 65}, 379 (1995);

  M.~Derrick {\it et al.}  [ZEUS Collaboration],
  Phys.\ Lett.\ B {\bf 345}, 576 (1995);

  C.~Adloff {\it et al.}  [H1 Collaboration],
  Eur.\ Phys.\ J.\ C {\bf 21}, 33 (2001).

  The neutrino cross-section inferred from HERA data is discussed in 
  D.~A.~Dicus, S.~Kretzer, W.~W.~Repko and C.~Schmidt,
  Phys.\ Lett.\ B {\bf 514}, 103 (2001)

\bibitem{WIMPaccept}
  C.~Barbot, M.~Drees, F.~Halzen and D.~Hooper,
  Phys.\ Lett.\ B {\bf 563}, 132 (2003).


\bibitem{losses}
  P.~Lipari and T.~Stanev,
  Phys.\ Rev.\ D {\bf 44} (1991) 3543;

  P.~Antonioli, C.~Ghetti, E.~V.~Korolkova, V.~A.~Kudryavtsev and
  G.~Sartorelli,
  Astropart.\ Phys.\  {\bf 7}, 357 (1997);

  I.~A.~Sokalski, E.~V.~Bugaev and S.~I.~Klimushin,
  Phys.\ Rev.\ D {\bf 64}, 074015 (2001).


\bibitem{DRSS}
  S.~I.~Dutta, M.~H.~Reno, I.~Sarcevic and D.~Seckel,
  Phys.\ Rev.\ D {\bf 63}, 094020 (2001).


\bibitem{BS}
  E.~V.~Bugaev and Y.~V.~Shlepin,
  Phys.\ Rev.\ D {\bf 67}, 034027 (2003).


\bibitem{KLS}
  K.~S.~Kuzmin, K.~S.~Lokhtin and S.~I.~Sinegovsky,
  hep-ph/0412377.


\bibitem{Bertou1}
  X.~Bertou, P.~Billoir, O.~Deligny, C.~Lachaud and
  A.~Letessier-Selvon,
  Astropart.\ Phys.\  {\bf 17}, 183 (2002).


\bibitem{DHR05}
  S.~I.~Dutta, Y.~Huang and M.~H.~Reno,
  Phys.\ Rev.\ D {\bf 72}, 013005 (2005).


\bibitem{GQRS96}
  R.~Gandhi, C.~Quigg, M.~H.~Reno and I.~Sarcevic,
  Astropart.\ Phys.\  {\bf 5}, 81 (1996).


\bibitem{Aramo}
  C.~Aramo, A.~Insolia, A.~Leonardi, G.~Miele, L.~Perrone, O.~Pisanti
  and D.~V.~Semikoz,  
  Astropart.\ Phys.\  {\bf 23}, 65 (2005).


\bibitem{taupolzn}
  M.~Aoki, K.~Hagiwara, K.~Mawatari and H.~Yokoya,
  Nucl.\ Phys.\ B {\bf 727}, 163 (2005).


\bibitem{EUSO}
  ESA and EUSO Team, 2000, ESA/MSM-GU/2000.462/AP/RDA,
  \url{http://www.euso-mission.org};

   P.~Gorodetzky  [The EUSO Collaboration],
  Nucl.\ Phys.\ Proc.\ Suppl.\  {\bf 151}, 401 (2006);

  G.~D'Ali Staiti  [EUSO Collaboration],
  Nucl.\ Phys.\ Proc.\ Suppl.\  {\bf 136}, 415 (2004).


\bibitem{AJLclouds}
  T.~Abu-Zayyad, C.~C.~H.~Jui and E.~C.~Loh,
  Astropart.\ Phys.\  {\bf 21}, 163 (2004).


\bibitem{OWL}
  F.~W.~Stecker, J.~F.~Krizmanic, L.~M.~Barbier, E.~Loh,
  J.~W.~Mitchell, P.~Sokolsky and R.~E.~Streitmatter,
  Nucl.\ Phys.\ Proc.\ Suppl.\  {\bf 136C}, 433 (2004).


\bibitem{EEfluxes}

  O.~E.~Kalashev, V.~A.~Kuzmin, D.~V.~Semikoz and G.~Sigl,
  Phys.\ Rev.\ D {\bf 66}, 063004 (2002);

  D.~V.~Semikoz and G.~Sigl,
  JCAP {\bf 0404}, 003 (2004).


\bibitem{WB}

  E.~Waxman and J.~N.~Bahcall,
  Phys.\ Rev.\ D {\bf 59}, 023002 (1999);

  ibid.,
  Phys.\ Rev.\ D {\bf 64}, 023002 (2001).

  A modification of the WB flux, accommodating the apparent
  lower-energy crossover from galactic to extra-galactic cosmic-rays,
  is given in 

  M.~Ahlers, L.~A.~Anchordoqui, H.~Goldberg, F.~Halzen, A.~Ringwald
  and T.~J.~Weiler, 
  Phys.\ Rev.\ D {\bf 72}, 023001 (2005).

\bibitem{BZflux}
  G.~T.~Zatsepin,
  Phys.\ Lett.\ B {\bf 28}, 423 (1969);

 R.~Engel, D.~Seckel and T.~Stanev,
  Phys.\ Rev.\ D {\bf 64}, 093010 (2001).


\bibitem{waterenhanced}
The enhancement of the UAS rate over the ocean relative to over land
has been noted in

  D.~Fargion, P.~G.~De Sanctis Lucentini and M.~De Santis,
  Astrophys.\ J.\  {\bf 613}, 1285 (2004).


\bibitem{Feng1}
  J.~L.~Feng, P.~Fisher, F.~Wilczek and T.~M.~Yu,
  Phys.\ Rev.\ Lett.\  {\bf 88}, 161102 (2002).


\bibitem{HEtaus}
  J.~J.~Tseng, T.~W.~Yeh, H.~Athar, M.~A.~Huang, F.~F.~Lee and
  G.~L.~Lin,
  Phys.\ Rev.\ D {\bf 68}, 063003 (2003);

  N.~Gupta,
  Phys.\ Rev.\ D {\bf 68}, 063006 (2003);

  J.~Jones, I.~Mocioiu, M.~H.~Reno and I.~Sarcevic,
  Phys.\ Rev.\ D {\bf 69}, 033004 (2004).


\bibitem{MPPAuger} 
  G.~Miele, S.~Pastor and O.~Pisanti,
  astro-ph/0508038.


\bibitem{Zas}
  E.~Zas,
  New J.\ Phys.\  {\bf 7}, 130 (2005).


\bibitem{Bottai1}
  S.~Bottai and S.~Giurgola,
  Astropart.\ Phys.\  {\bf 18}, 539 (2003).


\bibitem{FargionRate}
  D.~Fargion, M.~Khlopov, R.~Konoplich, P.~G.~De Sanctis Lucentini,
  M.~De Santis and B.~Mele, 
  Recent Res.\ Dev.\ Astrophys. {\bf 1}, 395 (2003).

  D.~Fargion, P.~G.~De Sanctis Lucentini, M.~De Santis and M.~Grossi,
  Astrophys.\ J.\  {\bf 613}, 1285 (2004).


\end{thebibliography}
\end{document}